\documentclass{article}

\usepackage{natbib}
\usepackage{epubtk}
\usepackage{graphicx}
\usepackage{amsbsy}
\usepackage{amssymb}
\usepackage{amsmath}
\usepackage{bm}
\usepackage{upgreek}
\usepackage{xspace}

\showlistoftablesfalse

\makeatletter
\DeclareRobustCommand\citepos
  {\begingroup
   \let\NAT@nmfmt\NAT@posfmt
   \NAT@swafalse\let\NAT@ctype\z@\NAT@partrue
   \@ifstar{\NAT@fulltrue\NAT@citetp}{\NAT@fullfalse\NAT@citetp}}

\let\NAT@orig@nmfmt\NAT@nmfmt
\def\NAT@posfmt#1{\NAT@orig@nmfmt{#1's}}
\makeatother

\newcommand{\approxprop}{\mathrel{\vcenter{
  \offinterlineskip\halign{\hfil$##$\cr
    \propto\cr\noalign{\kern2pt}\sim\cr\noalign{\kern-2pt}}}}}
\newcommand{\derivp}[2]{\frac{\partial #1}{\partial #2}}

\newcommand{\eq}[1]{Eq.~(\ref{#1})}
\newcommand{\rd}[1]{\,\mathrm d#1}
\renewcommand*\vec[1]{\ensuremath{\boldsymbol{#1}}}
\def\tens#1{\ensuremath{\mathbf{#1}}}
\def\etal{\textit{et~al.}\xspace}
\def\ob{\overline}
\def\tdel{\hat\delta}  
\def\Ldel{\updelta}    

\def\di{\partial_i}
\def\dj{\partial_j}

\def\bx{\vec{\xi}}

\def\hseq{\hspace{20pt}}
\def\hsdw{\hspace{\displaywidth minus1fil}}

\def\mut{\tilde\mu}

\def\dror{\frac{\delta r}{ r_0}}
\def\dtot{\frac{\delta T}{ T_0}}
\def\dpop{\frac{\delta p}{ p_0}}
\def\dpgopg{\frac{\delta p_{\mathrm{g}}}{ p_{\mathrm{g0}}}}
\def\dptopt{\frac{\delta p_{\mathrm{t}}}{ p_{\mathrm{t0}}}}

\def\ddod{\frac{\delta\rho}{\rho_0}}

\def\dkok{\frac{\delta\varkappa}{\varkappa_0}}

\begin{document}

\title{Interaction Between Convection and Pulsation}

\author{\epubtkAuthorData{G{\"u}nter Houdek}{%
Stellar Astrophysics Centre \\
Aarhus University\\
8000 Aarhus C, Denmark}{%
hg@phys.au.dk}{%
http://phys.au.dk/~hg62/}%
\and
\epubtkAuthorData{Marc-Antoine Dupret}{%
Institut d'Astrophysique et de G{\'e}ophysique\\
Universit{\'e} de Li{\`e}ge\\
4000 Li{\`e}ge, Belgium}{%
MA.Dupret@ulg.ac.be}{%
http://www.astro.ulg.ac.be/~dupret/}
}

\date{}
\maketitle


\begin{abstract}
This article reviews our current understanding of modelling convection 
dynamics in stars. Several semi-analytical time-dependent convection models
have been proposed for pulsating one-dimensional stellar structures 
with different formulations for
how the convective turbulent velocity field couples with the global stellar oscillations. 
In this review we put emphasis on two, widely used, time-dependent convection 
formulations for estimating pulsation properties in one-dimensional stellar models. 
Applications to pulsating stars 
are presented with results for oscillation properties, such as the effects
of convection dynamics on the oscillation frequencies, or the stability of 
pulsation modes, in classical pulsators and in stars supporting solar-type
oscillations.
\end{abstract}

\epubtkKeywords{Stellar convection, Time-dependent convection, Mode physics}

\newpage
\tableofcontents
\newpage

\section{Introduction}
\label{sec:introduction}

Transport of heat (energy) and momentum by turbulent convection is a phenomenon that 
we experience on a daily basis, such as the boiling of water in a kettle, the 
circulation of air inside a non-uniformly heated room, or the formation of 
cloud patterns.
Convection may be defined as fluid (gas) motions 
brought about by temperature differences with gradients in any 
direction \citep{Koschmieder93}. 
It is not only important to engineering applications but also to a 
wide range of astrophysical flows, such as in galaxy-cluster plasmas, interstellar medium,
accretion disks, supernovae, and during several evolutionary stages of all stars
in the Universe.  
The transport of turbulent fluxes by convection is mutually affected by other physical
processes, including radiation, rotation, and any kind of mixing processes.
In stars turbulent convection affects not only their structure and evolution but 
also any dynamical processes with characteristic time scales 
that are similar to the characteristic time scale of convection in the overturning 
stellar layers. One such important process is stellar pulsation, the study of which
has become the field of asteroseismology. Asteroseismology and, when applied to the Sun,
helioseismology, is now one of the most important diagnostic tools for testing and improving the 
theory of stellar structure and evolution by analysing the observed pulsation properties. 
It is, therefore, the aim of this review to provide an up-to-date account on the most widely used
stellar convection models with emphasis on the formalisms that describe the interaction of
the turbulent velocity field with the stellar pulsation.
\smallskip

The temperature in a star is determined by the
balance of energy and its gradient depends on the details how energy 
is transported throughout the stellar interior. Red giants and
solar-like stars exhibit substantial convection zones in the 
outer stellar layers, which affect the properties of the
oscillation modes such as the oscillation frequencies and mode stability.
Among the first problems of this nature was the modelling 
of the red edge of the classical instability strip in the 
Hertzsprung--Russell diagram which, for intermediate-mass stars with about 
$1.5\mbox{\,--\,}2.0\,M_{\odot}$, is located approximately at surface temperatures
between 7200\,--\,6600~K. 
The first pulsation calculations 
of classical pulsators without any pulsation-convection modelling 
predicted red edges which were much too cool and which were at 
best only neutrally stable. What followed, were several attempts 
to bring the theoretically predicted location of the red edge in 
better agreement with the observed location by using time-dependent 
convection models in the pulsation analyses 
\citep[e.g.,][]{Deupree77a, BakerGough79, Gonczi82b, Stellingwerf84}. 
Later, several authors, e.g., 
\citet{BonoEtal95, BonoEtal99}, \citet{Houdek97, Houdek00}, 
\citet{XiongDeng01, XiongDeng07}, \citet{DupretEtal05b, DupretEtal05c}
were successful to model the red edge of the classical instability strip,
and mode lifetimes in stars supporting solar-like oscillations
\citep[e.g.,][]{Gough80, Balmforth92a, HoudekEtal99, XiongEtal00, HoudekGough02,
DupretEtal04a, ChaplinEtal05, DupretEtal06a, Houdek06, DupretEtal09, BelkacemEtal12}.

Thermal heat transport in convective regions is governed by turbulent motion of the
underlying fluid or gas. To determine the average of vertical
velocity, temperature and momentum fluctuations, the full structure of the turbulent 
flow is needed. This is until today not a tractable theoretical problem
without the introduction of some hypothetical assumptions in order
to close the system of equations describing the turbulent flow.
Such closure models can be classified basically into four
categories: (i) `algebraic models', including the mixing-length
approach \citep[e.g.,][]{Prandtl25, Vitense53, BohmVitense58}, 
(ii) `one-equation models', which use a modified turbulent kinetic 
energy equation with high-order moments closed
approximately by means of a locally defined mixing length 
\citep[e.g.,][]{Rodi76, Stellingwerf82}, (iii) `two-equation models', 
such as the $K-\epsilon_{\mathrm{t}}$ model, where $K$ denotes the turbulent 
kinetic energy and $\epsilon_{\mathrm{t}}$ the associated viscous dissipation 
of turbulent energy \citep[e.g.,][]{JonesLaunder72, JonesLaunder73}, 
and (iv) `Reynolds stress models', which use transport 
equations for all second-order moments (typically five) including 
the turbulent fluxes of heat and momentum, and appropriate approximation 
for the third-order moments to close the equations 
\citep[e.g.,][]{KellerFriedmann24, Rotta51, Castor68, Xiong77, Canuto92, Grossman96, 
CanutoDubovikov98, Kupka99, MontgomeryKupka04, XiongDeng07}.
 
Theories based on the mixing-length formalism \citep{Prandtl25}
still represent the main method for computing the
stratification of convection zones in stellar models.
An alternative convection formulation, based on the 
Eddy-Damped Quasi-Normal Markovian approximation to 
turbulence \citep[e.g.,][]{Orszag77}, was introduced by 
\citet{CanutoMazzitelli91} which, however, still requires a (local)
mixing length for estimating the convective heat (enthalpy) flux.
The Eddy-Damped Quasi-Normal Markovian approximation is characterized as 
a two-equation model and is sometimes referred to as two-point closure, 
because it describes correlations of two different points in space, 
or two different wave numbers $\bm{k}$ and $\bm{k^\prime}$ in
Fourier space. Although two-equation models have a reasonable
degree of flexibility, they are restricted by the assumption
of a scalar turbulent viscosity and that the
stresses are proportional to the rate of mean strain. The
Reynolds stress models are, in principle, free of these restrictions
and were discussed, for example, by \citet{Xiong89} and \citet{Canuto92, Canuto93} 
for the application in stellar convection. Xiong's model was applied 
successfully to various types of pulsating star, and Canuto's model was 
applied to non-pulsating stars with relatively shallow surface convection
zones \citep{KupkaMontgomery02, MontgomeryKupka04}.

Time-dependent convection models are required to describe the interaction between
the turbulent velocity field and the oscillating stellar background.
Semi-analytical models for pulsating stars were proposed, for example, by
\citet{Schatzman56}, \citet{Gough77a}, \citet{Unno67}, \citet{Xiong77}, 
\citet{Stellingwerf82}, \citet{Gonczi82a}, \citet{Kuhfuss86}, and \citet{Grossman96}.

The present unprecedented computer revolution enables us to
perform fully hydrodynamical simulations of large-scale
turbulent flows (large eddy simulation) of stellar surface 
convection \citep[e.g.,][]{SteinNordlund89, SteinNordlund00,
NordlundStein96, TrampedachEtal98, KimEtal96, ChanSofia96, 
FreytagEtal02, RobinsonEtal04, WedemeyerEtal04,MuthsamEtal10,
MagicEtal13, TrampedachEtal13, TrampedachEtal14a, MagicEtal15}.
A review of three-dimensional (3D) hydrodynamical simulations of the Sun, 
together with their shortcomings, was presented by \citet{Miesch05}.
Such numerical simulations represent a fruitful tool for 
investigating the accuracy and hence the field of application
of phenomenological prescriptions of convection such as the
mixing-length approach.

In this review, we summarize the two time-dependent convection
models by \citet{Gough65, Gough77a} and \citet{Unno67, Unno77} for 
estimating stellar stability properties
in classical pulsators and solar-type stars. In Section~\ref{sec:hydrodynamical_equations}, we 
start from the equations of fluid motion to derive first the mean and
fluctuating equations within the commonly adopted Reynolds separation
approach. Section~\ref{sec:local_mlm} discusses first the time-dependent
convection equations by \citet{Gough65, Gough77a} and \citet{Unno67, Unno77} 
for radially pulsating stellar envelopes, followed by a summary of
\citepos{Gough77b} nonlocal equations,
before embarking on a discussion on a generalization of 
\citepos{Unno67} 
model to nonradial stellar oscillations by \citet{GabrielEtal75}
and \citet{GrigahceneEtal05}.
A summary of Reynolds stress models adopted to stellar convection
is provided in Section~\ref{sec:reynolds_stress_models}. 
Applications of the two time-dependent convection models by \citet{Gough77a, Gough77b} and 
\citet{GrigahceneEtal05} are provided in
Sections~\ref{sec:frequency_effects}, \ref{sec:stability}, and
\ref{sec:multi-colour}, starting 
with the role of convection dynamics on the oscillation frequencies, the 
so-called surface effects, followed by a summary of our current understanding 
of mode physics in classical pulsators and stars supporting solar-like oscillations.
Final remarks and prospects are given in Section~\ref{sec:prospects}.
 
\newpage
\section{Hydrodynamical Equations}
\label{sec:hydrodynamical_equations}

For simplicity we neglect any symmetry-breaking agents such as
rotation or magnetic fields and adopt spherical geometry for which, for example, 
the velocity field $\vec{u}=(u_{\mathrm{r}}, u_\varphi, u_\theta)$.
In this approximation the fluid conservation equations for mass, momentum and
thermal energy equation, using vector notation, are 
\citep{LedouxWalraven58, LandauLifshitz59, Batchelor67}

\begin{equation}
\frac{\partial \rho }{\partial t}+\nabla \cdot \left( \rho \vec{v}\right) =0\,,
\label{eq4}
\end{equation}

\begin{equation}
\frac{\partial \left( \rho \vec{v}\right) }{\partial t}+\nabla \cdot \left(
\rho \vec{vv}\right) = \rho \vec{g} +\nabla \cdot 
   \bm{\uptau}
\,,
\label{eq5}
\end{equation}

\begin{equation}
\frac{\partial (\rho e)}{\partial t}+\nabla \cdot \left( \rho e\vec{v}\right)-
   \bm{\uptau}
{\,:\,}
\nabla \vec{v} = \rho \varepsilon - \nabla \cdot \vec{F}_{\mathrm{R}} \,,
\label{eq6}
\end{equation}
%
%
where $\rho$ is density, $\vec{v}$ the velocity vector,
$\vec{g}=(-g,0,0)$ with $g$ being the magnitude of the gravitational acceleration, 
$\bm{\uptau}
 := -p\,\tens{I}\,+\,\bm{\upsigma}$,
with $p$ being the sum of the gas and radiative pressures
($\tens{I}$ is the unity or identity matrix), 
and $\bm{\upsigma}$       
is the sum of the gaseous and radiative deviatoric (viscous) stress tensors;
$e$ is the specific internal energy, $\varepsilon $ is the rate of energy generation per
unit mass by nuclear reactions and $\vec{F}_{\mathrm{R}}$ is the
radiative flux. The dissipation of energy by internal stress and (reversible) 
interchange with strain energy is indicated by 
$\bm{\uptau}\,:\,\nabla \vec{v}=-p\,\nabla\cdot\vec{v}+\bm{\upsigma}:\nabla\vec{v}$, 
which is the dyadic notation for $\uptau_{ij}v_{i,j}$. 

\subsection{Mean equations}
\label{sec:meaneq}

We follow the standard Reynolds approach and separate all variables 
into an average (or mean) part, and into a fluctuating part. 
Thus 

\begin{equation}
y=\overline{y}+y',  \label{eq1}
\end{equation}

\begin{equation}
\vec{v}=\vec{U}+\vec{u},  \label{eq2}
\end{equation}
where $y$ is any of the variables $\rho $, $p$, $T$, etc., and 
$\overline{y}$ and $\vec{U}$ are 
the appropriately averaged mean values (i.e., typically horizontal averages). 
The fields $y'$ and $\vec{u}$ are convective (Eulerian) fluctuations.
The separation of the velocity into mean and fluctuating components must 
be carried out with some care. Because there is no mean transport of mass 
(mass flux) across layers with constant radius, we adopt
\citep[e.g.,][]{LedouxWalraven58, Unno67, Gough69, GabrielEtal75, Gough77a, 
UnnoXiong90, GrigahceneEtal05}
\begin{equation}
\ob{\rho\vec{u}}=0\,,\quad\ob{\vec{u}}\ne0\,.
\label{emf}
\end{equation}
{We further define the time derivative following the pulsational motion by
\citep[e.g.,][see also Section~\ref{sec:mean_boussinesq_equations}]{LedouxWalraven58,Unno67, Gough69}} 
\begin{equation}
\frac{\rd}{\rd t}=\frac{\partial }{\partial t}+\vec{U}\cdot
\nabla . 
\label{eq3}
\end{equation}
The mean equations of mass and momentum conservation are obtained from
taking averages of the Eqs.~(\ref{eq4}) and (\ref{eq5}). 
{Stellar turbulence is characterized by high Reynolds numbers $R_{\mathrm{e}}$, typically
in the order of $R_{\mathrm{e}}\geq10^{12}$. We, therefore, can neglect the viscous stress 
tensor $\ob{\bm{\upsigma}}$} in the momentum equation~(\ref{eq5})
and obtain for the averaged continuity and momentum equations

\begin{equation}
\frac{\partial \overline{\rho }}{\partial t}+\nabla \cdot \left( \overline{%
\rho }\vec{U}\right) =0 ,  \label{eq8}
\end{equation}

\begin{equation}
\overline{\rho }\frac{\rd\vec{U}}{\rd t}= \overline{\rho }%
\,\ob{\vec{g}} -\nabla 
(\ob{p}+p_{\mathrm{t}})
-\nabla \cdot 
\bm{\upsigma}_\mathrm{t}\,,  
\label{eq13}
\end{equation}
in which we neglected the perturbation to the gravitational acceleration. 
In the literature \citep[e.g.,][]{LedouxWalraven58, UnnoXiong90} it is common 
to adopt Boussinesq's suggestion for representing the turbulent stresses 
in a similar way as for the viscous stresses, i.e., to separate the
Reynolds stress tensor $\overline{\rho\,\vec{uu}}$ 
into an isotropic component $\hat p_{\mathrm{t}}$,
which is typically called the turbulent pressure,
and into a nonisotropic (deviatoric) part $\bm{\upsigma}_\mathrm{t}$, 
\begin{equation} 
\ob{\rho \vec{uu}}:=\tilde p_{\mathrm{t}}\,{ \tens{I}}+{\bm{\upsigma}_\mathrm{t}}\,.  
\label{eq11a}
\end{equation}
Here $\tilde p_{\mathrm{t}}:=\ob{\rho|\vec{u}|^{2}}/3$, 
and $\sigma_{ij,\mathrm{t}}:=
\mu_{\mathrm{t}}[(2/3)\partial_k\ob{v_k}\delta_{ij}-\partial_i\ob{v_j}-\partial_j\ob{v_i}]$,
where $\delta_{ij}$ is the Kronecker delta, 
and $\mu_{\mathrm{t}}$ represents a scalar turbulent (eddy) viscosity.
Equation~(\ref{eq11a}) is, however, strictly valid only for isotropic turbulence but 
the convective velocity field in stars is, in general, predominantly 
anisotropic \citep[e.g.,][]{Houdek12}. In this review we shall therefore 
follow \citet{Gough77a} and parametrize the anisotropy of the turbulent velocity field
$\vec{u}=(u_{\mathrm{r}}, u_\theta, u_\phi)$ by the 
anisotropy parameter
\begin{equation}
\label{anisotropyeq}
\ob{\Phi}:=\frac{\overline{\rho |\vec{u}|^{2}}}
{\overline{\rho u_{\mathrm{r}}^{2}}}\,,
\end{equation}
and define the turbulent pressure 
\begin{equation}
p_{\mathrm{t}}:=\ob{\rho{u_{\mathrm{r}}u_{\mathrm{r}}}}
\label{eq11}
\end{equation}
as the $(r,r)-$component of the Reynolds stress tensor $\ob{\rho \vec{uu}}$.
Note that $\Phi=3$ represents an isotropic velocity field. 

The mean equation for the turbulent kinetic energy is obtained by multiplying
Eq.~(\ref{eq5}) by $\vec{v}$ and Eq.~(\ref{eq13}) by $\vec{u}$, followed by
 averaging the difference between the resulting expressions. The outcome is
\begin{eqnarray}
\overline{\rho }\frac{\rd}{\rd t}
\left( \frac{1}{2}\frac{\overline{\rho \vec{u}^{2}}}{\overline{\rho }}\right)  
&=&-\overline{\rho \vec{uu}}:\nabla \vec{U} 
   -\overline{\vec{u}\cdot \nabla p}
   -\frac{1}{2}\nabla \cdot \left( \overline{\rho \vec{u}^{2}\vec{u}}\right) 
   -\overline{\bm{\upsigma}:\nabla \vec{u}}  
   +\nabla \cdot \overline{\left( \bm{\upsigma} \cdot \vec{u}\right) }\,,
\label{ecin0}
\end{eqnarray}
where the last term on the right-hand side is small and can therefore be neglected
\citep[e.g.,][]{LedouxWalraven58}. 
The first two terms on the right-hand side represent respectively the production of turbulent 
kinetic energy from the mean motion and from the gravitational potential energy. 
The third term on the right-hand side is the divergence of the turbulent kinetic energy flux and
the fourth term is the dissipation of kinetic energy (per unit volume) into heat.  
We emphasize here a significant difference with 
other studies. In this review we adopt Eq.~(\ref{emf}) for the averaging process, which 
leads to no buoyancy term in Eq.~(\ref{ecin0}). If, however, instead $\overline{\vec{u}}=0$ 
is assumed, the additional buoyancy term,
$\overline{\rho'\vec{u}\cdot\vec{g}}$, appears on the right-hand side of the kinetic energy 
equation \citep[e.g.,][]{Canuto92}, where $\vec{g}$ is the gravitational acceleration.

The mean equation of the thermal energy conservation is obtained by taking the average
of Eq.~(\ref{eq6}):
\begin{equation}
\ob{\rho}\frac{\mathrm{d\,\ob{e}}}{\rd t}
+\ob{p}\,\nabla\cdot\vec{U}
=\ob{\rho\varepsilon}
-\nabla\cdot\left(\ob{\vec{F_{\mathrm{r}}}}+\ob{\rho { h}\vec{u}}\right)
+{\ob{\vec{u}\cdot\nabla p}}
+\ob{\bm{\upsigma}:\nabla \vec{u}}\,,
\label{tec0}
\end{equation}
where $h=e+p/\rho$ is the (specific) enthalpy.
Essentially, all semi-analytical stellar convection models adopt the Boussinesq 
approximation to the equations of motion \citep{SpiegelVeronis60}. 
This approximation is valid for 
fluids for which the vertical dimension of the fluid is much less than
any scale height, and the motion-induced density and pressure fluctuations
must not exceed, in order of magnitude, the mean values of these quantities,
i.e., the Mach number of the fluid, which is the ratio of the fluid velocity
over the adiabatic sound speed, must remain small.
Part of the Boussinesq approximation is to neglect squares of
fluctuating thermodynamic quantities, 
and neglecting pressure 
fluctuations $p^\prime$ compared to $\rho^\prime$ or $e^\prime$
(see also Section~\ref{sec:gough_unno_mlm}).
These conditions are believed not to be satisfied everywhere in stars, in particular in
the superadiabatic boundary layer, yet we shall adopt it here in this review. Within the
Boussinesq approximation the mean equations for the conservation of mass and momentum
are identical to Eqs.~(\ref{eq8}) and (\ref{eq13}) respectively. The mean equations
for the conservation of turbulent kinetic energy (\ref{ecin0}), thermal energy
(\ref{tec0}) and anisotropy of the turbulent velocity field (\ref{anisotropyeq})
become:
 
\begin{eqnarray}
\frac{\overline{\rho }}{2}\frac{\mathrm{d\,\overline{|\vec{u}|^{2}}}}{\rd t}
 &=&-\ob{\rho \vec{uu}}{\,:\,}\nabla \vec{U}
    -\ob{\vec{u}}\cdot \nabla \ob{p}
    -\ob{\rho}\epsilon_{\mathrm{t}} \,,
\label{ecin}
\end{eqnarray} 

\begin{equation}
\overline{\rho }\,\overline{T}\,\frac{\mathrm{d}\overline{s}}{\rd t}%
= \ob{\rho }\,\ob{\varepsilon }%
 -\nabla \cdot \left( \ob{\vec{F}}_{\mathrm{R}}+\vec{F}_{\mathrm{c}}\right)
 +\ob{\vec{u}}\cdot \nabla\ob{p} 
 +\ob{\rho }\,{\epsilon_{\mathrm{t}}} \,,
\label{eq18}
\end{equation}
\begin{equation}
\label{anisotropyeq2}
\ob{\Phi}:=\frac{\overline{|\vec{u}|^{2}}}
{\ob{u_{\mathrm{r}}^{2}}}\,,
\end{equation}
where the vector field ${\vec{F}}_{\mathrm{c}}$ is the convective heat 
(enthalpy $h$) flux 
\begin{equation}
{\vec{F}}_{\mathrm{c}}:=\ob{\left(p+\rho e\right)^\prime\vec{u}}=\ob{\rho h'\vec{u}},
\label{e:enthalpy-flux}
\end{equation}
which can be further simplified within the Boussinesq approximation to
\begin{equation}
\vec{F}_{\mathrm{c}}\simeq\overline{\rho }\,\ob{T}\:\ob{\vec{u}s^\prime}
=\ob{\rho}\,\ob{c_{\mathrm{p}}}\ob{\vec{u}T^\prime}\,, 
\label{fconv}
\end{equation}
($s$ is specific entropy, $T$ is temperature, and $c_{\mathrm{p}}$ the specific heat at constant pressure), and $\ob{\rho}\epsilon_{\mathrm{t}}=\ob{\bm{\upsigma}:\nabla \vec{u}}$
is the viscous dissipation of turbulent kinetic energy (per unit volume) 
into heat (sink term in the kinetic energy equation). 
For an incompressible 
Newtonian fluid the viscous dissipation of turbulent kinetic energy 
$\ob{\rho}\epsilon _{\mathrm{t}}=2\mu\ob{e_{ij}e_{ij}}$,
where $e_{ij}:=(\partial_ju_i+\partial_iu_j)/2$ is the fluctuating strain rate 
and $\mu$ is the (constant) molecular (dynamic) viscosity. The penultimate term on 
the right-hand side of the turbulent kinetic energy equation~(\ref{ecin}) is the work of the 
pressure gradient transforming gravitational potential energy into 
kinetic energy of turbulence (source term).
Both terms 
$-\ob{\vec{u}}\cdot\nabla\ob{p}-\ob{\rho}\epsilon_{\mathrm{t}}$ are also present with 
opposite sign in the mean thermal energy equation~(\ref{eq18}). For the stationary
($\partial/\partial t = 0$) equilibrium state with no mean flow ($\vec{U}=\vec{0}$) the
turbulent kinetic energy is constant. Consequently these
two terms vanish and are therefore neglected in the mean thermal energy
equation~(\ref{eq18}) in the Boussinesq approximation 
\citep[see also][]{SpiegelVeronis60}. 
We shall, however, see later that its perturbation due to oscillations may not
necessarily vanish, even at second order.
The turbulent kinetic-energy flux [third term on the right-hand side of 
Eq.~(\ref{ecin0})] is not necessarily small everywhere. According to
three-dimensional hydrodynamical simulations of the outer atmospheric layers in the Sun,
the kinetic energy flux can be as large as $\sim$~15\% of the total energy flux 
\citep[e.g.,][]{TrampedachEtal14a}, yet it is typically ignored in semi-analytical 
convection models. We follow the same approximation
and omit this term in Eq.~(\ref{ecin}).
An expression for the kinetic energy flux within the mixing-length approach was recently provided
by \citet{Gough12a}.

\subsection{Boussinesq mean equations for radially pulsating atmospheres}
\label{sec:mean_boussinesq_equations}

One of the first questions to ask is how one would go about the separation of the
velocity field into a component that is associated with the stellar pulsation and into
another component that is related to the convection. The answer is not necessarily
straightforward \citep[for a recent discussion see, e.g.,][$\mathsection$3.1]{AppourchauxEtal10}. 
This separation of the velocity field is probably best known for radial
pulsations, for which the horizontal motion is uniform (the convective motion is not). 
By adopting Eq.~(\ref{emf}) for averaging the horizontal motion of the convective
velocity field the radial pulsations can be separated in an (mathematically) obvious way
\citep[e.g.,][]{Gough69}, in which the small-scale convective Eulerian fluctuations 
($\vec{u}$) are advected by the large-scale radial Lagrangian motion 
($U$) of the pulsation. 

Below, we follow the discussion by \citet{Gough77a} and summarize the mean Boussinesq
equations for radial pulsations adopting a mixed Lagrangian-Eulerian coordinate system
($q_i$) defined in terms of spherical polar coordinates $(r,\theta,\varphi)$
\begin{equation}
(\rd{q_1}, \rd{q_2}, \rd{q_3})=(r\rd{\theta}, r\sin\theta\rd{\varphi},{\ob\rho}\rd{r})\,,
\label{eq:coordinates_ba}
\end{equation}
where $q_3=0$ at $r=0$, and an overbar means, as before, an instantaneous average 
over a spherical surface with constant $r$ (i.e., horizontal average). 
As already mentioned before, the introduction of Eq.~(\ref{emf}) in this coordinate
system has the property that there is no mean mass flux across 
a surface with $q_3=\,$constant \citep{LedouxWalraven58, Gough69, Gough77a} and this
coordinate system describes the large-scale pulsational motion of a fluid layer in a
Lagrangian frame of reference, whereas inside this moving layer the convective motion is
described in an Eulerian frame. The time derivative at constant (Eulerian) $q_i$ is then 
related to that in spherical polar coordinates by [see also Eq.~(\ref{eq3})] 
\begin{equation}
\left(\frac{\partial}{\partial t}\right)_{q_i}=
\left(\frac{\partial}{\partial t}\right)_{r,\theta,\phi}+\,\ob{\rho}\,U\,\frac{\partial}{\partial q_3}\,,
\label{e:Gough-A6}
\end{equation}
where $U=(\partial r/\partial t)_{q_3}$.
Within this adopted coordinate system, the mean equations for the radial component of
the momentum equation and the thermal energy equation in the Boussinesq 
approximation become with $\varepsilon=0$ and $q=q_3$ \citep{Gough77a}:
\begin{equation}
\derivp{}{q}({\ob p}+p_{\mathrm{t}})+(3-{\ob\Phi})\frac{p_{\mathrm{t}}}{r{\ob\rho}}=
-\frac{{\mathrm G}m}{r^2}-\frac{\partial^2r}{\partial t^2}\,,
\label{eq:mean_momentum_ba}
\end{equation}
\begin{equation}
\ob{c_{\mathrm{p}}}\derivp{\ob{T}}{t}-\frac{\ob{\tdel}}{\ob{\rho}}\derivp{\ob{p}}{t}=
-\frac{1}{r^2}\derivp{}{q}\left[r^2\left(\ob{F_3}+F_{\mathrm{c}}\right)\right]\,,
\label{eq:mean_thermal_energy_ba}
\end{equation}
\begin{equation}
\frac{1}{r^2}\derivp{}{q}\left(r^2\ob{F_3}\right)=
4\pi\ob{\varkappa}\left(\ob{B}-\ob{J}\right)\,,
\label{eq:mean_eddington_F}
\end{equation}
\begin{equation}
\derivp{\ob{J}}{q}=-\frac{3}{4\pi}\ob{\varkappa}\ob{F_3}\,,
\label{eq:mean_eddington_J}
\end{equation}
where
\begin{equation}
\ob{\Phi}=\frac{\ob{u_iu_i}}{\ob{w^2}}\,,
\label{eq:Phi}
\end{equation}
\begin{equation}
p_{\mathrm{t}}:=\ob{\rho u_3u_3}\simeq\ob\rho\,\ob{ww}
\label{eq:turbpressure}\,,\ 
\end{equation}
and
\begin{equation}
F_{\mathrm{c}}:=\ob{\rho u_3h^\prime}\simeq\ob{\rho}\,\ob{c_{\mathrm{p}}}\,\ob{wT^\prime}\,,
\label{eq:convective_heat_flux}
\end{equation}
are respectively the anisotropy parameter [see also Eq.~(\ref{anisotropyeq2})],
the $(r,r)$-component of the Reynolds stress tensor $\ob{\rho u_iu_j}\,$,
with $u_i=(u,v,w)$, 
and the convective heat flux in the Boussinesq approximation;
$\tdel:=-(\partial\ln\rho/\partial\ln T)_p$ is the isobaric 
expansion coefficient, and $\varkappa$ is the Rosseland mean opacity.

The second term on the left-hand side of \eq{eq:mean_momentum_ba} results 
from taking the horizontal average of the radial component of the nonlinear 
advection term $\rho\vec{v}\cdot\nabla\vec{v}$: with the definition of
the Reynolds stress tensor~(\ref{eq11a}) and velocity anisotropy~(\ref{eq:Phi})
the last term of Eq.~(\ref{eq13}),
$(\nabla\cdot\vec{\sigma}_{\mathrm{t}})_{\mathrm{r}}=(3-\ob{\Phi})p_{\mathrm{t}}/r$, assuming 
axisymmetric turbulence about the radial direction. 
From a physical point of view this term
arises because horizontal motion, in spherical coordinates, transfers momentum
in the radial direction, 
resulting from a difference in the net radial force between the horizontal component
of $\ob{\rho u_iu_j}\,$ for $\ob\Phi\neq3$, and the radial component of magnitude 
$p_\mathrm{t}$ for $\ob\Phi=3$ \citep{Gough77a}.

The radiative flux $\ob{\vec{F}}=(0, 0, \ob{F_3})$ is typically treated in the diffusion 
approximation to radiative transfer. Here we adopt the general, grey Eddington
approximation by \citet{UnnoSpiegel66}, given by Eqs.~(\ref{eq:mean_eddington_F})
and (\ref{eq:mean_eddington_J}), where $B$ is the Planck function, $J$ is the mean
intensity, and $\varkappa$ is the Rosseland-mean opacity. Note that in
\eq{eq:mean_eddington_J} one should actually use the 
Planck-mean opacity, which is the more appropriate mean for optically thin layers
\citep[e.g.,][]{Mihalas78}, instead of the Rosseland-mean opacity.
In radiative equilibrium the radiative flux has zero divergence and 
consequently $\ob{J}=\ob{B}$, reducing the Eddington approximation
(\ref{eq:mean_eddington_F})\,--\,(\ref{eq:mean_eddington_J})
to the diffusion approximation
\begin{equation}
\ob{F_3}=-\frac{4ac\ob{T}^3}{3\ob{\varkappa}}\derivp{\ob{T}}{q}\,,
\end{equation}
where $a$ is the radiation density constant and $c$ is the speed of light. 

The mean equations for a Boussinesq fluid in a radially pulsating star are 
Eqs.~(\ref{eq:mean_momentum_ba})-(\ref{eq:convective_heat_flux}),
supplemented by the continuity equation
\begin{equation}
\frac{\rd m}{\rd q}=4\pi r^2\,.
\end{equation}
Note that mass $m$ is a Lagrangian coordinate (i.e., independent of time) in 
a radially moving atmosphere.

In the mean equations, the turbulent pressure (Reynolds stress) $p_{\mathrm t}$ and
convective heat flux $F_{\mathrm c}$ are the quantities that must be determined from
the equations for the convective fluctuations. To solve these equations, a model
for the convective turbulence is required, which is discussed in the next
section.

\newpage
\section{Time-dependent Mixing-Length Models}
\label{sec:local_mlm}

\subsection{Introduction}
\label{sec:intro_local_mlm}

The simplest closure model of turbulence is the early one by
\citet{Boussinesq1877}, who suggested that turbulent flow could be 
considered as having an enhanced viscosity, a turbulent (or eddy)
viscosity $\nu_\mathrm{t}$. Boussinesq assumed $\nu_\mathrm{t}$ to be constant,
in which case the equations of mean motion become identical in 
structure with those for a laminar flow. This assumption, however, 
does become invalid near the convective boundary layers, where
the turbulent fluctuations vanish, and so does $\nu_\mathrm{t}$,
at least in a local convection model.

The simplest turbulence model able to account for the variability
of the turbulent mixing with the use of only one empirical constant 
is the mixing-length idea, introduced independently by
\citet{Taylor15} and \citet{Prandtl25}. Based on Boussinesq's approach
and considering the turbulent fluid decomposed into so-called 
eddies, parcels or elements, Prandtl obtained, for the case of 
shear flow, from dimensional reasoning, an expression for the turbulent 
viscosity or exchange coefficient of momentum (``Austauschkoeffizient'').
This expression is in the form of a product of the velocity 
fluctuation perpendicular (transverse) to the mean motion of 
the turbulent flow and the mixing length $\ell$. 
The mixing length is characterized 
by the distance in the transverse direction which must be covered
by a fluid parcel travelling with its original mean velocity in
order to make the difference between its velocity and the velocity
in the new layer equal to the mean transverse fluctuation in the
turbulent flow. Inherent in this physical picture is the major
assumption that the momentum of the turbulent parcel is assumed
to be constant along the travel distance $\ell$, which is
analogous to neglecting the streamwise pressure forces and 
viscous stresses. Prandtl's concept of a mixing length may
be compared, up to a certain point, with the mean free path
in the kinetic theory of gases. A somewhat different result
was obtained by \citet{Taylor32} who assumed that the rotation
(vorticity) during the transverse motion of the parcel remains
constant, yielding a mixing length which is larger by a factor
$\sqrt 2$ compared with Prandtl's momentum-transfer picture.

Neglecting rotation and magnetic fields, thermal heat transport 
in stars corresponds to the case of free convection where there is 
no externally imposed velocity scale as in shear flow. Hence, it is 
necessary to consider the dynamics of the turbulent elements in 
greater detail. The imbalance between buoyancy forces, pressure 
gradients and nonlinear advection processes causes the turbulent
elements to accelerate during their existence. Ignoring different
combinations of these processes and approximating the remaining terms
in different ways, various phenomenological models can be established.
In the astrophysical community basically two physical pictures
have emerged, which were first applied to stellar convection by
\citet{Biermann32, Biermann38, Biermann43, Biermann48} and 
\citet{Siedentopf33, Siedentopf35}. 
In both physical pictures the turbulent element is considered 
as a convective cell with a characteristic vertical length $\ell$ 
as illustrated in Figure~\ref{fig:conv_cell}.

\epubtkImage{conv-cell_fig1.png}{%
\begin{figure}[htb]
  \centerline{\includegraphics[scale=1.00]{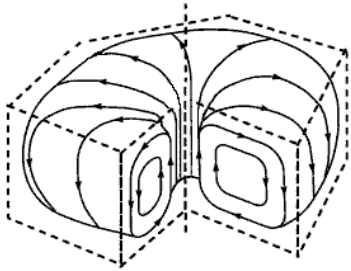}}
  \caption{Sketch of an overturning hexagonal (dashed lines) 
               convective cell with vertical extent $\ell$. Near the 
               centre the gas raises from the 
               hot bottom to the cooler top (surface) where it moves nearly 
               horizontally towards the edges, thereby loosing heat. The 
               cooled gas then descends along the edges to close the circular
               flow. Arrows indicate the direction of the flow pattern.
  Image adapted from \citet{Swenson97}.}
  \label{fig:conv_cell}
\end{figure}}

The first physical picture interprets the turbulent flow by direct 
analogy with kinetic gas theory. The motion is not steady and
one imagines the overturning convective element to accelerate from rest followed 
by an instantaneous breakup after the element's lifetime. 
Thus the nonlinear advection terms are neglected in the convective 
fluctuation equations but are taken to be responsible for the creation 
and destruction of the convective eddies \citep{Spiegel63, Gough77a, Gough77b}. 
By retaining only the acceleration terms the equations become linear
and the evolution of the fluid properties carried by 
the turbulent parcels can be approximated by linear growth rates.
The mixing length $\ell$ enters in the calculation of the 
eddy's survival probability
for determining the convective heat and momentum fluxes (see Appendix~\ref{app:A}).

In the second physical picture 
the fluid element maintains
exact balance between buoyancy force and turbulent drag by
continuous exchange of momentum with other elements and its 
surrounding \citep{Prandtl32}. Thus the acceleration terms are
unimportant in a static atmosphere and the evolution of the 
convective fluctuations are independent of the initial conditions.
The nonlinear advection terms (i.e., momentum exchange) provide dissipation 
(of kinetic energy) that balances the driving terms, 
and are approximated appropriately \citep[e.g.,][]{Kraichnan62, Unno67}, leading
to two nonlinear equations which need to be solved numerically together 
with the mean equations of the stellar structure.

The two physical pictures are complementary in envelopes
that do not pulsate \citep{Gough77a}. However, in a time-dependent 
formulation additional information is required how the initial state 
of a convective element depends on conditions at the time of its creation. 
Hence, the different versions of mixing-length models yield different formulae 
for the turbulent heat and momentum fluxes when applied to
pulsating stars \citep{Unno67, Gough77a, Gough12a}.

In the above discussed models, the overturning fluid parcels were
still considered to move adiabatically. \citet{Opik50} suggested
to treat radiative heat exchange between the element and the 
background fluid in a similar way as for the momentum exchange.
Based on this assumptions \citet{Vitense53} and \citet{BohmVitense58}
established a mixing-length description which is still
widely used for calculating the convective heat flux 
in stellar models.

The perhaps simplest description to model the temporal modulation of the convection
by the oscillations, put forward in the 1960s, is to presume that the convective fluxes 
simply relax exponentially on a timescale $\tau_{\mathrm{c}}$ towards the time-independent formula
$\rd F_{\mathrm{c}}/\rd t = (F_{\mathrm{c0}}-F_{\mathrm{c}})/\tau_{\mathrm{c}}$, where
$F_{\mathrm{c}}$ is a component of any turbulent flux and $F_{\mathrm{c0}}$ is the formula for
$F_{\mathrm{c}}$ in a statistically steady environment.
The constant $\tau_{\mathrm{c}}$ is a multiple
of $w/\ell$ with $\ell$ being the mixing length and $w$ a characteristic convective velocity.\\
In the past, various time-dependent convection models 
were proposed, for example, by \citet{Schatzman56}, \citet{Gough65, Gough77a}, 
\citet{Unno67, Unno77}, \citet{Xiong77, Xiong89}, \citet{Stellingwerf82}, 
\citet{Gonczi82a}, \citet{Kuhfuss86}, 
\citet{UnnoEtal89}, \citet{Canuto92}, \citet{Gabriel96}, \citet{Grossman96}, and 
\citet{GrigahceneEtal05}.
Here, we shall review and compare the basic concepts of two, currently in
use, convection models. The first model is that by \citet{Gough77a, Gough77b}, which 
has been used, for example, by \citet{BakerGough79}, \citet{Balmforth92a}, \citet{HoudekEtal95}, 
\citet{RosenthalEtal95}, \citet{Houdek97, Houdek00}, \citet{HoudekEtal99}, and 
\citet{ChaplinEtal05}. The second model 
is that by \citet{Unno67, Unno77}, upon which the generalized models 
by \citet{Gabriel96} and \citet{GrigahceneEtal05} are based, with applications by 
\citet{DupretEtal05a, DupretEtal05b, DupretEtal05c, DupretEtal06a, 
DupretEtal06b, DupretEtal06c, DupretEtal09}, 
\citet{BelkacemEtal08a, BelkacemEtal08b, BelkacemEtal12},
and \citet{GrosjeanEtal14}.



\subsection{Two time-dependent convection models for radially pulsating stars} 
\label{sec:gough_unno_mlm}

\citet{Unno67} and \citet{Gough65, Gough77a} generalized the mixing-length formulation 
for modelling the interaction of the turbulent velocity field 
with radial pulsation. Both authors adopted the Boussinesq approximation. 
The mean equation of motions were already
discussed in Section~\ref{sec:mean_boussinesq_equations} for a radially 
pulsating atmosphere. Therefore, we start here with the Boussinesq approximation
for the convective fluctuations.
This approximation is based on a careful scaling 
argument and an expansion in small parameters, i.e., the ratio of the maximum density
variation across the layer over the (constant) spatial density
average, and the ratio of the fluid layer height to the locally
defined smallest scale height \citep{SpiegelVeronis60, Mihaljan62, Gough69}.

\noindent
In this subsection we follow the discussion by \citet{Gough77a}.

\subsubsection{Boussinesq fluctuation equations}
The Boussinesq approximation results in (i) an incompressible fluid, which
renders the convective velocity field $\vec{u}$ in the continuity equation 
to be divergence-free, i.e., $\nabla\cdot\vec{u}=0$,
(ii) neglecting the density fluctuations $\rho^\prime$ in the momentum
equation, except when they are coupled to the gravitational acceleration
in the (driving) buoyancy force, (iii) neglecting squares of
fluctuating thermodynamic quantities, such as $\rho^\prime T^\prime$, where
$T^\prime$ is the temperature fluctuation, and neglecting pressure 
fluctuations $p^\prime$ compared to $\rho^\prime$ or $T^\prime$, thus
removing the acoustic energy flux in the momentum equation. The latter
assumption also leads to the Boussinesq equation of state
\begin{equation}
\frac{\rho^\prime}{\ob{\rho}}=-\ob{\tdel}\frac{T^\prime}{\ob{T}}\,,
\label{eq:eos_ba}
\end{equation}
where $\tdel:=-(\partial\ln\rho/\partial\ln T)_p$ is the isobaric 
expansion coefficient.
Also, under the restrictions outlined above, \citet{SpiegelVeronis60}
and \citet{Mihaljan62} demonstrated 
for the case for which $\vec{U}=\vec{0}$ 
that
the viscous dissipation term, $\ob{\rho}\epsilon_{\mathrm{t}}$, in the 
mean thermal energy equation is negligibly small compared to the 
other terms in the thermal energy equation, such as the term of 
convection of internal energy $\rho\vec{u}\cdot\nabla e$. Therefore, the last
two terms in Eq.~(\ref{eq18}) are neglected in Gough's convection formulation.

The fluctuation equations are obtained from subtracting the horizontally 
averaged Eqs.~(\ref{eq8}), (\ref{eq13}), and (\ref{eq18}) from the 
instantaneous Eq.~(\ref{eq4})\,--\,(\ref{eq6}).
Within the adopted coordinate system (\ref{eq:coordinates_ba}) the convective 
(Eulerian) fluctuation equations are then \citep{Gough77a}:
\begin{equation}
\partial_iu_i=0\,,
\label{eq:fluctuating_cont_ba}
\end{equation}
\begin{equation}
\derivp{u_i}{t}+\left(u_j\partial_ju_i-\ob{u_j\partial_ju_i}\right)
 -w\derivp{\ln(r^2\ob{\rho})}{t}\delta_{i3}
=-\frac{1}{\ob{\rho}}\,\partial_ip^\prime
 +\frac{g\ob{\tdel}}{\ob{T}}T^\prime\,\delta_{i3}\,,
\label{eq:fluctuating_momentum_ba}
\end{equation}
\begin{equation}
\derivp{T^\prime}{t}+\left(u_j\partial_jT^\prime-\ob{u_j\partial_jT^\prime}\right)
+\left[\left(\ob{c_{pT}}-\ob{\tdel}\right)\derivp{\ln\ob{T}}{t}
-\ob{\nabla_{\mathrm{ad}}}\,{\ob{\tdel_T}}\,\derivp{\ln\ob{p}}{t}\right]T^\prime
-\beta w=-\frac{1}{\ob{\rho}\,\ob{c_p}}\,\partial_jF_j^\prime\,,
\label{eq:fluctuating_thermal_energy_ba}
\end{equation}
\begin{equation}
\partial_jF_j^\prime=4\pi\ob{\rho}\,\ob{\varkappa}
\left[B^\prime-J^\prime+\left(\ob{\varkappa_T}-\ob{\tdel}\right)
\left(\ob{B}-\ob{J}\right)\frac{T^\prime}{\ob{T}}\right]\,,
\label{eq:fluc_eddington_F}
\end{equation}
\begin{equation}
\partial_iJ^\prime=-\frac{3}{4\pi}\ob{\rho}\,\ob{\varkappa}
\left[F_i^\prime+\left(\ob{\varkappa}-\ob{\tdel}\right)
\ob{F_3}\,\frac{T^\prime}{\ob{T}}\,\delta_{i3}\right]\,,
\label{eq:fluc_eddington_J}
\end{equation}
\begin{equation} 
\partial_i:=\left(\derivp{}{q_1},\derivp{}{q_2},\ob\rho\derivp{}{q_3}\right)\,,
\label{eq:part_deriv_ba}
\end{equation} 
where (with $q=q_3$)
\begin{equation}
\beta:=-\left(\ob{\rho}\derivp{\ob{T}}{q}-
\frac{\ob{\tdel}}{\ob{c_p}}\derivp{\ob{p}}{q}\right)\,,
\label{eq:beta}
\end{equation}
is the superadiabatic temperature gradient (or superadiabatic lapsrate),
and
\begin{equation}
g=\frac{Gm}{r^2}+\left(\frac{\partial^2r}{\partial t^2}\right)_{\!q_3}\,, 
\end{equation}
is the effective magnitude of the gravitational acceleration;
$\ob{\nabla_{\mathrm{ad}}}=\ob{p}\,\ob{\tdel}/\ob{c_p}\,\ob{\rho}\,\ob{T}$ is the 
adiabatic temperature gradient, 
and $c_{pT}$, $\tdel_T$, and $\varkappa_T$ are the logarithmic derivatives of
the specific heat at constant pressure, $c_p$, isobaric expansion coefficient, $\tdel$,
and opacity, $\varkappa$, with respect to ${T}$ at constant $p$; 
$\delta_{ij}$ is the Kronecker delta.


In these fluctuation equations geometrical terms, which distinguish Cartesian from 
the spherical coordinates $q_i$, are neglected, i.e., it is assumed that the convective
velocity field is located in stellar layers where $\ell\ll r$. It is also assumed, in 
accordance with the Boussinesq approximation, that $\ell\ll H$, where $H$ represents any
locally-defined scale height .

The third term on the left-hand side of \eq{eq:fluctuating_momentum_ba} comes
from substituting the mean continuity equation 
into the mean radial component of the nonlinear advection term of 
the mean momentum equation. With the help of \eq{e:Gough-A6}, to relate time-derivatives in
Eulerian convective fluctuations to the Lagrangian coordinates $q_i$,
one obtains $\partial_3U=-\partial(\ln r^2\ob{\rho})/\partial t$. 
The third term of the left-hand side of \eq{eq:fluctuating_thermal_energy_ba} 
is a result of having taken into account the pulsationally induced time dependence of
the mean temperature $\ob T$ and gas pressure $\ob p$ in a pulsating atmosphere.

\subsubsection{Local mixing-length models for static atmospheres}
\label{sec:static_LMLM}

Linear pulsation calculations perturb the stellar structure equations around a 
time-independent (on a dynamical time scale) equilibrium model, which must be 
constructed first from, e.g., stellar evolutionary calculations. We
start the discussion of two versions of the mixing-length formulation first
for a static stellar envelope before embarking on the model description for radially
pulsating envelopes. 

The convective Eulerian fluctuation equations for a 
Boussinesq fluid are obtained from setting the time derivatives of the mean 
(equilibrium) quantities to zero in the Eqs.~(\ref{eq:fluctuating_momentum_ba}) 
and (\ref{eq:fluctuating_thermal_energy_ba}), leading to
\begin{equation}
\derivp{u_i}{t}+\left(u_j\partial_jw-\ob{u_j\partial_jw}\right)
=-\frac{1}{\ob{\rho}}\,\partial_ip^\prime
 +\frac{g\ob{\tdel}}{\ob{T}}T^\prime\,\delta_{i3}\,,
\label{eq:static_fluc_momentum}
\end{equation}
\begin{equation}
\derivp{T^\prime}{t}+\left(u_j\partial_jT^\prime-\ob{u_j\partial_jT^\prime}\right)
-\beta w=-\frac{1}{\ob{\rho}\,\ob{c_p}}\,\partial_jF_j^\prime\,,
\label{eq:static_fluc_thermal_energy}
\end{equation}
which must be supplemented by Eqs.~(\ref{eq:fluctuating_cont_ba}), 
(\ref{eq:fluc_eddington_F}) and (\ref{eq:fluc_eddington_J}).
The pressure gradient in the first term on the right-hand side of the fluctuating
momentum equation~(\ref{eq:static_fluc_momentum}) couples the vertical to the
horizontal motion, and the third term on the left hand side of the fluctuating
thermal energy equation~(\ref{eq:static_fluc_thermal_energy}) describes the
deformation of the mean temperature field $\ob{T}$ by the turbulent heat transport.

\medskip
In Section~\ref{sec:intro_local_mlm} we introduced two physical pictures of 
mixing-length models, both of which are based on the picture of an overturning 
convective cell (see Figure~\ref{fig:conv_cell}). In both pictures, the convective 
cell is created as a result of instability with
the same average properties than its immediate surroundings. 
The overturning motion of a convective cell is then accelerated by the imbalance 
between buoyancy forces, nonlinear advection processes, pressure gradients, and 
heat losses by radiation. Various guises of convection models can be obtained by
approximating these processes in different ways and even neglecting some of it.
Also, different assumptions about the geometry of the turbulent flow does lead to
different results in the turbulent fluxes. Two of the convection
models will be described below which, to some extent, make different assumptions 
about the dynamics of the turbulence.

\paragraph{Convection model 1: Kinetic theory of accelerating eddies}
\label{sec:static_LMLM_1}
\mbox{}\\

\noindent
The first model, which was generalized by \citet{Gough65, Gough77a, Gough77b} to 
the time-dependent case, interprets the turbulent flow by indirect analogy with 
kinetic gas theory. The motion is not steady and one imagines an overturning 
convective cell to accelerate and grow exponentially with time from a small 
perturbation according to the linearized version of the fluctuation 
equations~(\ref{eq:static_fluc_momentum})\,--\,(\ref{eq:static_fluc_thermal_energy}).
During this growth the nonlinear advection terms are neglected but taken entirely into 
account by the cell's subsequent instantaneous destruction (annihilation) by internal stresses after its finite lifetime. The linearized equations thus become
\begin{equation} 
\derivp{u_i}{t}=-\frac{1}{\rho}\,\partial_ip^\prime
 +\frac{g\tdel}{T}T^\prime\,\delta_{i3}\,,
\label{eq:lin_static_fluc_momentum}
\end{equation} 
\begin{equation}
\derivp{T^\prime}{t}-\beta w=-\frac{1}{\rho\,c_p}\,\partial_jF_j^\prime\,,
\label{eq:lin_static_fluc_thermal_energy}
\end{equation}
where from now on overbars will be omitted from mean variables to simply the notation.
The equations can be further simplified if we can eliminate the pressure fluctuations.
This can be obtained by taking the double curl of \eq{eq:lin_static_fluc_momentum}
leading to
\begin{equation}
\derivp{}{t}\left(\partial_i\partial_iw\right)
-\frac{g\tdel}{T}\left(\partial^2_1+\partial^2_2\right)T^\prime=0\,.
\label{eq:double_curl_momentum}
\end{equation}
Eqs.~(\ref{eq:double_curl_momentum}) and (\ref{eq:lin_static_fluc_thermal_energy}) 
describe a linear stability problem for the vertical component of the convective 
velocity $w$ and the convective temperature fluctuation $T^\prime$. With the 
assumption that the coefficients are
constant over the vertical extent of the convective cell the solutions to
Eqs.~(\ref{eq:double_curl_momentum}) and (\ref{eq:lin_static_fluc_thermal_energy})
are separable \citep{Chandrasekhar61}
\begin{equation}
T^\prime=\Theta(q,t)f(q_1, q_2)\,,
\label{e:temp-expand}
\end{equation}
\begin{equation}
w=W(q,t)f(q_1, q_2)\,,
\label{e:velo-expand}
\end{equation}
with a horizontal flow structure $f(q_1, q_2)$ satisfying the Helmholtz equation
\begin{equation}
\left(\partial^2_1+\partial^2_2\right)f=-k^2_{\mathrm{h}}f\,,\qquad\ob{f^2}=1\,.
\end{equation}
The separation constant $k_{\mathrm h}$ represents the horizontal
wavenumber of the motion.
With the advent of the horizontal 
wavenumber there is no longer only one single length scale
associated with the fluid parcel, which brings its shape
into play. This coupling between vertical and horizontal motion
is due to the inclusion of the pressure fluctuations $\partial_ip^\prime$ in 
the momentum equation, diverting the vertical motion into horizontal flow and
thus reducing the efficacy with which the motion might otherwise
have released potential energy gained by the buoyancy forces \citep{Gough77a}.
The vertical motion in a convective cell near the central axis, as illustrated 
in Figure~\ref{fig:conv_cell}, is governed by buoyancy; the horizontal flow 
across the top of the cell to its edge, however, experiences only damping forces 
due to dissipative processes without any compensation. Hence, the horizontal 
motion is considerably wasteful. It is related to the vertical velocity 
field through the anisotropy or shape parameter $\Phi$~(\ref{eq:Phi}). 
Because the velocity field $u_i$, described by the linear 
Eqs.~(\ref{eq:lin_static_fluc_thermal_energy}) and (\ref{eq:double_curl_momentum}),
has no vertical component of vorticity \citep[e.g.,][]{LedouxEtal61} the resulting flow
geometry~\citep{Platzman65} allows $\Phi$ to be related to $k_{\mathrm{h}}$ as 
\begin{equation}
\Phi=1+\frac{k^2_{\mathrm v}}{k^2_{\mathrm h}}\,,
\label{eq:shape-param}
\end{equation}
where $k_{\mathrm v}$ is the vertical wave number of an convective cell (eddy)
with vertical extent $\ell$
\begin{equation}
k_{\mathrm v}:=\frac{\pi}{\ell} \,.
\label{eq:vertical-wavenumber}
\end{equation}
In this view, convective motion becomes most 
efficient for eddies with a geometry of tall thin needles, for which
$\Phi\rightarrow1$. 
The differential equation for the horizontal structure of the convective
fluctuations (\ref{e:temp-expand}) and (\ref{e:velo-expand}) 
can be solved subject to proper periodic boundary conditions in the domain
described by the planform $f(q_1,q_2)$, which is defined on the
surface of a sphere \citep{Spiegel63}. Thus the horizontal 
wavenumber $k_{\mathrm h}$ can take any value 
from an infinite discrete set of eigenvalues. Assuming the eigenvalue
spectrum to be dense for relatively high harmonics and since the motion
is unlikely to be coherent over the whole spherical surface, 
it might be a reasonable approximation to consider $\Phi$ as continuous.
Within this approximation, \citet{Gough77a} has chosen a value for $k_{\mathrm h}$ 
that maximizes the convective velocity at fixed $k_{\mathrm v}$. This is 
equivalent to selecting the most rapidly growing mode in the theory of linear 
stability \citep{Spiegel63} and corresponds to an anisotropy value 
$\Phi=5/3$ \citep{Gough78}.

Since the assumption of constant coefficients in the linear stability equations, which 
had led to the separation of the solutions (\ref{e:temp-expand}) 
and (\ref{e:velo-expand}), may not be satisfied in the very upper, optically thin,
region of the convection zone, the last term in brackets of
\eq{eq:fluc_eddington_J} may be neglected without the introduction of a 
larger assumption \citep{Gough77a}, leading to 
\begin{equation} 
\partial_iJ^\prime=-\frac{3}{4\pi}{\rho}\,{\varkappa}F_i^\prime\,.
\end{equation}
The radiative heat loss of the convective eddy is then
given in the general Eddington approximation to the radiative transfer
\citep{UnnoSpiegel66} as
\begin{equation}
\partial_jF^\prime_j=\upphi Kk^2\Theta \,,
\label{eq:heatflux-eddappr}
\end{equation}
where
\begin{equation}
\upphi=\frac{1+1/4(\varkappa_{\mathrm T}-\tdel)(1-J/B)}
          {1+\Phi\Sigma/(\Phi-1)}
\,,\qquad 
\Sigma=\frac{1}{3}\frac{\pi^2}{\left(\rho\varkappa\ell\right)^2}\,,
\label{eq:optical-trans-parameter}
\end{equation}
provide a smooth transition between the optically thin and thick regions of the star,
with $K=4acT^3/(3\rho\varkappa)$ being the radiative conductivity and
$k^2=k^2_{\mathrm h}+k^2_{\mathrm v}$.
The linearized fluctuating equations of motion 
(\ref{eq:double_curl_momentum}) and (\ref{eq:lin_static_fluc_thermal_energy})
can then be reduced to
\begin{equation}
\Phi\derivp{W}{t}=\frac{g\tdel}{T}\Theta\,,
\label{eq:dog-lmlt-moment-expand}
\end{equation}
\begin{equation}
\left(\derivp{}{t}+\frac{\upphi K}{\rho c_p}k^2\right)\Theta=\beta W\,,
\label{eq:dog-lmlt-engy-expand}
\end{equation}
where the vertical derivatives have been replaced by ${\mathrm i}k_{\mathrm v}$ 
(e.g., $\partial_3^2W=-k^2_{\mathrm v}W$) for harmonic solutions.
The shape parameter $\Phi$ effectively increases the inertia of 
the vertically moving fluid, without changing the functional form
of the equation of motion. 
These linear equations can be solved with the ansatz
$W\propto\exp(\sigma_{\mathrm c} t)$ and $\Theta\propto\exp(\sigma_{\mathrm c} t)$
where the convective linear growth rate $\sigma_{\mathrm c}$ is obtained from
the characteristic equation
\begin{equation}
\sigma^2_{\mathrm c}+\frac{\upphi K}{\rho c_p}k^2\sigma_{\mathrm c}-
\frac{g\tdel\beta}{\Phi T}=0\,,
\label{eq:dog-lmlt-growth-rate}
\end{equation}
with the solution
\begin{equation}
\sigma_{\mathrm c}=\eta^{-1}S^{-1/2}\left(\frac{g\tdel\beta}{\Phi T}\right)^{1/2}
       \left[\left(1+\eta^2S\right)^{1/2}-1\right]
\,,\qquad
S=\frac{g\left(\tdel/T\right)\beta\ell^4}
                        {\left(\upphi K/\rho c_p\right)^2}\,,
\label{eq:dog-lmlt-conv-sigma}
\end{equation}
where $\eta=2\pi^{-2}\Phi^{-3/2}(\Phi-1)$ is a geometrical factor.
The last term on the right-hand side of \eq{eq:dog-lmlt-growth-rate} 
can be interpreted as $N^2/\Phi$, where
\begin{equation}
N^2:=-\frac{\tdel}{T}g\beta
\label{eq:brunt-vaisala}
\end{equation}
is the Brunt--V{\"a}is{\"a}l{\"a} frequency (in a homogeneous medium), which is negative for
convective instability, and consequently $1/|N|$ represents
a characteristic time scale of the convection (buoyancy time scale). The coefficient of
the second term on the right-hand side of \eq{eq:dog-lmlt-growth-rate}
accounts for the radiative cooling time. The squared ratio of these two 
time scales defines the convective efficacy $S=-(N\ell^2/\upphi\kappa)^2$
(where $\kappa=K/\rho\,c_\mathrm{p}$ is the thermal diffusivity), 
with which the convection transports the heat flux and which 
can be interpreted as the product of the molecular Prandtl number and 
the locally defined Rayleigh number.
For efficient convection ($S \gg 1$) the linear convective growth 
rate $\sigma_{\mathrm{c}}\propto|N|$,
i.e., convection is dominated by the buoyancy time scale. For inefficient convection
($S \ll 1$) the growth rate $\sigma_{\mathrm{c}}\propto|N|S^{1/2}$ and is therefore dominated 
by the thermal diffusion time scale. 
\citet{GoughWeiss76} demonstrated that every local mixing-length formulation can be
interpreted as an interpolation formula between efficient ($S \gg 1$) and inefficient
($S \ll 1$) convection. In solar-type stars and stars hotter than the Sun the
transition between these two limits occurs in a very thin layer at the top of the 
bulk of the convection zone where the temperature gradient is substantially
superadiabatic.

The turbulent fluxes are then obtained from the
eddy annihilation hypothesis by means of an eddy survival 
probability \citep{Spiegel63, Gough77a, Gough77b}, which is discussed
in more detail in Appendix~\ref{app:A}. Within this hypothesis the turbulent fluxes are
\begin{equation}
F_{\mathrm c}=\frac{1}{4}\frac{\rho c_p\Phi T\ell^2}{g\tdel}\,\sigma_{\mathrm c}^3\,,
\label{eq:heat-flux-sigma}
\end{equation}
\begin{equation}
p_{\mathrm t}=\frac{1}{4}\rho\ell^2\sigma_{\mathrm c}^2 \,.
\label{eq:momentum-flux-sigma}
\end{equation} 

\paragraph{Convection model 2: Balance between buoyancy and turbulent drag}
\label{sec:static_LMLM_2}
\mbox{}\\

\noindent
In the second convection model, adopted by \citet{Unno67, Unno77}, the turbulent
element (eddy) 
maintains exact balance between buoyancy force and turbulent drag
by continuous exchange of momentum with other elements and its surroundings. Thus
the acceleration terms $\partial_tu_i$ and $\partial_tT^\prime$ 
in Eqs.~(\ref{eq:fluctuating_momentum_ba}) and (\ref{eq:fluctuating_thermal_energy_ba})
are omitted in a static atmosphere, and the nonlinear advection terms 
provide dissipation of kinetic energy that balances the driving terms. The nonlinear
advection terms are approximated by
\begin{equation}
u_j\partial_ju_i-\ob{u_j\partial_ju_i}\simeq\frac{2w^2}{\ell}\,
\label{eq:nonlinear_approximation_velocity}
\end{equation}
\begin{equation}
u_j\partial_jT^\prime-\ob{u_j\partial_jT^\prime}\simeq\frac{2wT^\prime}{\ell}\,
\label{eq:nonlinear_approximation_temperature}
\end{equation}
which is based on \citepos{Prandtl25} mixing-length 
idea of scaling the shear stress by
means of a turbulent viscosity $\nu_{\mathrm t}\simeq w\ell$, i.e.,
$-\ob{\rho u_iu_j}\simeq\rho\nu_{\mathrm{t}}\,\partial_3w\simeq2w^2/\ell$, with 
$\partial^{-1}_3\simeq\ell^{-1}$ \citep{Kraichnan62}.

Unno assumes for the velocity field a structure similar to 
\eq{e:velo-expand} with a vanishing vertical vorticity component,
so that 
the pressure fluctuations in the momentum \eq{eq:static_fluc_momentum} can be eliminated
by proper vector operations. With the nonlinear terms retained and the time
derivatives omitted in (\ref{eq:static_fluc_momentum}) and 
(\ref{eq:static_fluc_thermal_energy})
the fluctuating equations of motion in Unno's model are
\begin{equation}
\Phi\frac{2w^2}{\ell}
=\frac{g\tdel}{T}T^\prime\,,
\label{eq:unno_static_fluc_momentum}
\end{equation}
\begin{equation}
\frac{2wT^\prime}{\ell}
-\beta w=-\frac{\tilde\upphi K}{\rho c_p}\,k^2T^\prime\,.
\label{eq:unno_static_fluc_thermal_energy}
\end{equation}
Unno chooses $k^2_{\mathrm h}=k^2_{\mathrm v}$, which implies $\Phi=2$, a
value which is also adopted in \citepos{BohmVitense58} mixing-length model. 
For radiative losses Unno chooses
\begin{equation}
\tilde\upphi=\frac{2\pi^{-2}}{1+4\pi^{-2}\Sigma}
\label{eq:unno_optical-trans-parameter}
\end{equation}
for describing the transition between optically thin and thick regions,
which is similar to \eq{eq:optical-trans-parameter} if $J=B$, i.e., 
for radiative equilibrium.

The nonlinear Eqs.~(\ref{eq:unno_static_fluc_momentum}) and 
(\ref{eq:unno_static_fluc_thermal_energy}) are solved numerically
for $w$ and $T^\prime$ from which the turbulent fluxes 
$F_{\mathrm c}\simeq\ob{\rho cp}\,\ob{wT^\prime}$ and 
$p_{\mathrm t}\simeq\ob{\rho}\,\ob{ww}$ are constructed. 
\citet{Unno67} neglects, however, the turbulent pressure $p_{\mathrm{t}}$
in the mean momentum equation (\ref{eq:mean_momentum_ba}). 


\subsubsection{Local mixing-length models for radially pulsating atmospheres}
\label{sec:time-dependent_LMLM}

In the previous section, we discussed two mixing-length models in a static atmosphere.
In a static atmosphere the (mean) coefficients $\rho$, $c_p$, and $\tdel$ are
independent of time, which had led to Eqs.~(\ref{eq:dog-lmlt-moment-expand}) 
and (\ref{eq:dog-lmlt-engy-expand}) for the convection model of accelerating eddies.
What follows is a discussion of the time-dependent treatment of the two convection models
in a radially pulsating atmosphere.

\paragraph{Convection model 1: Kinetic theory of accelerating eddies}
\label{sec:time-dependent_LMLM_1}
\mbox{}\\

\noindent
In order to study the coupling between convection and a pulsating atmosphere 
the time-dependence of the mean values (coefficients) needs to be considered, 
i.e., all that is necessary is to restore the time derivatives in 
Eqs.~(\ref{eq:fluctuating_momentum_ba}) and
(\ref{eq:fluctuating_thermal_energy_ba}) 
with the nonlinear advection terms neglected. 
We obtain (overbars for mean values are omitted)
\begin{equation}
\left[\derivp{}{t}-\frac{1}{\Phi}\derivp{\ln(r^2\rho)}{t}\right]W
-\frac{g\tdel}{\Phi T}\Theta=0\,,
\label{eq:fluc-moment-puls}
\end{equation}
\begin{equation}
\left[\derivp{}{t}+\left(c_{pT}-\tdel\right)\derivp{\ln T}{t}
      -\tdel_T\nabla_\mathrm{ad}\derivp{\ln p}{t}\right]\Theta
-\beta W+\frac{\upphi K}{\rho c_p}k^2\Theta=0\,,
\label{eq:fluc-engy-puls}
\end{equation}
where, as before, $c_{pT}$, and $\tdel_T$ are the logarithmic derivatives of
$c_p$ and $\tdel$ with respect to $T$ at constant gas pressure $p$. 

In a static atmosphere the evolution process of a convective element
is described by the linear growth rate, and the element itself is
characterized by its wave number $\vec{k}$ ($k^2=k^2_{\mathrm h}+k^2_{\mathrm v}$), 
and thus by the constant values of the 
mixing-length $\ell$, \eq{eq:vertical-wavenumber}, 
and shape-parameter $\Phi$, \eq{eq:shape-param}, at each point in the atmosphere. 
These latter parameters, however, are no longer constant in a pulsating 
atmosphere, because of the locally changing environment. The eddies are 
advected by the pulsating flow and, in a Lagrangian frame moving with the 
pulsation, they deform as they grow. Thus the evolution of the convective
elements becomes influenced by the temporal behaviour of the atmosphere.
\citet{Gough77a} adopted the theory of rapid distortion of turbulent shear flow
\citep[e.g.,][]{Townsend76} to describe the shape distortion of a convective element
advected by the pulsation. In this theory the eddy size varies approximately with 
the pressure scale height $H_{\mathrm{p}}$ if the lifetime of the eddy is short compared to the
pulsation period, and with the local Lagrangian scale of the mean
flow (pulsation) if the eddy lifetime is large compared to the pulsation period.
Moreover, one has also to consider the initial conditions
at the time $t_0$ the element was created.
If the time dependence of the fluctuating quantities $W$ and $\Theta$
is taken to be proportional to $\exp(-{\mathrm i}\omega t)$, where
$\omega$ denotes the complex pulsation frequency, 
their evolution with time is independent from the initial conditions 
at the time $t_0$. 
In a moving atmosphere, however, the phase between pulsation and the 
convective perturbations at the instant $t_0$ substantially influences the 
stability of pulsation. The dependence on the initial conditions at $t_0$
can be taken into account by linearizing the variation of the atmosphere
about its equilibrium state and defining this state at the instant $t_0$, 
which provides
the following expression for the linearized form of the vertical wave number
as a function of the (complex) pulsation frequency $\omega$ \citep{Gough77a}
\begin{equation}
k_\mathrm{v}=\frac{\pi}{\ell}\left[1+
 \left(2\frac{\Ldel r}{r_0}+\frac{\Ldel\rho}{\rho_0}\right)
 \left(\mathrm{e}^{\mathrm{i}\omega t}-\mathrm{e}^{\mathrm{i}\omega t_0}\right)
-\frac{\Ldel H_{\mathrm{p}}}{H_{\mathrm{p0}}}\,\mathrm{e}^{\mathrm{i}\omega t_0}\right]
:= k_\mathrm{v0}
\left(1+k_\mathrm{v10}\mathrm{e}^{\mathrm{i}\omega t_0}+k_{v11}
                       \mathrm{e}^{\mathrm{i}\omega t}\right)\,,
\label{eq:verti-wavenumber-puls}
\end{equation}
\begin{equation}
k_\mathrm{h}=k_\mathrm{h0}
\left(1+k_\mathrm{h10}\mathrm{e}^{\mathrm{i}\omega t_0}+
k_\mathrm{h11}\mathrm{e}^{\mathrm{i}\omega t}\right)\,,
\label{eq:hori-wavenumber-puls}
\end{equation}
where $\Ldel X(q)$ is the pulsational perturbation of
$X(q,t)=X_0(q)\left[1+\Ldel X(q)/X_0(q)\exp({\mathrm i}\omega t)\right]$ 
in a Lagrangian frame of reference, and
a subscript zero denotes the value in the static equilibrium model.
The coefficient $k_\mathrm{v0}$ and $k_\mathrm{h0}$ are the wave numbers 
characterizing a convective element in a static atmosphere, and
$k_\mathrm{v10}$, $k_\mathrm{v11}$, $k_\mathrm{h10}$ and $k_\mathrm{h11}$ are the
linearized pulsational perturbations of it.

Combining Eqs.~(\ref{eq:fluc-moment-puls}) and (\ref{eq:fluc-engy-puls})
and linearizing the result leads to a second-order differential equation
for the evolving velocity fluctuations with coefficients depending on 
$k_\mathrm{h}$ and $k_\mathrm{v}$. The coupling of this equation with the 
pulsation is achieved by expressing these coefficients in the form such as 
given here for the shape parameter $\Phi$ 
\begin{equation}
\Phi= \Phi_0\left(1+\Phi_{10}\mathrm{e}^{\mathrm{i}\omega t_0}
      +\Phi_{11}\mathrm{e}^{\mathrm{i}\omega t}\right)\,,
\label{eq:shape-param-puls}
\end{equation}
where we used Eqs.~(\ref{eq:shape-param}),
(\ref{eq:hori-wavenumber-puls}) and (\ref{eq:verti-wavenumber-puls}).
Equation~(\ref{eq:shape-param-puls}) represents the influence of the pulsating 
atmosphere on the shape of the eddy. The resulting expression for the vertical component of the velocity field becomes, to first order in pulsational perturbations 
\begin{equation}
\frac{\partial^2W}{\partial t^2}
+2\kappa\upphi k^2\left(1+\kappa_{10}\mathrm{e}^{\mathrm{i}\omega t_0}
                         +\kappa_{11}\mathrm{e}^{\mathrm{i}\omega t}
                  \right)\derivp{W}{t}
+\frac{N^2}{\Phi}\left(1+2\mu_{10}\mathrm{e}^{\mathrm{i}\omega t_0}
                        +2\mu_{11}\mathrm{e}^{\mathrm{i}\omega t}
                 \right)W=0\,,
\label{eq:fluc-velocity-puls-ODE}
\end{equation}
where $N^2=-g\beta\tdel/T$ is, as already defined earlier, 
the squared Brunt--V{\"a}is{\"a}l{\"a} frequency. The coefficients $\kappa_{10}$, etc,
are given in the Appendix~\ref{app:Gough-coefficients}.
This equation can be solved exactly (for constant coefficients) and can be written for 
the convective elements travelled about one mixing length approximately as
\begin{equation}
W=W_0\left[1 +W_{10}\mathrm{e}^{\mathrm{i}\omega t_0}
             +W_{11}\mathrm{e}^{\mathrm{i}\omega t}
+\sigma_{\mathrm c}(t-t_0)W_{12}\mathrm{e}^{\mathrm{i}\omega t_0}
      \right]\,,
\label{eq:fluc-velocity-puls}
\end{equation}
where $W_0\simeq\widehat W_0\exp\left[\sigma_{\mathrm c}(t-t_0)\right]$ represents the evolving 
convective velocity fluctuation in a static atmosphere as given by
\eq{eq:flucquant-timedependent}. A similar expression results for the
convective temperature fluctuation $\Theta$. Thus the pulsationally induced 
perturbations of the convective fluxes may be obtained, with the help of
\eq{eq:survival-probability}, by substituting these solutions into the 
integral expressions (\ref{eq:heat-flux-integral}) and
(\ref{eq:momentum-flux-integral}), which become to first order in
relative perturbations
\begin{equation}
\frac{\Ldel F_\mathrm{c}}{F_\mathrm{c,0}}=
\frac{\Ldel\rho}{\rho_0}+\frac{\Ldel c_p}{c_{p0}}+W_{11}+\Theta_{11}+W_{21}
+(W_{10}+\Theta_{10}){\cal F}+(W_{12}+\Theta_{12}){\cal FG}+{\cal H}\,,
\label{eq:heat-flux-perturb}
\end{equation}
\begin{equation}
\frac{\Ldel p_\mathrm{t}}{p_\mathrm{t,0}}=
\frac{\Ldel\rho}{\rho_0}+2W_{11}+W_{21}
+2(W_{10}{\cal F}+W_{12}{\cal FG})+{\cal H}\,.
\label{eq:momentum-flux-perturb}
\end{equation}
For the linearized perturbation of the shape parameter $\Phi$ 
one obtains
\begin{equation}
\frac{\Ldel\Phi}{\Phi_0}=\Phi_{11}-\Phi_{10}{\cal F} \,.
\label{eq:shape-param-perturb}
\end{equation}
%
The coefficients $W_{1i}$, $\Theta_{1i}$ and $W_{12}$, 
as well as the functional expressions ${\cal F, G}$ and ${\cal H}$ are
given in Appendix~\ref{app:Gough-coefficients}.
The expressions ${\cal F, G}$ and ${\cal H}$ account for a
statistical averaging of the convective fluctuations at the
instant $t_0$ in form of a quadratic distribution function,
because the mixing-length formulation provides information only about
the largest scale of the turbulent spectrum.
Thus those terms in Eqs.~(\ref{eq:heat-flux-perturb}) and
(\ref{eq:momentum-flux-perturb}) which include the expressions
${\cal F, G}$ and ${\cal H}$ 
significantly influence the phases between the convective fluctuations
and the pulsating environment of the background fluid and hence, 
the pulsational stability of a star.

The properties of any local time-dependent convection model leads
to serious failure when applied to the problem of solving
the linearized pulsation equations. It fails to treat properly the
convective dynamics across extensive eddies. In deeper parts of the
convection zone, where the stratification is almost adiabatic,
convective heat transport is very efficient, thus radiative diffusion becomes
unimportant and the perturbation of the heat flux is dominated
by the advection of the temperature fluctuations. In this
limit the convective elements grow very slowly compared to the 
pulsationally induced changes of the local stratification, 
i.e., $\omega/\sigma_{\mathrm c}\gg 1$, and local theory predicts
\begin{equation}
\frac{\Ldel F_\mathrm{c}}{F_\mathrm{c,0}}\sim
\frac{\Ldel\Theta}{\Theta_0}\sim
-\mathrm{i}\frac{\sigma_{\mathrm c}}{\omega}\frac{\Ldel\beta}{\beta_0}\,,
\label{eq:imaginary-diffusivity}
\end{equation}
hence, the perturbation of the heat flux is
described by means of a diffusion equation with an imaginary diffusivity
\citep{BakerGough79}. This gives rise to
rapid spatial oscillations of the eigenfunctions, introducing a
resolution problem which is particularly severe in layers where the
stratification is very close to being adiabatic [see also discussion about Eq.~(\ref{pclose1})]. 
This and the other drawbacks of a local theory, discussed before, may be 
obliterated by using a nonlocal convection model such as that discussed
in Section~\ref{sec:nonlocal_mlm}. 

\paragraph{Convection model 2: Balance between buoyancy and turbulent drag}
\label{sec:time-dependent_LMLM_2}
\mbox{}\\

\noindent
In the model adopted by \citet{Unno67} the nonlinear advection terms representing the
interaction of the convective elements with the small-scale turbulence
are retained and approximated by the nonlinear terms 
$2W^2/\ell$ and $2W\Theta/\ell$, but the mean values
$\rho$, $c_p$, and $\tdel$ are considered to be independent of time leading
to the following fluctuation equations
\begin{equation}
\left[\derivp{}{t}-\frac{2W}{\ell}\right]W
-\frac{g\tdel}{2T}\Theta=0\,,
\label{eq:unno_fluc-moment-puls}
\end{equation}
\begin{equation}
\left[\derivp{}{t}+\frac{2W}{\ell}\right]\Theta
-\beta W+\frac{\tilde\upphi K}{\rho c_p}k^2\Theta=0\,,
\label{eq:unno_fluc-engy-puls}
\end{equation}
which may be compared with Eqs.~(\ref{eq:fluc-moment-puls}) and 
(\ref{eq:fluc-engy-puls}). These equations are perturbed to first order
in the relative pulsational (Lagrangian) variations ($\Ldel X/X_0$) the 
outcome of which is \citep{Unno67}
\begin{equation}
\label{eq:perturb_unno_fluc-moment-puls}
\left({\mathrm i}\omega+\frac{2}{\tau_\mathrm{c}}\right)\frac{\Ldel W}{W_0}
-\frac{1}{\tau_\mathrm{c}}\frac{\Ldel\Theta }{\Theta_0}=\frac{1}{\tau_\mathrm{c}}
\left(\derivp{\Ldel p}{p_0}+2\frac{\Ldel r}{r_0}
      -\frac{\Ldel T}{T_0}+\frac{\Ldel\tdel}{\tdel_0}
      +\frac{\Ldel\ell}{\ell_0}
\right)\,,
\end{equation}
\begin{eqnarray}
\left({\mathrm i}\omega + \frac{1}{\tau_\mathrm{c}} + 
      \frac{1}{\tau_{\mathrm{R}}}
\right)
\frac{\Ldel\Theta}{\Theta_0}-\frac{1}{\tau_{\mathrm{R}}}
\frac{\Ldel W}{W_0}&=& 
-\frac{3}{\tau_{\mathrm{R}}}\frac{\Ldel T}{T_0}
 + \left(\frac{1}{\tau_\mathrm{c}} + 
           \frac{1}{\tau_{\mathrm{R}}}
     \right)\frac{\Ldel\beta}{\beta_0}
 + \left(\frac{1}{\tau_\mathrm{c}} + \frac{2}{\tau_{\mathrm{R}}}\right)
 \frac{\Ldel\ell}{\ell_0}\nonumber \\
 &&+ \frac{1}{\tau_{\mathrm{R}}}\left(\frac{\Ldel c_p}{c_{p0}}
                                 + \frac{\Ldel\kappa}{\kappa_0}
                                 + \frac{2\Ldel\rho}{\rho_0}
                               \right)\,,
\label{eq:perturb_unno_fluc-engy-puls}
\end{eqnarray}
where $\tau_\mathrm{c}:=\ell_0/2W_0$ defines the dynamical and 
$\tau_{\mathrm{R}}:= c_{p0}/2\pi^2ac\tilde\upphi_0\kappa_0T^3_0$ 
the radiative time scales of the convection, and $\Ldel\beta/\beta_0$ is 
obtained from perturbing \eq{eq:beta}. 
The time dependence of the fluctuating quantities $W$ and $\Theta$
is taken to be proportional to $\exp(-{\mathrm i}\omega t)$,
but the evolution of the turbulent fluctuations is independent 
from any initial conditions. 

As for the convection model 1 there is still need for finding a prescription how 
to perturb the mixing length $\ell$. \citet{Unno67} chooses
(see also Section~\ref{s:mlt-perturbation})
\begin{equation}
\frac{\Ldel\ell}{\ell_0}\simeq
\left\{
 \begin{array}{lll}
   \Ldel H_{\mathrm{p}}/H_{{\mathrm{p}}0}&\quad\mathrm{for}\ \omega\tau_{\mathrm c}\le1\,,\\ 
   \\
   \Ldel r/r_0&\quad\mathrm{for}\ \omega\tau_{\mathrm c}>1\,,
 \end{array}
\right.
\label{eq:unno_var_ell}
\end{equation}
which is for the limit $\omega\tau_{\mathrm c}>1$ similar to the
result for rapid distortion theory,
i.e., $\ell$ would vary with the local Lagrangian 
scale of the mean flow, but only if the pulsations were homologous. It is also 
assumed that the eddy shape $\Phi$ does not change with the pulsating
atmosphere. Later, \citet{Unno77} and \citet{UnnoEtal79, UnnoEtal89} 
improved on \eq{eq:unno_var_ell} [see Eq.~(\ref{eq:unnodl})].

From solving numerically Eqs.~(\ref{eq:perturb_unno_fluc-moment-puls})\,--\,(\ref{eq:unno_var_ell})
the pulsationally perturbed convective heat flux is then obtained from 
\begin{equation}
\frac{\Ldel F_\mathrm{c}}{F_\mathrm{c,0}}=
\frac{\Ldel\rho}{\rho_0}+\frac{\Ldel c_p}{c_{p0}}
                        +\frac{\Ldel W}{W_0}+\frac{\Ldel\Theta}{\Theta_0}\,,
\label{eq:unno_heat-flux-perturb}
\end{equation}
which may be compared with \eq{eq:heat-flux-perturb}.

\subsection{A nonlocal mixing-length model}
\label{sec:nonlocal_mlm}

One of the major assumptions in the above described local
mixing-length theory is that the characteristic
length scale $\ell$ must be shorter than any scale length associated with 
the structure of the star. This condition is violated, however, for solar-like 
stars and red giants where evolution calculations reveal a typical value for 
the mixing-length parameter $\alpha=\ell/H_{\mathrm{p}}$ of order unity, where 
$H_{\mathrm{p}}$ is the pressure scale height. This implies that fluid
properties vary over the extent of a convective element and the
superadiabatic gradient can vary on a scale much shorter than $\ell$.

The nonlocal theory takes some account of the finite size of a
convective element and averages the representative value of a variable
throughout the eddy. 
\citet{Spiegel63} proposed a nonlocal description based
on the concept of an eddy phase space and derived an equation for the
convective flux which is familiar in radiative transfer theory. The
solution of this transfer equation yields an integral expression which
would convert the usual ordinary differential equations of stellar model
calculations into integro-differential equations. An approximate solution
can be found by taking the moments of the transfer equation and using the
Eddington approximation to close the system of moment equations at second
order \citep{Gough77b}. The next paragraphs provide a brief overview
of the derivation of the nonlocal convective fluxes, following
\citet{Gough77b} and \citet{Balmforth92a}.

\subsubsection{Formulation for stationary atmospheres}
\label{s:nonlocal-mlm}
In the generalized mixing-length model proposed by \citet{Spiegel63},
the turbulent convective elements are described by a distribution
function $\psi(x_i, u_i,t)$ representing the number density
of elements within an ensemble in the six-dimensional phase 
space $(x_i, u_i)$, where $u_i$ is the velocity vector of an eddy at
the position $x_i$. The conservation of the eddies within the ensemble 
gives rise to an equation for the evolution of $\psi$
\begin{equation}
\frac{\partial\psi}{\partial t}
+\frac{\partial}{\partial x_i}(u_i\psi)
+\frac{\partial}{\partial u_i}(\dot u_i\psi)
=q-\frac{u\psi}{\ell}\,,
\label{phase-space-conservation}
\end{equation}
where a dot denotes an Eulerian time derivative. The source term $q$ describes the
local creation of convective elements, whereas the last term on the
right hand side accounts for their annihilation after having 
travelled a distance of one mixing length.

In the static mean atmosphere the eddy ensemble is described by a 
statistically steady-state distribution function with vanishing
time-derivative and the conservation equation in a plane parallel geometry
becomes
\begin{equation}
\mu\frac{\rd\Psi}{\rd z}+\frac{\Psi}{\ell}=\frac{{\cal Q}}{\ell}
\,,
\label{phase-space-conservation-ppa}
\end{equation}
where $\Psi=u\psi$ and $\mu=\cos\theta$, with $\theta$ being 
the angle between the vertical co-ordinate $z$ and the direction of 
fluid element trajectories.
The nonlinear term $\partial(\dot u_i\psi)/\partial u_i$ in
\eq{phase-space-conservation} describes the
acceleration of the elements through buoyancy and pressure forces, and
has been absorbed into the source function ${\cal Q}$, which changes
\eq{phase-space-conservation-ppa} into a form like the
radiative transfer equation in a grey atmosphere. Thus the equation can be
formally solved for $\Psi$ as function of ${\cal Q}$ 
\citep[e.g.,][]{Chandrasekhar50}, where the first moment, obtained by
multiplying \eq{phase-space-conservation-ppa} by $h^\prime\ell$ and
integrating with respect to $\mu$, 
can be interpreted as the convective heat flux written as
\begin{equation}
{\cal F}_{\mathrm{c}}=\int_{-1}^{1}|h^\prime|\Psi\mu \rd\mu=
       \int_0^\infty|h^\prime|{\cal Q}(\xi_0)E_2(|\xi_0-\xi|) \rd\xi_0
\,,
\label{nlmlt-heat-flux-int}
\end{equation}
where $E_2$ denotes the second exponential integral and we assumed 
symmetry for upward and downward moving elements (compatible to the Boussinesq 
equations, which are up-down symmetric, but not necessarily their solutions) 
each having a specific enthalpy fluctuation of $h^\prime$. The 
vertical displacement of an element from its initial position has been 
redefined by the more natural variable
\begin{equation}
\rd\xi=-\frac{\rd z}{\ell}
\,.
\label{nlmlt-idependent-variable}
\end{equation}

There is still to define the source function $|h^\prime|{\cal Q}(\xi_0)$
for which \citet{Spiegel63} chooses in the limit for small mixing lengths to set it
equal to the convective heat flux $F_{\mathrm{c}}(\xi_0)$,
as it would be computed in a purely local way, i.e., as given by 
\eq{eq:heat-flux-sigma}.
Thus we still have in this formulation inherent the approximation
that the mixing length has to be small compared to any scale length 
in the star. The convective heat flux is proportional to the
cube of the eddy growth rate $\sigma_{\mathrm{c}}$, and $\sigma_{\mathrm{c}}$ is proportional
to the superadiabatic temperature gradient $\beta$ (\ref{eq:beta}).
In order to account for the case where the trajectories 
of the eddies are in the order or larger than the local scale height 
of the envelope, \citet{Spiegel63} used variational
calculations to suggest that $\beta$ in \eq{eq:dog-lmlt-conv-sigma} 
should be replaced by its average value
\begin{equation}
{\cal B}(z)=\frac{2}{\ell}\int_{z-\ell/2}^{z+\ell/2}\beta(z_0)\cos^2
          \left[\frac{\pi}{\ell}(z_0-z)\right] \rd z_0
\,.
\label{average-beta} 
\end{equation}

We are now faced with integral expressions, which would convert
the ordinary differential equations of stellar structure into
integro-differential equations, increasing considerably the
complexity of the 
numerical treatment. Fortunately, an approximate solution of 
\eq{phase-space-conservation-ppa} for $\Psi$ and thus for the 
convective fluxes may be obtained by taking moments of the transfer 
equation and using the Eddington approximation to close the system of
equations at second order \citep{Gough77b}. This reduces the integral equation to
\begin{equation}
\frac{1}{a^2}\frac{\rd^2{\cal F}_{\mathrm{c}}}{\rd\xi^2}={\cal F}_{\mathrm{c}}-F_{\mathrm{c}}
\,,
\label{nltflx}
\end{equation}
when using for the source function $|h^\prime|{\cal Q}=F_{\mathrm{c}}$ 
and for the additional parameter $a=\sqrt 3$. The exact solution of
this equation is
\begin{equation}
{\cal F}_{\mathrm{c}}(\xi)=\int_{-\infty}^{\infty}F_{\mathrm{c}}(\xi_0){\cal K}(\xi,\xi_0) \rd\xi_0
\,,
\label{nlmlt-heat-flux-approx}
\end{equation}
where the kernel ${\cal K}$ is given by
\begin{equation}
{\cal K}(\xi,\xi_0)=\frac{1}{2}a\exp(a|\xi-\xi_0|)
\,.
\label{approx-kernel}
\end{equation}
Thus, the approximation~(\ref{nltflx}) is equivalent to replacing the kernel
$E_2(|\xi_0-\xi|)$ in (\ref{nlmlt-heat-flux-int}) by the simpler
form of \eq{approx-kernel}. This suggests, however, a different
value for the coefficient $a$, which can be determined by demanding
that terms in the Taylor expansions about $\xi$ of ${\cal K}$ and $E_2$
differ only at fourth order, which yields $a=\sqrt 2$. 

The expression for the averaged superadiabatic temperature gradient ${\cal B}$,
\eq{average-beta}, may be obtained in a similar way. The integration
limits can be formally set to $\pm\infty$, if contributions to ${\cal B}$
from beyond the trajectory of the eddy are assumed to vanish. By 
approximating the kernel, which may be written as $2\cos^2[\pi(\xi_0-\xi)]$,
by ${\cal K}$, one obtains
\begin{equation}
\frac{1}{b^2}\frac{\rd^2{\cal B}}{\rd\xi^2}={\cal B}-\beta
\,,
\label{nltbbb}
\end{equation}
where $b\simeq\sqrt{61}$ using the Taylor-expansion technique
described above.

The momentum flux of the eddies within the ensemble can be treated using
exactly the same approach as for the convective heat flux, where one 
obtains a similar expression for the turbulent pressure written as
\begin{equation}
\frac{1}{a^2}\frac{\rd^2{\cal P}_{\mathrm{t}}}{\rd\xi^2}={\cal P}_{\mathrm{t}}-p_{\mathrm{t}}
\,.
\label{nlttpr}
\end{equation}

The nonlocal equations discussed above were derived in the 
physical picture in which the convective elements are accelerated 
from rest and whose evolutions along their trajectories are
described by linear growth rates, as already discussed
in the local theory. Obviously the nonlocal equations may also
be discussed in the view of the second picture, where the eddies are
regarded as cells with the size of one mixing-length and centred at
some fixed height, again, similar as in the local treatment of mixing 
length theory. This is the picture in which \citet{Gough77b} discussed the 
derivation of the nonlocal equations, which
corresponds to treating the finite extent of the eddy and
the nonlocal transfer of heat and momentum across it by using the
averaging idea which had led to the equation for ${\cal B}$ described above.
The integral expression~(\ref{average-beta}) may then be interpreted such
that an eddy centred at $z_0$ samples ${\cal B}$ over the range 
determined by the extend of the eddy, i.e.$\;(z_0-\ell/2,z_0+\ell/2)$.
Moreover, the averaged convective fluxes ${\cal F}_{\mathrm{c}}$ and ${\cal P}_t$ are
constructed not only by eddies located at $z_0=z$, but by all the eddies
centred between $z_0-\ell/2$ and $z_0+\ell/2$.
Hence the two additionally parameters $a$ and $b$ (three, if the 
kernels for the convective heat flux and turbulent pressure are 
treated differently) control the spatial coherence of the ensemble of eddies
contributing to the total heat and momentum flux ($a$), and the degree to which
the turbulent fluxes are coupled to the local stratification ($b$).
Theory suggests values for these parameters, but the quoted values are
approximate and to some extent these parameters are free.
These parameters control the degree of ``nonlocality'' of convection,
where low values imply highly nonlocal solutions and in the limit
$a,b\rightarrow\infty$, the system of equations reduces to the local theory.
\citet{Balmforth92a} explored the effect of $a$ and $b$ on the turbulent fluxes
in the solar case very thoroughly and \citet{ToothGough88} calibrated
$a$ and $b$ against laboratory convection.

\smallskip
\citet{DupretEtal06a} proposed to calibrate the nonlocal convection parameters $a$ and $b$ 
against 3D large-eddy simulations (LES) in the convective overshoot regions.

\subsubsection{Formulation for radially pulsating atmospheres}
\label{sec.TNLMLT}
In order to derive expressions for the pulsationally induced
perturbations of the nonlocal turbulent fluxes, the time
derivative in \eq{phase-space-conservation} has to
be taken into account. One can then proceed as in the static 
atmosphere and take moments of this equation to derive an 
expression for the convective flux. This expression may be
linearized which provides a differential form for the perturbations
to the turbulent fluxes including the term of the time-dependent
source function $\cal Q$. The time-dependence of $\cal Q$, as 
introduced in the static discussion, can be described as the
instantaneous creation of elements, whereas the
term $\partial\psi/\partial t$ in \eq{phase-space-conservation}
accounts for the phase delay between the source function and
the response of the distribution function $\psi$. In the local
prescription of Gough, the distortion of the mean eddy size between
creation and annihilation with the mean environment is accounted for 
appropriately and thus also the phase lag between the deformation of the
mean environment and the response of the turbulent fluxes. Hence, the
phase lag due to the $\partial\psi/\partial t$ term is already
taken into account by the source function $\cal Q$, when it is set
to Gough's locally computed turbulent fluxes. Thus
\eq{phase-space-conservation} becomes essentially the form of 
\eq{phase-space-conservation-ppa} and the perturbations to the
turbulent fluxes are obtained by perturbing linearly equations
(\ref{nltflx}), (\ref{nltbbb}) and (\ref{nlttpr}) leading to \citep{Balmforth92a}
\begin{equation}
\frac{1}{\epsilon^2}\frac{\partial}{\partial\xi}
\left[\frac{\partial}{\partial\xi}(\delta{\cal T})-
 \frac{\partial}{\partial\ln p_{0}}\left(\frac{\Ldel p}{p_0}\right)
 \frac{\partial{\cal T}}{\partial\xi}
\right]=
\delta{\cal T}-\delta {\cal S}+({\cal T}-{\cal S})
\frac{\partial}{\partial\ln p_{0}}\left(\frac{\Ldel p}{p_0}\right)
\,,
\label{nlmlt-conv-fluxes-perturb}
\end{equation}
where ${\cal T}$ is either of ${\cal F}_{\mathrm{c}}$, ${\cal P}_{\mathrm{t}}$ or
${\cal B}$, and $\cal S$ is the corresponding source function or $\beta$.
The Lagrangian perturbations to the pressure and the quantity ${\cal T}$
are represented by $\delta p$ and $\delta {\cal T}$, respectively, and
the parameter $\epsilon$ is either $a$ or $b$. The corresponding
perturbation to the local quantities $\cal S$ are obtained from Gough's 
local, time-dependent convection formulation, as discussed in 
Section~\ref{sec:time-dependent_LMLM}.

\smallskip
Gough's nonlocal generalization was adopted, in a simplified form, by \citet{DupretEtal06c} 
for \citepos{GrigahceneEtal05} convection model. It was implemented only in the 
pulsation calculations and, instead of perturbing the nonlocal
equations as shown by Eq.~(\ref{nlmlt-conv-fluxes-perturb}), \citet{DupretEtal06c}
replaced the turbulent fluxes, $(F_{\mathrm{c}}, {\cal{F}_{\mathrm{c}}})$ and 
$(p_{\mathrm{t}},{\cal{P}_{\mathrm{t}}})$ in Eqs.~(\ref{nltflx}) and (\ref{nlttpr})
by their Lagrangian perturbations.

\subsection{Unno's convection model generalized for nonradial oscillations} 
\label{sec:unno_gabriel_mlm}

In this section, we summarize the model by \citet{GrigahceneEtal05}, who adopted and generalized
\citepos{Unno67} description for approximating the nonlinear terms in the
fluctuating convection equations and Gabriel \etal's (\citeyear{GabrielEtal75}, see also \citealp{Gabriel96})
approach for describing time-dependent convection in nonradially pulsating stellar models.

For completeness we summarize the nonradial pulsation equations of the stellar
mean structure in Appendix~\ref{app:meanpert}.

\subsubsection{Equations for the convective fluctuations}
\label{sec:conveq}

As in the previous section of radially pulsating stars, we also adopt the 
Boussinesq approximation to the convective fluctuation equations for
nonradially pulsating stars. The detailed discussion of the derivation
is presented in Appendix~\ref{app:G05_conv}. Here, we introduce and discuss the final equations
that are used in the stability computations.

The continuity equation is the same as for the radial case. 
The fluctuating momentum and thermal energy equations for a nonradially pulsating star,
describing the (Eulerian) convective velocity field $\vec{u}$ and (Eulerian) entropy fluctuation
$s^\prime$ are
\begin{equation}
\nabla \cdot \vec{u}=0\,,
\label{eq*}
\end{equation}


\begin{equation}
\ob{\rho }\,\frac{\rd \vec{u}}{\rd t}
          +\Lambda \frac{\overline{\rho }\vec{u}}{\tau_{\mathrm{c}}}
          +\ob{\rho} (\vec{u}\cdot \nabla) \vec{U}
=-\nabla p^\prime
 -\frac{\ob{\rho}\,\ob{\tdel}}{\ob{c_p}}s^\prime\left(\vec{{g}}-\frac{\rd \vec{U}}{\rd t}\right),
\label{eq23}
\end{equation}

\begin{equation}
 \ob{\rho}\,\ob{T}\frac{\rd s^\prime}{\rd t}
+\ob{\rho}\,\ob{T}\frac{s^\prime}{\tau_{\mathrm{c}}}
+{\left( \rho T\right)^\prime }\frac{\rd\ob{s}}{\rd t}
+\ob{\rho}\,\ob{T} \nabla \ob{s} \cdot\vec{u}
= -\nabla\cdot \vec{F}^\prime_{\mathrm{R}}\,,
\label{eq:fluc_thermal_nonradial}
\end{equation}
in which we set the nuclear reaction rate $\varepsilon=0$, because we consider only
convective envelopes, and $\ob{\tdel}:=(\partial\ln\ob{\rho}/\ln\ob{T})_p$ is, as before, 
the isobaric expansion coefficient. 

Note that the radial component 
$(\ob{\rho}\,\ob{T} \nabla \ob{s} \cdot\vec{u})_{r}$ = $-\ob{\rho}\,\ob{c_p}\beta w$, 
where the superadiabatic lapsrate $\beta$ is defined by Eq.~(\ref{eq:beta}) and $w$ is
the vertical component of $\vec{u}$.

The second term on the left-hand side of both Eqs.~(\ref{eq23}) and 
(\ref{eq:fluc_thermal_nonradial}) approximate the nonlinear terms [see
also Eqs.~(\ref{eq:nonlinear_approximation_velocity}) and
(\ref{eq:nonlinear_approximation_temperature})]
\begin{equation}
\nabla\cdot(\rho\vec{u}\vec{u}-\ob{\rho\vec{u}\vec{u}}\,)
\simeq\ob{\rho}\left[(\vec{u}\cdot\nabla)\vec{u}-\ob{(\vec{u}\cdot\nabla)\vec{u}}\right]
\simeq\Lambda \frac{\ob{\rho }\vec{u}}{\tau_{\mathrm{c}}}\,,
\label{eq20}
\end{equation}
and
\begin{equation}
\ob{\rho}\left[\vec{u}\cdot(T\nabla s)^\prime-\ob{\vec{u}\cdot(T\nabla s)^\prime}\right]
-\bm{\upsigma}:\nabla\vec{u}+\ob{\bm{\upsigma}:\nabla\vec{u}}
\simeq\ob{\rho }\ob{T}\frac{s^\prime}{\tau_{\mathrm{c}}}\,,
\label{closen}
\end{equation}
respectively, as suggested by \citet{Gabriel96}, following the approximation
by \citet[][see also Section~\ref{sec:static_LMLM_2}]{Unno67},
to close the system of equations.
This approximation of the nonlinear terms by means of a scalar drag coefficient assumes that the
nonlinear terms are parallel to $\vec{u}$, which corresponds in a sense to
\citepos{Heisenberg46} formulation.
Note that $\ob{\bm{\upsigma}:\nabla\vec{u}}=:\ob{\rho}\epsilon_{\mathrm{t}}$.

The parameter $\Lambda$ is a dimensionless constant of order unity, 
and $\tau_{\mathrm{c}}$ 
is the characteristic turn-over time of the convective elements. It can be related to
the mixing length $\ell=-\alpha (\rd \ln\ob{p}/\rd r)^{-1}$ and 
the radial component of mean turbulent velocity 
\begin{equation}
\label{turnovertime}
\tau_{\mathrm{c}}=\ell / \sqrt{\overline{u_r^2}}.
\end{equation}
The gradient of the pressure fluctuations in Eq.~(\ref{eq23}) can be eliminated by 
adopting for the velocity field a structure similar to Eq.~(\ref{e:velo-expand}), as
discussed also in Section~(\ref{sec:time-dependent_LMLM_2}). 
These are the equations adopted by \citet{GrigahceneEtal05} for nonradially
pulsating stars with convective envelopes and may be compared with
Gough's Eqs.~(\ref{eq:fluctuating_cont_ba})\,--\,(\ref{eq:fluctuating_thermal_energy_ba})
for radially pulsating atmospheres. 

For the radiative transfer we adopt the diffusion approximation and express 
$\nabla \cdot \vec{F}^\prime_{\mathrm{R}}$ as
\begin{equation}
\nabla \cdot \vec{F}^\prime_{\mathrm{R}}=\omega _{%
\mathrm{R}}s'\overline{\rho}\,\ob{T},\label{eqrad}
\end{equation}
with 
\begin{equation}
\label{omr}
\omega _{\mathrm{R}}=\frac{1}{\tau _{%
\mathrm{R}}}=\frac{4ac}{3}\frac{\overline{T}^{3}}{c_{\mathrm{p}}\overline{%
\kappa }\overline{\rho }^{2}\mathcal{L}^{2}}\,,
\end{equation}
where $\tau_{\mathrm{R}}$ is the characteristic cooling time of
turbulent eddies due to radiative losses. $\mathcal{L}$ is the
characteristic length of the eddies. It is related to the mixing length $\ell$
by $\mathcal{L}^2=(2/9)\ell^2$ to recover the mixing-length formulation
by \citet{BohmVitense58}. 
With these definitions the energy \eq{eq:fluc_thermal_nonradial} can be rewritten as:
\begin{equation}
\frac{\left( \rho T\right)' }{\overline{\rho T}}\frac{\rd %
\overline{s}}{\rd t}+\frac{\rd s'}{\rd t}+\vec{u}%
\cdot \nabla \overline{s}=-\frac{\omega _{\mathrm{R}}\tau
_{\mathrm{c}}+1}{\tau _{\mathrm{c}}}\,s'. 
\label{eq28}
\end{equation}


As shown by \citet{Unno67}, 
searching for stationary solutions of Eqs.~(\ref{eq*}), (\ref{eq23}) and 
(\ref{eq28}), 
leads to the 
classical mixing-length equation 
with the well-known cubic equation 
for the radiative dimensionless temperature
gradient $\nabla_{\mathrm{rad}}:=(\rd\ln T/\rd\ln p)_{\mathrm{rad}}$
\begin{equation}
\label{mixcub}
\frac{9}{4}\;{\tilde S}^3\;+\;{\tilde S}^2\;+\;{\tilde S}
\;=\;A\;(\nabla_{\mathrm{rad}}-\nabla_{\mathrm{ad}})\;,
\end{equation}
where $\nabla_{\mathrm{ad}}$ is the dimensionless adiabatic temperature gradient,
${\tilde S}=\tau_{\mathrm{R}}/\tau_{\mathrm{c}}=:S^{1/2}$ 
is the (square root of the) convective efficacy 
[see also Eq.~(\ref{eq:dog-lmlt-conv-sigma})], and
\begin{equation}
A:=\frac{\ob{p}+p_{\mathrm{t}}}{\Phi H_{\mathrm{p}}^2}\frac{\tau_{\mathrm{r}}^2\ob{\tdel}}{\Lambda \ob{\rho}}\,.
\label{e:cubic-a}
\end{equation}
For models with turbulent pressure included, $\nabla_{\mathrm{ad}}$ in Eq.~(\ref{mixcub}) 
needs to be multiplied by $\rd \ln\ob{p}/\rd \ln(\ob{p}+p_{\mathrm{t}})$. This is, however,
neglected in most stellar evolutionary codes because it leads to convergence problems
in any local convection formulation \citep[e.g.,][]{Gough77b}.
For isotropic turbulence and $\Lambda=8/3$ Eqs.~(\ref{mixcub}) and (\ref{e:cubic-a}) 
represent the cubic equation as found, for example, in the probably 
most commonly adopted mixing-length formulations by \citet{BohmVitense58} 
and \citet{Paczynski69}.


\subsubsection{Perturbation of the convection}
\label{convpert}

In order to determine the pulsational perturbations of the terms linked to convection we
proceed as follows. We perturb Eqs.~(\ref{eq*}), (\ref{eq23}) and (\ref{eq28}). 
Then we search for solutions of the form $\delta \left( X'\right)
=\delta \left( X'\right) _{\vec{k}}\,\mathrm{e}^{i\,\vec{k}\cdot \vec{r}%
}\,\mathrm{e}^{-{\mathrm{i}}\omega \,t}$, assuming constant coefficients, where $\delta$
denotes a linear pulsational perturbation in a Lagrangian frame of reference,
and $\omega$ is the (complex) eigenfrequency of the pulsations. 
These particular solutions are integrated over all wavenumber values of $k_{\theta }$ 
and $k_{{\varphi} }$ such that $k_{\theta }^{2}+k_{{\varphi} }^{2}=k_{r}^{2}/({\Phi}-1)$, 
assuming ${\Phi}$ to be constant, and
that every direction of the horizontal component of $\vec{k}$ has the same
probability. 
We have to introduce this distribution of $\vec{k}$ values to
obtain an expression for the perturbation of the Reynolds stress tensor which
allows the proper separation of the variables in the equation of motion 
\citep{Gabriel87}.

Horizontal averages are computed on a scale larger than the size of 
the eddies but smaller than the horizontal wavelength of the nonradial
oscillations ($r/l$).

The perturbed Eq.~(\ref{eq*}) becomes, for a given $\vec{k}$,
\begin{equation}
\vec{k\cdot }\delta \vec{u}=0.
\end{equation}
The perturbation of Eq.~(\ref{eq28}) gives: 
\begin{eqnarray}
\label{enconv0}
&&\left( \frac{\rho' }{\overline{\rho }}+\frac{T'}{\overline{T}}%
\right) \frac{\rd \delta \overline{s}}{\rd t}+\frac{\rd %
\left( \delta s'\right) }{\rd t}+\delta \vec{u}\cdot \nabla 
\overline{s}+\vec{u}\cdot \delta\left( \nabla \overline{s}\right) \nonumber
\\
&&=-\omega_{\mathrm{R}}\,\delta s' - \delta \omega_{%
\mathrm{R}}s' -\delta\left(\frac{s'}{\tau _{%
\mathrm{c}}}\right).
\end{eqnarray}

We recall that the term $s' / \tau_{\mathrm{c}}$ corresponds to the
closure approximation adopted in the mixing-length formulation by \citet{Unno67} 
for the energy equation [Eq.~(\ref{closen})]. When $\omega\,\tau_{\mathrm{c}} \ll 1$, 
convection instantaneously adapts to the changes due to oscillations and we could expect
that its perturbation behaves like: 
\begin{equation}
\label{pclose0}
\delta\left(\frac{s'}{\tau_{\mathrm{c}}}\right)=\frac{s'}{\tau_{%
\mathrm{c}}} \left(\frac{\delta s'}{s'}-\frac{\delta\tau_{\mathrm{%
c}}}{\tau_{\mathrm{c}}}\right).
\end{equation}
This is the treatment adopted by \citet{Gabriel96}. 

However, many complex physical process, including the whole cascade of energy are extremely 
simplified in this approach. Therefore, it is clear that much uncertainty is
associated to the perturbation of this term. A point to emphasize is that
the occurrence of the non-physical spatial oscillations 
[see discussion about Eq.~(\ref{eq:imaginary-diffusivity})]
is directly linked to the perturbation of this closure term. When these
oscillations occur ($\omega\,\tau_{\mathrm{c}} \gg 1$), the radial
derivatives of $\delta\overline{s}$ and $\delta s'$ are of the order of 
$(\omega\tau_{\mathrm{c}}/\ell)\delta\overline{s}$ and $(\omega\tau_{\mathrm{c}%
}/\ell)\delta s'$ respectively. Therefore, if we take equation (\ref{pclose0}),
we see that the order of magnitude of the perturbation on the right-hand
side of Eq.~(\ref{closen}) is $\omega\tau_{\mathrm{c}}$ times larger
than the left-hand side. To have the same order of magnitude, the
perturbation of the left-hand side should rather be given by
\begin{equation}
\label{pclose1}
\delta\left(\frac{s'}{\tau_{\mathrm{c}}}\right)={\hat\beta}\,\omega\,{\delta s'}%
 -s'\frac{\delta\tau_{\mathrm{c}}}{\tau_{\mathrm{c}}^2},\ 
\end{equation}
where ${\hat\beta}$ is a (complex) coefficient of order unity. 
For a continuous transition between $\omega\,\tau_{\mathrm{c}} \ll 1$ [Eq.~(\ref{pclose0})]
and $\omega\,\tau_{\mathrm{c}} \gg 1$ Eq.~(\ref{pclose1})] 
\citet{GrigahceneEtal05} proposed the expression 
\begin{equation}
\delta\left(\frac{s'}{\tau_{\mathrm{c}}}\right)=\frac{s'}{\tau_{%
\mathrm{c}}} \left[(1+{\hat\beta}\omega\tau_{\mathrm{c}})\frac{\delta s'}{%
s'} -\frac{\delta\tau_{\mathrm{c}}}{\tau_{\mathrm{c}}}\right]\,.
\end{equation}

We note, however, that a nonlocal treatment of the convective fluxes, such as 
the model discussed in Section~\ref{sec:nonlocal_mlm}, does not necessarily need the 
ad-hoc introduction of the additional parameter $\hat\beta$.

Perturbing Eq.~(\ref{turnovertime}) leads to 
\begin{equation} 
\label{plifetime}
\frac{\delta\tau_{\mathrm{c}}}{\tau_{\mathrm{c}}}=\frac{\delta\ell}{\ell}-\frac{%
\overline{u_r\delta u_r}}{\ob{u_r^2}}\,,
\end{equation}
and perturbing equation (\ref{eq23}) provides 
\begin{eqnarray}
-{\mathrm{i}}\omega \overline{\rho} \delta \vec{u} &=&\delta \left( \frac{\rho' }{%
\overline{\rho }}\right) \nabla \overline{p}+\frac{\rho' }{\overline{%
\rho }}\delta \left( \nabla \overline{p}\right)-\delta \left( \nabla { p^\prime}\right)\nonumber\\
&-&\overline{\rho} \vec{u}\cdot \delta \nabla \vec{U}-\frac{\Lambda
\overline{\rho} \vec{u}}{\tau _{\mathrm{c}}}\left( \frac{\delta \overline{%
\rho} }{\overline{\rho} }-\frac{\delta \tau _{\mathrm{c}}}{\tau _{\mathrm{c}}%
}\right) -\frac{\Lambda \overline{\rho} \delta \vec{u}}{\tau _{\mathrm{c}}}.
\label{eq34}
\end{eqnarray}
Taking the divergence of this equation leads to an expression for $\delta p'$.
Another approach is to take the curl of these expression, which eliminates the
pressure fluctuations (see also discussion in Section~\ref{sec:static_LMLM_1}).

The expressions for the remaining perturbed quantities are listed in 
Appendix~\ref{sec:nonradial-confluc}.
 

\subsubsection{Perturbation of the convective heat flux}
\label{convfluxpert}

We see in Eq.~(\ref{eq32}) the appearance of the 
convective flux perturbation.
To obtain it, we perturb Eq.~(\ref{fconv}) leading to (overbars of mean values are omitted)
\begin{equation}
\delta \vec{F}_{\mathrm{c}}=\vec{F}_{\mathrm{c}}\left( \frac{\delta \rho }{%
\rho}+\frac{\delta T}{T}\right) +{\rho }{T}\left( 
\overline{\delta s'\vec{u}}+\overline{s'\delta \vec{u}}\right).
\label{F}
\end{equation}
The radial component of this equation is
\begin{equation}
\frac{\delta F_{\mathrm{c}, \mathrm{r}}}{F_{\mathrm{c, \mathrm{r}}}}=\left( \frac{\delta \rho }{%
\rho} +\frac{\delta T}{T}\right)+\frac{\overline{u_{\mathrm{r}}\delta s'}}{\overline{u_{\mathrm{r}}s'}}+%
\frac{\overline{u_{\mathrm{r}}\delta u_{\mathrm{r}}}}{\overline{u_{\mathrm{r}}^2}},
\end{equation}
and the horizontal component
\begin{equation}
\frac{\delta F_{\mathrm{c, h}}}{F_{\mathrm{c, r}}}=\frac{\overline{u_{\mathrm{h}}\delta
s'}}{\overline{u_{\mathrm{r}}s'}}+\frac{\overline{u_{\mathrm{r}}\delta u_{\mathrm{h}} }}{\overline{u_{\mathrm{r}}^2}}.
\end{equation}

From Eqs.~(\ref{dsh}), (\ref{plifetime}) and (\ref{pomr}) we obtain the
explicit form for the radial component of the perturbation of the convective
flux
\begin{eqnarray}
\frac{\delta F_{\mathrm{c, r}}}{F_{\mathrm{c, r}}} &=&\frac{\delta \rho }{\rho%
}+\frac{\delta T}{T} {\mathrm{i}}\omega \tau _{\mathrm{c}} D 
\left( 1-\delta\right)
\frac{\delta s}{c_{\mathrm{p}}} +C \left[\frac{%
\rd \delta s}{\rd s}-\frac{\rd \xi_{\mathrm{r}}}{\rd r}%
\right] \nonumber \\
&&-\omega _{\mathrm{R}}\tau _{\mathrm{c}} D\left( 3\frac{%
\delta T}{T}-\frac{\delta c_{\mathrm{p}}}{c_{\mathrm{p}}}-\frac{\delta
\kappa }{\kappa }-2\frac{\delta \rho }{\rho }\right)  \nonumber \\
&&+ (-{\mathrm{i}}\omega \tau _{\mathrm{c}}+2\omega _{\mathrm{R}}\tau
_{\mathrm{c}}+1)D 
\frac{\overline{u_{\mathrm{r}}\delta u_{\mathrm{r}}}}{\overline{u_{\mathrm{r}}^2}}  \nonumber \\
&&+(2\omega_{\mathrm{R}}\tau _{\mathrm{c}}+1)D\,\frac{%
\delta\ell}{\ell},  \label{dfcr}
\end{eqnarray}
where the expressions $B$, $C$, $D$ and $\overline{u_{\mathrm{r}}\delta u_{\mathrm{r}}/u_{\mathrm{r}}^2}$ are 
given in Appendix~\ref{sec:nonradial-confluc}. 
From Eqs.~(\ref{dsh}) and (\ref{vhdvr}), and using the notation (\ref{dfcdef}) 
for the convective heat flux, we find after some algebra that

\begin{eqnarray}
\frac{\delta F_{\mathrm{c, h}}}{F_{\mathrm{c, r}}} &=&\frac{({\Phi}-1)C\left(
B+1\right) }{2\left( B-C\right) }\frac{\delta s}{\rd s/\rd \ln
r}  \nonumber \\
&&+\:\frac{{\Phi}-1}{2 B}\left[ \frac{C\left( B+1\right) }{B-C}
+\frac{2{\Phi}-1}{\Phi-1}\right] \frac{%
\delta p}{\rd p/\rd \ln r}  \nonumber \\
&&+\left[\frac{{\Phi}-1}{{\Phi}}\left(\frac{C(B+1)}{B-C}+\frac{B-1}{2 B}
+({\Phi}-1)\frac{C(B+1)^2}{2 B(B-C)}\right)\right.  \nonumber \\
&&\;\;\;\left.+\frac{2{\Phi}-1}{2 B}\right]\,\left( \frac{\xi_{\mathrm{h}}}{r}-\frac{\xi_{\mathrm{r}}}{r}\right)  \nonumber\\
&&-\:\frac{({\Phi}-1)(B-1)}{2 {\Phi} B }\left[ \frac{C\left( B+1\right) }{B-C}%
+\frac{2{\Phi}-1}{{\Phi}-1}\right] \frac{\rd \xi_{\mathrm{h}}}{\rd r}.  \label{dfch}
\end{eqnarray}

\subsubsection{Perturbation of the turbulent pressure}

\label{pturbpert}

The perturbed turbulent pressure (appearing explicitly in Eqs.~\ref{pmm}, \ref{pmmt}, \ref{eq44}, \ref{vhdvr}, \ref{dvh})
 is directly obtained by perturbing Eq.~(\ref{eq11}) leading to
\begin{equation}
\frac{\delta p_{\mathrm{t}}}{ p_{\mathrm{t}}}=\frac{\delta \rho }{\rho}+2%
\frac{\overline{ u_{\mathrm{r}}\delta u_{\mathrm{r}}}}{\overline{u_{\mathrm{r}}^2}}\,,  \label{pturbperteq}
\end{equation}
where $\overline{u_{\mathrm{r}}\delta u_{\mathrm{r}}}/\overline{u_{\mathrm{r}}^2}$ is 
given by Eq.~(\ref{eq44}).
Since a term proportional to $\rd \delta s/\rd s$ is present in 
Eq.~(\ref{eq44}), 
the order of the system of differential equations is increased by one
with the inclusion of the turbulent pressure in the mean equation 
of motion \citep[see also][]{Gough77b}.

\subsubsection{Perturbation of the rate of dissipation of turbulent kinetic
energy into heat}
\label{ecinpert} 

We consider now the perturbation of the last term appearing in the perturbed
energy equation~(\ref{eq32}), $\delta \left( 
\epsilon _{\mathrm{t}}+\overline{\vec{u}\cdot \nabla p}/\overline{\rho} \right) $. This term
also appears in the conservation equation of kinetic turbulent energy. We
can thus determine it by perturbing Eq.~(\ref{ecin}), and obtain
\begin{equation}
\delta \left( \rho \epsilon _{\mathrm{t}}+\overline{\vec{u}\cdot \nabla p%
}\right) ={\mathrm{i}}\omega \rho\: \delta \left( \frac{\overline{\rho \vec{u}^{2}}}{%
2\rho }\right) +{\mathrm{i}}\omega \overline{\rho \vec{uu}}\,:\, \nabla \vec{\xi}%
.  \label{pecin0}
\end{equation}
The evaluation of the first term gives, using equation (\ref{dvhh}), 
\begin{eqnarray}
{\mathrm{i}}\omega \rho \,\delta \left( \frac{\overline{\rho \vec{u}^{2}}}{2\rho }%
\right)  &=&{\mathrm{i}}\omega p_{\mathrm{t}}\left[ 
\frac{\overline{u_{\mathrm{r}}\delta u_{\mathrm{r}}}}{\overline{u_{\mathrm{r}}^2}}
+\frac{\overline{\delta u_{\theta }u_{\theta }}}{\overline{u_{\mathrm{r}}^2}}+\frac{\overline{%
\delta u_{{\varphi} }u_{{\varphi} }}}{\overline{u_{\mathrm{r}}^2}}\right]   \nonumber \\
&=&{\mathrm{i}}\omega p_{\mathrm{t}}\,{\varphi}\:\frac{\overline{u_{\mathrm{r}}\delta u_{\mathrm{r}}}}{\overline{u_{\mathrm{r}}^2}}.
\end{eqnarray}
The second term of equation (\ref{pecin0}) leads to 
\begin{equation}
\overline{\rho \vec{uu}}\,:\, \nabla \vec{\xi}{=}p_{\mathrm{t}}\left[ 
\frac{\rd \xi_{\mathrm{r}}}{\rd r}+\frac{{\Phi}-1}{2}\left( 2\frac{\xi_{\mathrm{r}}}{r}%
-l \left( l +1\right) \frac{\xi_{\mathrm{h}}}{r}\right) \right] .
\end{equation}
We finally obtain 
\begin{equation}
\delta \left(\rho \epsilon _{\mathrm{t}}+ \overline{\vec{u}\cdot \nabla p}\right)  =\mathrm{i}\omega p_{\mathrm{t}}\left[ \frac{{\Phi}}{2}\left( \frac{
\delta p_{\mathrm{t}}}{p_{\mathrm{t}}}-\frac{\delta \rho }{\rho }\right)
+ \frac{\rd \xi_{\mathrm{r}}}{\rd r} + \frac{{\Phi}-1}{2}\left( 2\frac{
\xi_{\mathrm{r}}}{r}-l \left( l +1\right) \frac{\xi_{\mathrm{h}}}{r}\right) \right] .
\label{ecinperteq}
\end{equation}

As shown by \citet{LedouxWalraven58} and \citet{GrigahceneEtal05}, it is important to 
emphasize that the turbulent pressure
variation and the turbulent kinetic energy dissipation variations have an opposite
effect on the driving and damping of the modes. This can be seen clearly by considering
the contributions of these terms to the work integral.

\subsubsection{Perturbation of the mixing length}
\label{s:mlt-perturbation}


In Section~\ref{sec:time-dependent_LMLM}, we discussed two descriptions
for the pulsationally distorted convective eddy shape and therefore also 
for the pulsationally modulated mixing length. 
One of the earliest suggestions was provided by \citet{Cowling34}, who proposed 
$\delta\ell/\ell = \xi_{\mathrm{r}}/r$. Cowling's
suggestion was adopted by \citet{BouryEtal75}, and by \citet{Unno67}
in the limit $\omega\tau_{\mathrm{c}}\gg1$ [see also Eq.~(\ref{eq:unno_var_ell})],
where $\tau_{\mathrm{c}}$ is the convective turn-over time scale.
In this limit $\ell$ would vary with the local Lagrangian scale of the mean flow,
a result similar to rapid distortion theory 
(see Section~\ref{sec:time-dependent_LMLM_1}). 

Based on the definition $\ell=\alpha H_{\mathrm{p}}$,
\citet{Schatzman56}, \citet{Kamijo62}, and \citet{Unno67}
adopted the expression
\begin{equation}
\frac{\delta\ell}{\ell}=\frac{\delta H_{\mathrm{p}}}{H_{\mathrm{p}}}=\frac{\delta
(p+p_{\mathrm{t}})}{p+p_{\mathrm{t}}} -\frac{\rd \delta (p+p_{\mathrm{t}})}{\rd (p+p_{\mathrm{t}})}
+\frac{\rd \xi_{\mathrm{r}}}{\mathrm{%
d}r}\,,  \label{icof2}
\end{equation}
in the limit $\omega\tau_{\mathrm{c}}\ll 1$ 
[see also Eq.~(\ref{eq:unno_var_ell})].


Assuming that the convective element, with a constant convective 
turn-over time $\tau_{\mathrm{c}}$, has at the time of its creation the 
vertical extent of the locally defined pressure scale height, 
$\ell=\alpha H_{\mathrm{p}}$, and that $\rho\ell^3=$ constant
during $\tau_{\mathrm{c}}$, \citet{Unno77} and \citet{UnnoEtal89} suggested
\begin{equation}
\frac{\delta\ell}{\ell}=\frac{1}{1-{\mathrm{i}}\omega\tau_{\mathrm{c}}}\left[\frac{\delta
H_{\mathrm{p}}}{H_{\mathrm{p}}} + \frac{{\mathrm{i}}\omega\tau_{\mathrm{c}}}{3}\frac{%
\delta \rho}{\rho}\right].  \label{eq:unnodl}
\end{equation}
By comparing this expression with Eq.~(\ref{eq:verti-wavenumber-puls}) from 
rapid distortion theory, adopted by \citet{Gough77a}, 
we conclude that \citet{UnnoEtal89} implicitly assumed that the eddy shape $\Phi$
is invariant ($\delta\Phi=0$) during the pulsation, irrespective of 
the distortion of the background state.
 
\citet{GrigahceneEtal05} adopts Unno's expression~(\ref{eq:unnodl}) under
the assumption that the perturbation of the mixing length becomes 
negligible in the limit
$\omega\tau_{\mathrm{c}}\ll1$, 
i.e.,
\begin{equation}
\frac{\delta\ell}{\ell}=\frac{1}{1+\left( \omega \tau _{\mathrm{c}}\right) ^{2}}%
\frac{\delta H_{\mathrm{p}}}{H_{\mathrm{p}}}.  \label{icof3}
\end{equation}
Leaving aside the hypothesis $\rho\ell^3= \mathrm{const}$, the
real part of Eq.~(\ref{eq:unnodl}) leads to Eq.~(\ref{icof3}).

Expressions for the perturbation of the non-diagonal
components of the Reynolds stresses were reported by
\citet[][see also \citealt{HoudekGough01}, and \citealt{SmolecEtal11}]{Gabriel87}.

\subsection{Differences between Gough's and Unno's local convection models}
\label{sec:comp_gough_unno_mlm}

Section~\ref{sec:gough_unno_mlm} discussed the detailed equations of
\citepos{Gough65, Gough77a} 
and \citepos{Unno67} local, time-dependent mixing-length models.

Gough's model puts much attention on the dynamics of the linearly growing convective 
elements by means of an eddy creation and annihilation model. In particular, the phase between
the pulsating background state and the convective fluctuations are considered by
adopting a quadratic distribution function for the convective temperature fluctuations
at the time of the eddy creation (zero velocity of the eddies) in order to describe more
realistically the initial conditions of the convective elements. This turns out to be 
crucial for the
damping and driving of the stellar pulsations and consequently for their stability properties.
Although the nonlinear effects are taken into account by the instantaneous eddy 
disruption (annihilation) 
after the eddy's mean lifetime $\tau\propto\sigma_{\mathrm{c}}^{-1}$, where $\sigma_{\mathrm{c}}$ is the 
linear convective growth rate [see Eq.~(\ref{eq:dog-lmlt-conv-sigma}) and 
Appendix~\ref{app:A}], the 
continuous damping effects of the small-scale turbulence are omitted, which are expected
to limit both the velocity and the temperature fluctuations of an eddy and consequently the 
convective velocity. 

Unno's convection model includes the nonlinear advection terms, though in a 
simplified manner, by means of a scalar turbulent viscosity 
[Eqs.~(\ref{eq:nonlinear_approximation_velocity}) and 
(\ref{eq:nonlinear_approximation_temperature}), see also discussion in 
Section~\ref{sec:reynolds_stress_models}], 
but the evolution of the turbulent fluctuations is independent from any initial conditions.
Additional simplifications in Unno's model are the omission of the time-derivatives of the
mean quantities in the fluctuating convection equations, i.e., the third terms on the 
left-hand side
of Eqs.~(\ref{eq:fluctuating_momentum_ba}) and (\ref{eq:fluctuating_thermal_energy_ba}),
and of the (mean) turbulent pressure $p_{\mathrm{t}}$ in the equation of hydrostatic 
support~(\ref{eq:mean_momentum_ba}). 

Another substantial difference between the two convection models by Gough and 
Unno is the treatment of the anisotropy of the turbulent velocity field (or eddy shape)
in both the static and pulsating stellar model. The way how Unno eliminates the fluctuating 
pressure gradient $\nabla p^\prime$ in Eq.~(\ref{eq:fluctuating_momentum_ba}) 
leads to $k_{\mathrm{v}}^2=k_{\mathrm{h}}^2$, i.e., to an (fixed) anisotropy parameter $\Phi=2$ 
(this is also the value adopted by \citealt{BohmVitense58}).
Gough parametrizes $\Phi$, i.e., how the pressure fluctuations couple the horizontal 
to the vertical motion.
The most important difference is, however, the modelling of the pulsationally
modulated eddy shape $\Phi$ and consequently also mixing length $\ell$. 
While Gough adopts rapid distortion theory for describing the variation
of both $\Phi$ and $\ell$ [Eq.~(\ref{eq:verti-wavenumber-puls}), see also discussion in 
Sections~\ref{sec:time-dependent_LMLM} and \ref{s:mlt-perturbation}], 
assumes Unno the eddy shape to be invariant despite of an pulsating background, and 
adopts Eq.~(\ref{eq:unno_var_ell}) for describing the pulsational
variation of $\ell$ (see also Section~\ref{s:mlt-perturbation}). These differences 
affect the stability of the pulsations \citep{Gough77b, Balmforth92a}.

Radiative losses of the convective elements play also a role in determining convective
efficacy and dynamics. Unno adopts the diffusion approximation to radiative transfer 
[Eqs.~(\ref{eq:unno_static_fluc_thermal_energy}) and (\ref{eq:unno_optical-trans-parameter})]. 
Gough describes the radiative losses by means of the Eddington approximation by 
\citet{UnnoSpiegel66} [Eqs.~(\ref{eq:fluc_eddington_F}) and (\ref{eq:fluc_eddington_J}); see also Eqs.~(\ref{eq:heatflux-eddappr} and (\ref{eq:optical-trans-parameter})].

It should also be noted that Gough's model has only been applied to
linear radial pulsations. Efforts to generalize this model to nonradial oscillations
have been reported by \citet{HoudekGough01}, \citet{GoughHoudek01}, 
and \citet{SmolecEtal11, SmolecEtal13}.


\subsection{Differences between Unno's and Grigahc{\`e}ne \etal's local convection models}
\label{sec:unno_gabriel_differences}

\citet{Gabriel96}, and \citepos{GrigahceneEtal05} (G96-05) models are a generalization of
\citepos{Unno67} approach to nonradial oscillations. They treat the momentum 
equation in 
its fully vectorial form for the oscillations and for the convection.

There are also more subtle improvements in the treatment of the closure terms
in the momentum and energy equations for convection.
In the momentum equation for the convective fluctuations,
the linear part of the advection term is treated rigorously in
G96-05 [term $-\rho \vec{u}\cdot\nabla \vec{U}$ in Eq.~(\ref{eq23})], 
while it is neglected in \citet{Unno67}.
G96-05 models use an approximation for the nonlinear terms similar to 
\citet{Unno67} [Eq.~(\ref{eq20})],
but G96-05 use $\Lambda=8/3$ instead of $2$ in order to be consistent with their 
equilibrium structure models. Concerning the energy equation for the convective 
fluctuations~(\ref{eq28}), the first term 
$(\rho T)'/\overline{\rho T}\rd \overline{s}/\rd t$
is included in G96-05 while it is neglected in \citet{Unno67}.
G96-05 models use the same approximation for the nonlinear terms as \citet{Unno67} 
[Eqs.~(\ref{closen}) and (\ref{eqrad})] but again with different dimensionless 
geometrical factors 
in order to be consistent with their equilibrium structure models. 
Finally, \citet{GrigahceneEtal05}
introduced an additional parametrization for the perturbation of the closure 
term of the energy equation~(\ref{pclose1}). The complex parameter ${\hat\beta}$ introduced in this 
last approach allows to avoid
unphysical short wavelength oscillations of the mean entropy perturbation. 
It also requires calibration in order to fit the solar damping rates 
as discussed in more detail in Section~\ref{s:solar_damping}.

We now consider the pulsation equations. In addition to the fact that the 
G96-05 theory can deal
with nonradial oscillations, it includes several improvements. 
The perturbation of the
turbulent pressure [Eq.~(\ref{pturbperteq})] is included in the momentum 
equation for the pulsations 
[Eqs.~(\ref{pmm}) and (\ref{pmmt})].
The perturbation of the non-diagonal components of the Reynolds stress 
can also be obtained
\citep{Gabriel87}. The perturbation of the dissipation rate of 
turbulent kinetic energy into heat 
[Eq.~(\ref{ecinperteq})]
is included in the energy equation for the pulsations [Eq.~(\ref{eq32})]. 
All these terms are neglected in \citet{Unno67}.

\newpage
\section{Reynolds Stress Models}
\label{sec:reynolds_stress_models}

In the previous section, we derived two pictures of mixing-length models starting
from the Eqs.~(\ref{eq:static_fluc_momentum})\,--\,(\ref{eq:static_fluc_thermal_energy})
for the (Eulerian) velocity and temperature fluctuations $\vec{u}$ \& $T^\prime$ 
or from Eqs.~(\ref{eq23})\,--\,(\ref{eq:fluc_thermal_nonradial}) for $\vec{u}$ and the 
entropy fluctuation $s^\prime$. 
These equations were obtained from subtracting the averaged (mean) equations of motion
from the instantaneous equations of motion.
The most general averaging process in the Reynolds stress approach is an
ensemble average. In stellar astrophysics, however, it is the practice to 
adopt horizontal averages. 
The resulting fluctuating equations describe the evolution of the first-order 
moments, $\vec{u}$, $T^\prime$ or $\vec{u}$, $s^\prime$ in time and space and can only be
closed, and therefore solved, if appropriate expressions for the nonlinear, 
second-order moments, e.g., $\ob{\vec{u}\vec{u}}$ or $\ob{\vec{u}T^\prime}$
are found.
In the mixing-length approach these second-order moments are either neglected
during the linear growth of the convective fluctuations but taken into account
in the subsequent instantaneous annihilation of the convective eddies (see Appendix~\ref{app:A}),
or approximated in terms of a (constant) scalar turbulent viscosity 
$\nu_{\mathrm{t}}:=(\ob{u_{\mathrm{r}}u_{\mathrm{r}}})^{1/2}\hat\ell$, where $\hat\ell$ is a 
length scale, typically of the order of the mixing length $\ell$, such as [see also 
Eqs.~(\ref{eq:nonlinear_approximation_velocity})\,--\,(\ref{eq:nonlinear_approximation_temperature}) 
or (\ref{eq20})\,--\,(\ref{closen})]
\begin{equation}
-\,\nabla\cdot\left(\ob{\vec{u}\vec{u}}\right)\simeq\,\nabla\cdot\left(\nu_{\mathrm{t}}\nabla\vec{u}\right)
      \approxprop\vec{u}\tau_{\mathrm{c}}^{-1}\,,
\label{e:down-grade-appo}
\end{equation}
where $\tau_{\mathrm{c}}:=\hat\ell/(\ob{u_{\mathrm{r}}u_{\mathrm{r}}})^{1/2}$ is a characteristic turn-over
time~(\ref{turnovertime}) of the convection.
The terms in parentheses, $-\ob{\vec{u}\vec{u}}\simeq\,\nu_{\mathrm{t}}\nabla\vec{u}$,
represent the so-called diffusion or `down-gradient' approximation.

Instead of adopting approximations for the second-order moments, dedicated transport equations 
can be constructed, for example for $\ob{\vec{u}T^\prime}$, from multiplying Eq.~(\ref{eq:static_fluc_momentum}) by $T^\prime$, Eq.~(\ref{eq:static_fluc_thermal_energy})
by $\vec{u}$, summing the results followed by averaging, in a similar way as we did
for the averaged, turbulent kinetic energy equation~(\ref{ecin0}).
The so-constructed transport equations for the second-order moments
constitute the Reynolds stress approach, as proposed first by \citet{KellerFriedmann24}, and
first completely derived by \citet{Chou45}.
The transport equation for the second-order moments, however, include terms of
third-order moments which need, as discussed above for the second-order moments, to be 
represented by appropriate approximations or by additional transport equations, which will
contain terms of fourth-order moments. This can, in principle, be continued to ever higher-order
moments, but there will always be more variables (higher-order moments) than equations, 
representing the so-called closure problem of turbulence. 

\citet{Xiong77, Xiong89} and \citet{XiongEtal97} applied the Reynolds stress approach to stellar 
convection by constructing, for a Boussinesq fluid, four transport equations for the second-order 
moments $\ob{u_{\mathrm{r}}u_{\mathrm{r}}}$, $\ob{u_{\mathrm{r}}T^\prime}$, $\ob{T^\prime T^\prime}$, and for the 
$(r,r)$-component of the deviatoric part $\tens{\Sigma}$ of the velocity correlation tensor
$\ob{\vec{u}\vec{u}}:=\ob{u_{\mathrm{r}}u_{\mathrm{r}}}\,\tens{I}+\tens{\Sigma}$ [see also
Eq.~(\ref{eq11a})]. In order to form a closed system, these transport equations need 
to be supplemented by approximations for the second-order moments of the rate of turbulent 
kinetic energy dissipation, $\epsilon_{\mathrm{t}}$ [see Eq.~(\ref{eq18})], 
the dissipation rate $\epsilon_{T^\prime}$ 
of thermal potential energy (temperature variance $\ob{T^\prime T^\prime}$), and for the 
third-order moments, such as for the turbulent kinetic energy flux.
Xiong adopts the local approximation 
${\epsilon_{\mathrm{t}}}=2\sqrt{3}\chi(c_1\,H_{\mathrm{p}})^{-1}\,(\ob{u_{\mathrm{r}}u_{\mathrm{r}}})^{3/2}$,
where $\chi=0.45$ is the Heisenberg eddy coupling coefficient \citep{Heisenberg46} 
and $c_1$ is one out of three
closure coefficients of order unity, which needs to be calibrated in a similar way as the 
mixing-length parameter $\alpha$ in the models discussed in the previous chapter.
A similar local expression was adopted for the dissipation rate $\epsilon_{T^\prime}$.
For the third-order moments Xiong adopts the (local) down-gradient 
approximation [see also Eq.~(\ref{e:down-grade-appo})]
\begin{equation}
-\,\ob{u_{\mathrm{r}}xy}\approxprop\,\nu_{\mathrm{t}}\nabla\ob{xy}\,,
\end{equation}
in which $x$ and $y$ represent either $u_{\mathrm{r}}$ or $T^\prime$, 
thereby introducing the length scale $\hat\ell$ in the scalar turbulent viscosity
$\nu_{\mathrm{t}}:=(\ob{u_{\mathrm{r}}u_{\mathrm{r}}})^{1/2}\hat\ell$
which is, similar to the mixing length $\ell$, proportional to the locally-defined
pressure scale height $H_{\mathrm{p}}$, i.e., $\hat\ell=c_2H_{\mathrm{p}}$, where $c_2$ is of order unity
and needs, as $c_1$ before, to be calibrated. A third closure coefficient, $c_3$, is 
introduced by Xiong
to parametrize the anisotropy of the turbulent velocity field similarly to the anisotropy parameter
$\Phi$ [see Eq.~(\ref{anisotropyeq})], as a result of the coupling of the vertical 
to the horizontal
motion by the pressure fluctuations (pressure correlations; see also discussion in 
Section~\ref{sec:static_LMLM_1}).

Xiong's model was applied to stability computations of solar oscillations \citep{XiongEtal00} and
of classical pulsators \citep{XiongEtal98b, XiongDeng07}. These calculations could successfully
reproduce the location of the cool edge of the classical instability strip (see discussion in
Section~\ref{s:classpuls}), but report for a solar model 
overstable (unstable) radial modes 
with radial order $n=11 - 23$, which is in disagreement with the observed finite mode 
lifetimes discussed in Section~\ref{s:solar_damping}.

\citet{Canuto92, Canuto93} went beyond Xiong's treatment by proposing, additionally to the
second-order transport equations, including also nonlocal expressions for $\epsilon_{\mathrm{t}}$
and $\epsilon_{T^\prime}$,
separate transport equations for the third-order moments, which imply fourth-order moments.
Canuto adopts the Eddy-Damped Quasi-Normal approximation \citep{Orszag77, HanjalicLaunder76},
which is based on the quasi-normal approximation by \citet{Millionshtchikov41}, to close 
the fourth-order moments. This approximation assumes the fourth-order moments to be 
Gaussian random variables, leading to an expression of products and sums of second-order moments. 
The fourth-order pressure correlation terms are approximated by third-order damping terms.
In this approximation, all six third-order terms are expressed by six, partial differential
equations which now include only second- and third-order moments with 
five closure coefficients \citep[see][Eq.~37g]{CanutoDubovikov98}.
For the stationary case ($\partial/\partial t=0$) the third-order terms form a set of six linear
algebraic equations from which the third-order moments can be solved analytically as functions of
low-order moments. 
If the dissipation rate $\epsilon_{T^\prime}$ of thermal potential energy is approximated by
a local expression, the whole turbulent convection problem is described by five
coupled partial differential equations for the second-order moments
$\ob{wT^\prime}$, $\ob{T^\prime T^\prime}$, $\ob{ww}$, $\ob{\vec{u}\cdot\vec{u}}$ 
and $\epsilon_{\mathrm{t}}$,
where the velocity field $\vec{u}=(u,v,w)$ in a plane-parallel geometry.
The five transport equations for the second-order moments use five empirical closure 
coefficients \citep[see][Eqs.~(13)\,--\,(16)]{CanutoJCD98}, additionally to the five closure
coefficients for the third-order moments.

\citet{CanutoDubovikov98} extended \citepos{Canuto93} Reynolds 
stress model by deriving improved
expressions for the dissipation terms $\epsilon_{\mathrm{t}}$ and $\epsilon_{T^\prime}$, and for the
empirical constants that were used in \citepos{Canuto93} model, using renormalization group techniques.
Canuto and Dubovikov's model, together with a simplified version of the third-order moments in the
stationary limit, was implemented by \citet{Kupka99} and applied to non-pulsating 
(stationary) envelope 
models of A-stars and white dwarfs by \citet{KupkaMontgomery02} and \citet{MontgomeryKupka04}.

\newpage
\section{Convection Effects on Pulsation Frequencies}
\label{sec:frequency_effects}

Convection affects not only the structure of stars but also the properties of the
global oscillation modes. The most prominent observable is perhaps the oscillation
frequency, which can be compared with stellar models using different guises of
convection treatment. 
Today's observation techniques of stellar oscillation frequencies have reached 
an accuracy that allows us to interpret the differences between observed
and computed stellar eigenfrequencies to stem solely from the incomplete physics
describing the equilibrium and pulsation models.
We should, however, remain aware that in addition to convection, effects due to opacity, 
rotation, magnetic fields, equation of state, and element diffusion also play an important 
role in modelling stellar pulsation frequencies.
The latter effects have been investigated by various authors, e.g., 
\citet{JCD-Dappen92}, \citet{Guzik-Cox93}, \citet{Guenther94}, \citet{Tripathy-JCD96}, 
\citet{Guzik-Swenson97}, \citet{JCDEtal09}, and \citet{SuarezEtal13}. 
In particular the increase of low-temperature opacities and the use of more sophisticated
thermodynamics have reduced the discrepancy between computed and observed 
solar frequencies \citep[e.g.,][]{JCD-Dappen92, GuzikEtal96}. 
Although these improvements in stellar physics have brought the models closer to
the observations, we still have to explain the remaining 
frequency differences between observations and
computed adiabatic eigenfrequencies, particularly for modes with high-radial order. 
For solar frequencies
these differences between a `standard' solar model \citep[e.g.,][]{JCDEtal96}
and the Sun are as large as $\sim 13~\mu\mathrm{Hz}$. 
Figure~\ref{fig:frequ-diff-obs-LMLT}a shows the inertia-scaled differences between the 
solar and model frequencies. It demonstrates that adiabatically computed
frequencies not only overestimate severely the eigenfrequencies for modes with 
frequencies $\nu\gtrsim 2.5\mathrm{\ mHz}$, but also that these frequency differences are 
predominantly a function of frequency alone with little dependence on mode degree $l$ 
\citep[e.g.,][]{JCD84, JCD-Berthomieu91}.
This indicates that the effects are essentially confined to the very surface layers, where
the modelling details depend crucially on the functional form of the acoustic cutoff frequency 
$\nu_{\mathrm{ac}}$ (as it affects the acoustic potential) with radius.
For an isothermal atmosphere $\nu_{\mathrm{ac}}=c/4\pi H$, where $c$ is the adiabatic 
sound speed and $H$ the pressure scale height.
In the Sun $\nu_{\mathrm{ac}}\simeq5.5\mathrm{\ mHz}$.
The cutoff frequency determines the
location at which an incident acoustic wave is reflected back into the stellar interior, 
and the lower the frequency $\nu$, the deeper the location at which this reflection takes
place. For modes with frequencies $\nu$ much less than $\nu_{\mathrm{ac}}$ reflection takes 
place so deep
in the star that the modes are essentially unaffected by the near-surface structure.
When $\nu$ is comparable with $\nu_{\mathrm{ac}}$, however, the inertia, in units of mass, of 
the near-surface layers is a considerable fraction of the total mass above the 
reflecting layer, leading
to a greater modification to the phase shift in the spatial oscillation eigenfunctions,
and, through the dispersion relation, also to a change in frequency \citep{JCD-Gough80}. 
The inertia of the 
essentially hydrostatically moving near-surface layers depends on mass and consequently 
on the equilibrium pressure near the photosphere. 
Moreover, these upper layers are dominated by the convection dynamics and
treatment of the radiation field, which crucially influence the shape of the 
eigenfunctions of high-order modes and consequently many aspects of mode physics.
These effects have become known as ``surface effect'' 
\citep[e.g.,][]{JCD-Gough80, Gough84, JCD84, JCD-Berthomieu91, Balmforth92b, RosenthalEtal95, 
Houdek96, RosenthalEtal99, KjeldsenEtal08, Houdek10, GrigahceneEtal12}.
\citet{Gough84}, for example, used the local
convection model~1 in Section~\ref{sec:local_mlm} and a simplified analytical 
approximation of the eigenfunctions in the atmosphere. 
\citet{Balmforth92b} studied these effects 
with the more sophisticated nonlocal time-dependent 
mixing-length model introduced in Section~\ref{sec:nonlocal_mlm}.
Both authors concluded that the correction of the stratification 
of the superadiabatic boundary layers due to the inclusion of the 
mean turbulent pressure substantially decreases the adiabatic
frequency residuals.

\epubtkImage{fig_2_col_B-nu-diff_3D-MS.png}{%
\begin{figure}[htb]
  \centerline{(a)\parbox[t]{6cm}{\vspace{0pt}\includegraphics[width=6cm]{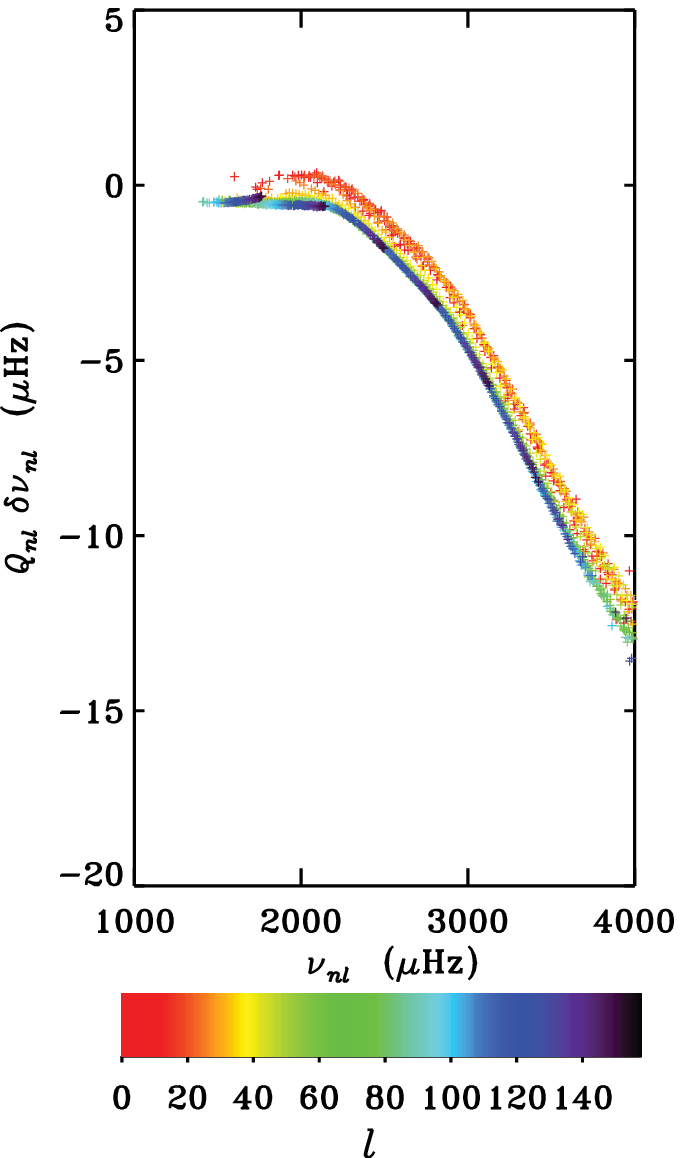}}\qquad
              (b)\parbox[t]{5.85cm}{\vspace{0pt}\includegraphics[width=5.85cm]{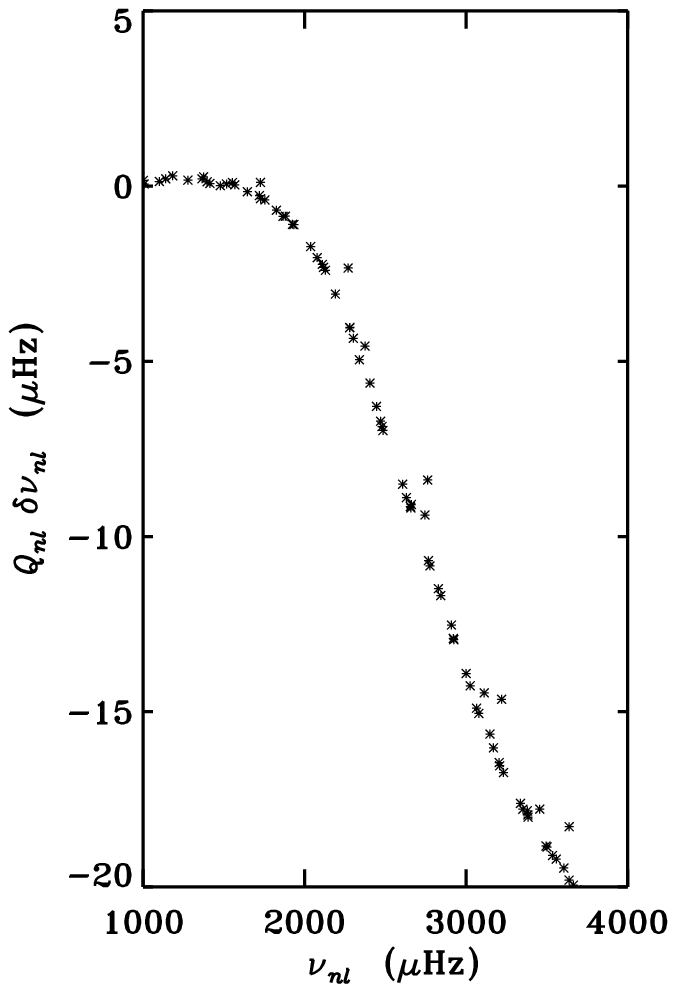}}}
  \caption{(a)~Scaled differences between observed GONG frequencies and
                   adiabatically computed frequencies of the `standard' solar Model~S.
                   The degree $l$ of the oscillation modes is indicated by the colour bar.
                   Image reproduced with permission from \citet{JCDEtal96}, copyright by AAAS.
               (b)~Scaled adiabatic frequency differences between a model
                   for which the near-surface layers were represented
                   by a hydrodynamical simulation, and the `standard' 
                   solar Model~S, which does not include the turbulent pressure. 
                   Image adapted from \citet{RosenthalEtal95}.}

  \label{fig:frequ-diff-obs-LMLT}
\end{figure}}

Convection modifies pulsation properties of stars principally through two effects: 
(i) dynamical effects through the additional turbulent pressure term $p_{\mathrm{t}}$~(\ref{eq11})
in the mean momentum equation~(\ref{eq:mean_momentum_ba}), and its
perturbation $\delta p_{\mathrm{t}}$~(\ref{eq:momentum-flux-perturb}), (\ref{nlmlt-conv-fluxes-perturb}),
where $\delta$
denotes here a linear perturbation in a Lagrangian-mean frame of reference, in the 
pulsationally perturbed mean momentum equation; 
(ii) nonadiabatic effects, additional to the perturbed radiative heat flux 
$\delta\vec{F}_{\mathrm{r}}$, through the perturbed convective heat (enthalpy) flux 
$\delta\vec{F}_{\mathrm{c}}$~(\ref{eq:heat-flux-perturb}), (\ref{nlmlt-conv-fluxes-perturb}) 
in the pulsationally perturbed mean thermal heat (energy) equation.
  
\subsection{The effect of the Reynolds stress in the equilibrium stellar model}

From the discussion before we conclude that the details of modelling the 
hydrostatic equilibrium structure in the near-surface layers play an important 
role in describing the residuals between observed and modelled oscillation
frequencies, particularly for modes with $\nu$ close to $\nu_{\mathrm{ac}}$. 
Almost all stellar model calculations consider only the gradient of the
gas pressure $p$ in the equation of hydrostatic support. In the
convectively unstable surface layers, however,
the additional contribution from the turbulent pressure $p_{\mathrm{t}}$ (\ref{eq11})
to the hydrostatic support can be significant
[see, e.g., Eq.~(\ref{eq:mean_momentum_ba})].
Hydrodynamical simulations of stellar convection have enabled us to estimate this
contribution from the turbulent pressure $p_{\mathrm{t}}$.
Figure~\ref{fig:sun_ptvz} shows $p_{\mathrm{t}}$ for three different 
simulations and models of the Sun. It indicates that $p_{\mathrm{t}}$
can be as large as 15\% of the total pressure $\hat p=p+p_{\mathrm{t}}$.

\epubtkImage{plot_ptvz_sun_colour_v2.png}{%
\begin{figure}[htb]
  \centerline{\includegraphics[width=9cm]{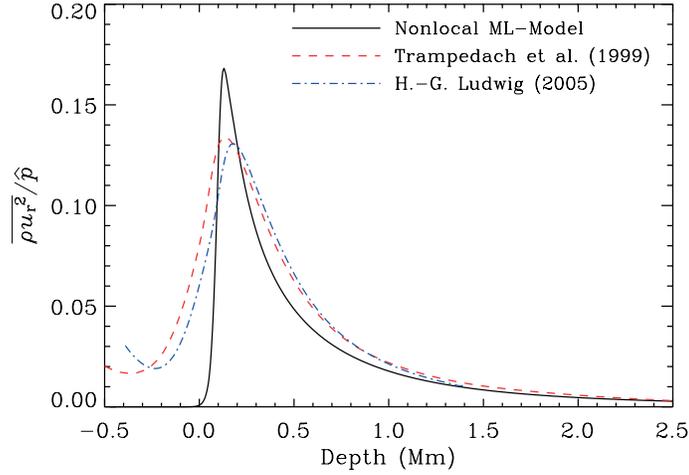}}
  \caption{Turbulent pressure $p_{\mathrm{t}}:=\ob{\rho u^2_r}$ over total pressure 
               $\hat p=p+p_{\mathrm{t}}$ as a function of the depth variable $R_{\odot}-r$ for 
               various solar simulations and models ($R_{\odot}$ is the photospheric solar radius). 
               Results are shown for the nonlocal mixing-length model of Section~\ref{s:nonlocal-mlm} 
               by \citet[][solid curve]{Gough77a, Gough77b} and from hydrodynamical 
               simulations by Regner Trampedach (1999, personal communication, 
               dashed curve) and H.-G. Ludwig (2005, personal communication, dot-dashed curve).}
  \label{fig:sun_ptvz}
\end{figure}}
 
\citet{RosenthalEtal95}, for example, investigated the effect of the contribution 
that $p_{\mathrm{t}}$ makes to the mean hydrostatic stratification on the adiabatic 
solar eigenfrequencies. They examined a hydrodynamical simulation by 
\citet{SteinNordlund91} of the outer 2\% by radius of the Sun, matched 
continuously in sound speed to a model envelope calculated, as in a `standard' solar model, 
with a local mixing-length formulation without $p_{\mathrm{t}}$. 
The resulting frequency shifts of adiabatic oscillations between the simulations and the
`standard' solar reference model, Model S, are illustrated in Figure~\ref{fig:frequ-diff-obs-LMLT}b. 
The frequency residuals behave similarly to the solar data depicted in the left panel of that figure,
though with larger frequency shifts at higher oscillation frequencies.

If turbulent pressure is considered in the mean structure, however,
additional assumptions have to be made about the turbulent pressure
perturbation in the adiabatic pulsation equations. The first adiabatic exponent 
$\gamma_1=(\partial\ln p/\partial\ln\rho)_s$ ($s$ being the specific entropy), 
is a purely thermodynamic quantity
and is expressed by means of the gas pressure $p$. 
Consequently, in the presence of turbulent pressure 
$p_{\mathrm{t}}$, such that the total pressure $\hat p=p+p_{\mathrm{t}}$ 
satisfies the equation of hydrostatic equilibrium, 
$\gamma_1$ experiences a modification of the form
\begin{equation}
\gamma_1=\left(\frac{\partial\ln\hat p}{\partial\ln\rho}\right)_{\!\!s}=
\frac{1}{\hat p}\left[\left(\frac{\partial p}{\partial\ln\rho}\right)_{\!\!s}+
                \left(\frac{\partial p_{\mathrm{t}}}{\partial\ln\rho}\right)_{\!\!s}
                \right]
\label{eq:gamma_tilde}
\end{equation}
and the relative Lagrangian perturbation in the total pressure is
\begin{equation}
\frac{\delta\hat p}{\hat p}=\frac{\delta p}{\hat p}+
\frac{\delta p_{\mathrm{t}}}{\hat p}=
\gamma_1\,\frac{\delta\rho}{\rho}
\,.
\label{eq:adiabat-press-fluc}
\end{equation}
Nonadiabatic pulsation calculations with
the inclusion of the turbulent pressure perturbation $\delta p_{\mathrm{t}}$ \citep{Houdek96}, 
and hydrodynamical simulation results \citep{RosenthalEtal95, RosenthalEtal99} indicate
that $\delta p_{\mathrm{t}}$ varies approximately in quadrature 
with the other terms in the linearized momentum equation, and hence contributes
predominantly to the imaginary part of the frequency shift, i.e., to the 
linear damping rate. The Lagrangian gas-pressure perturbation $\delta p$, however, 
responds adiabatically. Therefore, $\delta p_{\mathrm{t}}$ can be neglected
in Eq.~(\ref{eq:adiabat-press-fluc}), i.e., in the calculation of the 
(real) adiabatic eigenfrequencies it is assumed that 
$\delta\hat p/\hat p\simeq\delta p/\hat p\simeq\tilde\gamma_1\delta\rho/\rho$. With this assumption
$\tilde{\gamma}_1\simeq(p/\hat p)\gamma_1$, and the only modification to 
the adiabatic oscillation equations is the replacement of $\gamma_1$ by $\tilde\gamma_1$
\citep{RosenthalEtal95}.

\subsection{The effects of nonadiabaticity and momentum flux perturbation}

The effects of nonadiabaticity and convection dynamics on the pulsation 
frequencies were, for example, studied by \citet{Balmforth92b,
  RosenthalEtal95} and \citet{Houdek96}. In these studies, the nonlocal, time-dependent generalization 
of the mixing-length formulation by \citet{Gough77a,Gough77b} was adopted to model the heat 
and momentum flux consistently in both the equilibrium envelope model and in the 
nonadiabatic stability analysis. \citet{Houdek96} considered the following models:

\begin{quote}
\begin{itemize}
\item[\hfill L.a] A local mixing-length formulation without turbulent pressure $p_{\mathrm{t}}$
           was used to construct the mean envelope model. Frequencies were
           computed in the adiabatic approximation assuming $\delta p_{\mathrm{t}}=0$.
\item[\hfill NL.a] \citepos{Gough77a,Gough77b} nonlocal, mixing-length model, including turbulent
            pressure, was used to construct the mean envelope model. 
            Frequencies were computed in the adiabatic approximation assuming 
            $\delta p_{\mathrm{t}}=0$.
\item[\hfill NL.na] The mean envelope model was constructed as in NL.a. 
             Nonadiabatic frequencies were computed including consistently the 
             Lagrangian perturbations of the convective heat flux $\delta\vec{F}_{\mathrm{c}}$,
             additionally to $\delta\vec{F}_{\mathrm{r}}$, and turbulent momentum 
             flux $\delta p_{\mathrm{t}}$.
\end{itemize}
\end{quote}

Additional care was necessary when frequencies between
models with different convection treatments were compared, such as in the models L.A and NL.a.
In order to isolate the effect of the near-surface structures on the oscillation frequencies
the models had to posses the same stratification in their deep interiors. This was 
obtained by requiring the models to lie on the same adiabat near the base of the
(surface) convection zone and to have the same convection-zone depth.
Varying the mixing-length parameter $\alpha=\ell/H_{\mathrm{p}}$ 
and hydrogen abundance by iteration in model L.a, the same values for 
temperature and pressure were found at the base of the convection zone as those in
models NL.a and NL.na. The radiative interior of the nonlocal models NL.a and NL.na were 
then replaced by the solution of the local model L.a, and the convection-zone depth was
calibrated to $0.287\,R_{\odot}$ \citep{JCD-DOG-MJT91}.
Further details of the adopted physics 
in the model calculations can be found in \citet{HoudekEtal99}.

The outcome of these calculations is shown in Figure~\ref{fig:sun-frequ-diff}a.
As for the hydrodynamical simulations (Figure~\ref{fig:frequ-diff-obs-LMLT}b) 
the effect of the Reynolds stresses in the mean structure decreases the adiabatic frequencies 
(NL.a-L.a, solid curve) for frequencies larger than about 2~mHz, though the maximum deficit
of about $12~\mu\mathrm{Hz}$ is smaller than in the hydrodynamical simulations. The effects
of nonadiabaticity ($\delta\vec{F}_{\mathrm{r}}+\delta\vec{F}_{\mathrm{c}})$ and 
$\delta p_{\mathrm{t}}$ (NL.na-NL.a, dashed curve), however, lead to an 
increase of the mode frequencies by as much as $\sim 9~\mu\mathrm{Hz}$, nearly cancelling 
the downshifts from the
effect of $p_{\mathrm{t}}$ in the mean structure, as illustrated by the dot-dashed 
curve (NL.na-L.a).

\epubtkImage{sun-frequ-diff_v3-etaBoo_frequ_diff_v2.png}{%
\begin{figure}[htb]
  \centerline{(a)\parbox[t]{6.26cm}{\vspace{0pt}\includegraphics[width=6.26cm]{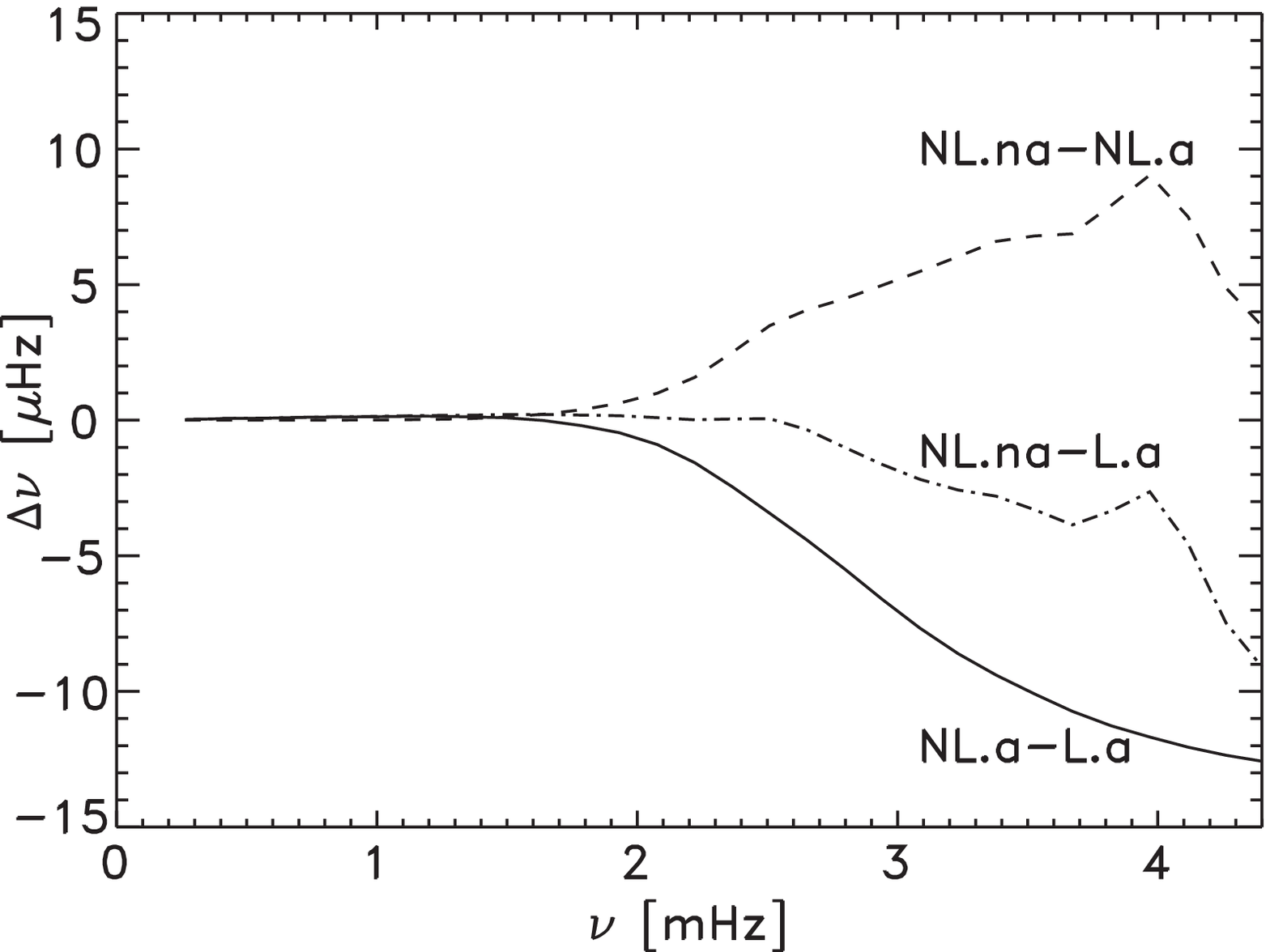}}\qquad
              (b)\parbox[t]{5.74cm}{\vspace{1pt}\includegraphics[width=5.74cm]{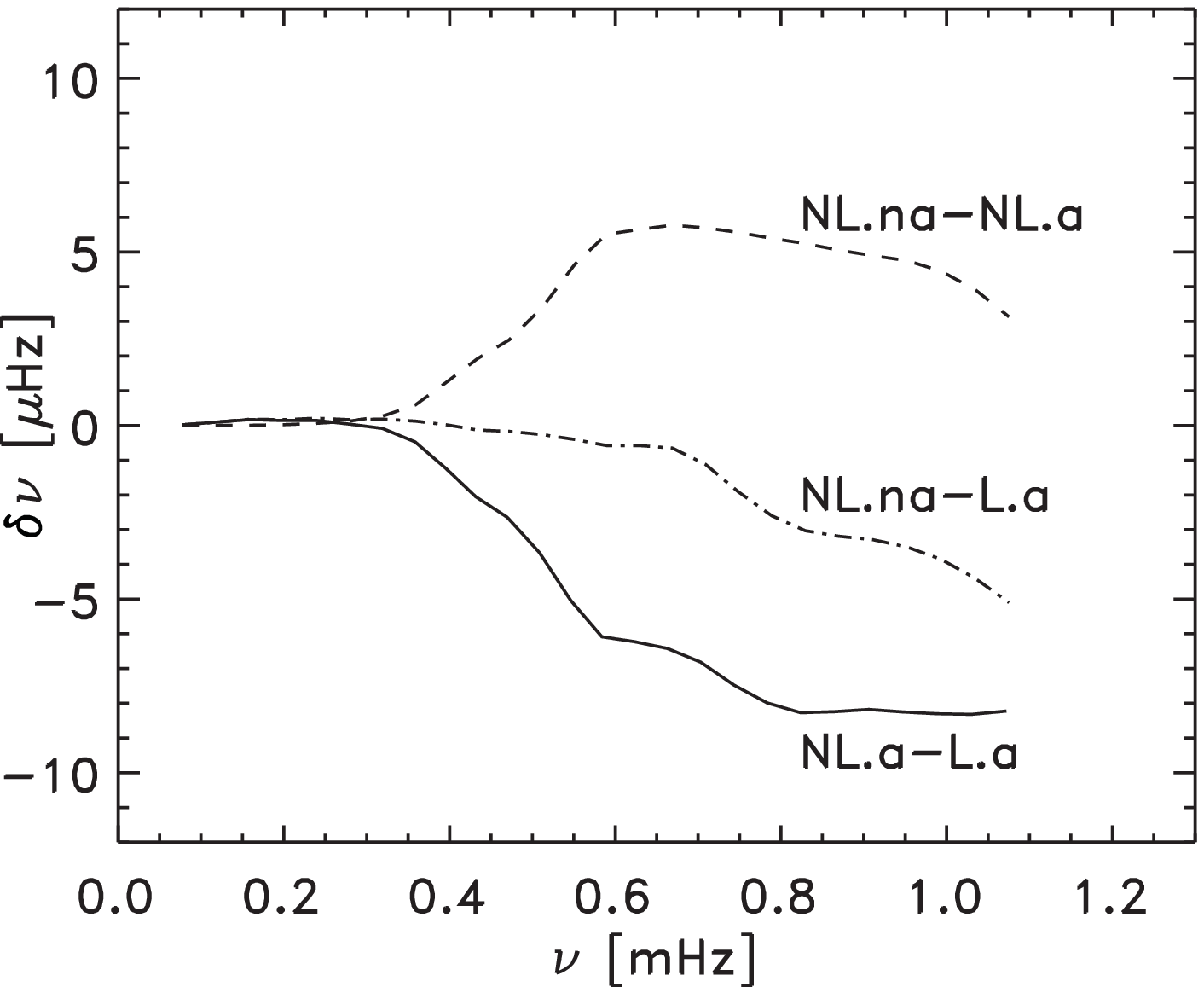}}}
  \caption{Frequency residuals between stellar models calculated with Gough's nonlocal
               time-dependent mixing-length formulation (Section~\ref{sec:nonlocal_mlm}).
               (a) Results are shown for solar models:
               the solid curve (NL.a-L.a) shows the adiabatic radial oscillation frequency shifts, 
               under the assumption $\delta p_{\mathrm{t}}=0$, caused by the turbulent 
               pressure $p_{\mathrm{t}}$ in the mean envelope. The dashed curve (NL.na-NL.a) 
               is the frequency shift caused by nonadiabaticity and effects of including 
               consistently $\delta p_{\mathrm{t}}$.
               The overall frequency shift (NL.na-L.a) is plotted by the dot-dashed curve.
               (b) Results are shown for models for the solar-like star $\eta$~Boo. The model
               calculations and line styles are as in panel (a). 
Images reproduced from \citet{Houdek96}.}
  \label{fig:sun-frequ-diff}
\end{figure}}

If, however, the positive frequency shifts between models NL.na and NL.a (dashed curve)
are interpreted as the nonadiabatic and momentum flux corrections to the oscillation
frequencies then their effects are to bring the frequency residuals of the hydrodynamical 
simulations (Figure~\ref{fig:frequ-diff-obs-LMLT}b) in better agreement with the data plotted in
Figure~\ref{fig:frequ-diff-obs-LMLT}a. 

The effects of the near-surface regions in the Sun were
also considered by \citet{RosenthalEtal99} and \citet{LiEtal02} 
based on hydrodynamical simulations. 

A similar conclusion as in the solar case was found for the solar-like star $\eta\,$Boo 
by \citet{JCDEtal95} and \citet{Houdek96}, demonstrated in 
Figure~\ref{fig:sun-frequ-diff}b, and more recently by \citet{StrakaEtal06b}.
\citet{GrigahceneEtal12} studied the surface effects in the Sun and in three 
solar-type stars with the conclusion that the use of the local time-dependent convection treatment 
of Section~(\ref{sec:unno_gabriel_mlm}) reduces the frequency residuals between 
observations and stellar models. In these calculations, however, the hydrostatic
equilibrium model was corrected a posteriori by the effect of the mean turbulent pressure with
some consequent inconsistencies in the thermal equilibrium structure.   


The near-surface frequency corrections also affect the determination of the modelled 
mean large frequency separation $\Delta\nu:=\langle\nu_{n+1l}-\nu_{nl}\rangle$ (angular 
brackets indicate an average over $n$ and $l$). In both models for the Sun \citep{JCD-Gough80}
and for $\eta\,$Boo the resulting
corrections to $\Delta\nu$ are about $-1~\mu\mathrm{Hz}$. Although this correction is less than 1\% it
does affect the determination of the stellar radii and ages from the observed values of
$\Delta\nu$ and small frequency separation $\delta\nu_{02}$ in distant stars.
 
A simple procedure for estimating the near-surface frequency corrections was suggested 
by \citet{KjeldsenEtal08}. It is based on the ansatz that the frequency shifts can be scaled
as $a(\nu/\nu_0)^b$ \citep{JCD-Gough80}, where $\nu_0$ is a suitable reference frequency,
$b$ is obtained from solar data, and the surface-correction amplitude $a$ is determined 
from fitting this expression to the observed frequencies.
  
Kjeldsen \etal's empirical power-law has been applied to the modelling of a large number 
of solar-type \textit{Kepler} stars \citep[e.g.,][]{MetcalfeEtal12, MathurEtal12, GruberbauerEtal13, MetcalfeEtal14}. \citeauthor{MathurEtal12}, 
for example, determined statistical properties of the surface-correction amplitude $a$ from 
22 \textit{Kepler} stars. The model frequencies were, however, obtained in the adiabatic approximation
neglecting, as did \citet{MetcalfeEtal12} and \citet{MetcalfeEtal14}, any convection 
dynamics in both the equilibrium and pulsation calculations.
\citet{MathurEtal12} concluded that the surface-correction amplitude $a$ is nearly a constant 
fraction of the mean large-frequency separation $\Delta\nu$.
Information like this could provide additional insight into the physical processes responsible
for the high radial-order frequency shifts between observations and stellar models.
\citet[][see also \citealp{GruberbauerGuenther13}]{GruberbauerEtal13} analysed the surface effects
in 23 \textit{Kepler} stars with a Bayesian approach neglecting, however, convection dynamics in both the
equilibrium and in the nonadiabatic eigenfrequency calculations.

\citet{JCD12} 
suggested an improved functional form for the high-order frequency shifts between observations and
stellar models. This improved functional form can be determined for the Sun from the surface term in
Duvall's differential form for the asymptotic expression for frequencies using a large 
range of mode degrees~$l$. This leads to a better
representation of the solar frequency residuals brought about by the very surface layers. 
By adopting the acoustic cutoff frequency as the relevant frequency scale the scaled solar-surface
functional form can also be applied to other stars that are not too dissimilar to the Sun.
\citet{JCD12} applied it to \textit{Kepler} data for the solar-type star 16~Cyg~A and 
reported a better
representation of the frequency surface correction compared to the empirical 
power law by \citet{KjeldsenEtal08}.
Another empirical approach was recently reported by \citet{BallGizon14}, which is based on
the scaling relation for mode inertia by \citet{Gough90}.

Although these empirical approaches offer some description of the surface effects, they
do not provide the much needed insight for describing the relevant physical processes.
The most promising approach today
for a better understanding of these surface effects is the use of the latest implementations 
of three-dimensional (3D) hydrodynamical simulations, and their results, for developing 
improved one-dimensional (1D) convection models. Several international groups are now 
pursuing this approach.

\newpage
\section{Driving and Damping Mechanisms}
\label{sec:stability}

The question of whether the amplitude (or energy) of a particular oscillation mode in a star
is growing or declining with time is related to the problem of vibrational stability.
Because vibrational stability (or instability) is characterized by the existence of a 
periodicity in the temporal behaviour of the perturbations, a reasonable useful criteria is
the sign of the total energy change (thermal and mechanical) over one pulsation 
period assuming that the system returns precisely to its original state at the end of the period.
This is the definition of the work integral $W$.

\subsection{The work integral}

\subsubsection{Expressions for radial pulsations} 

It was \citet{Eddington26}, who recognized first that a non-vanishing expression for the
work integral $W$, which is of second order in the pulsation quantities, can be obtained 
from the vanishing expression of the integrated (specific) entropy, $s$, over one 
pulsation period, i.e., from $\oint\rd s=0$, because $s$ is a state variable. 
For the nonadiabatic contributions 
$W$ is obtained from the (linearly) perturbed thermal energy equation (\ref{eq6}).
Recognizing that $\rho\rd e/\rd t+p\nabla\cdot\vec{v}=\rho T\rd s/\rd t$, 
where $\rd /\rd t$ is the substantial (or material) derivative in a frame with (total) 
velocity $\vec{v}$, we obtain
\begin{equation}
W=\int_0^M \rd m\oint\delta T\frac{\partial\delta s}{\partial t} \rd t\,,
\label{e:Eddington-W}
\end{equation}
for a star with zero gas pressure $p$ at the stellar surface $r=R$ and mass $M$.
The boundary condition, $p=0$ at the surface $R$,
expresses that no work ($\int _Ap\,\vec{U}\cdot\rd \vec{A}=0$, $\vec{A}$ is the star's 
surface area, and $\vec{U}$ is the pulsation velocity) is exerted from outside, i.e., 
from layers with radius $r>R$. If $W>0$ the energy
(pulsation amplitude) increases with time and consequently the pulsation mode is unstable or
overstable. For $W<0$ the mode is said to be (intrinsically) stable or damped.

Alternatively, one can also interpret the work integral $W$ as the amount of energy that is
required to be added to (damped, stable mode) or subtracted from (unstable mode) the pulsating 
system to maintain exact periodicity. This concept was adopted by 
\citet{Baker-Kippenhahn62, Baker-Kippenhahn65}, who started from the expression for the work done
\emph{on the (stellar) sphere} with volume $V$ during one pulsation period, i.e.,
\begin{equation}
W=-\oint p\,\frac{{\partial}V}{{\partial}t} \rd t
 =-\oint p\,\vec{U}\cdot\rd \vec{A} \rd t\,.
\end{equation}
From this expression they derived the work integral 
\begin{equation}
W_{\mathrm{g}}=\pi\int_0^M\frac{1}{\rho^2}\Im\left({\delta p^\ast\delta\rho}\right) \rd m\,,
\label{e:DOG-Wg}
\end{equation} 
associated with the gas pressure $p$,
for radial pulsations without convection dynamics and under the assumption that the imaginary 
part, $\omega_{\mathrm{i}}=\Im\left({\omega}\right)$, of the complex pulsation eigenfrequency, $\omega$,
vanishes; the asterisk denotes complex conjugate.

\citet{BakerGough79} considered the fully nonadiabatic linearized stability problem for 
radial modes with the inclusion of convection dynamics by adopting the local time-dependent
convection model of Section~(\ref{sec:time-dependent_LMLM}). Their derivation of the work integrals
starts from the pulsationally linearized form of the mean momentum equation~(\ref{eq13}) 
which, after multiplication by the
complex conjugate of the radial component $\xi_{\mathrm{r}}=:\delta r$ of the displacement
vector $\vec{\xi}$, leads to the work-integral contribution from the turbulent pressure $p_{\mathrm{t}}$
\begin{eqnarray}
W_{\mathrm{t}}(m)
&=&\pi\int_{m_{\mathrm{b}}}^{M_\star}\Im(\delta p_{\mathrm{t}}\,^\ast\,\delta\rho)\frac{\rd m}{\rho^2}\cr
&+&\pi\int_{m_{\mathrm{b}}}^{M_\star}
         \Biggl\{(3-\Phi)
            \left[\Im(\delta r\,^\ast\,\delta p_{\mathrm{t}})
             -\frac{p_{\mathrm{t}}}{\rho}\Im(\delta r\,^\ast\,\delta\rho)
            \right]
            -p_{\mathrm{t}}\Im(\delta r\,^\ast\,\delta\Phi)
         \Biggr\}\frac{\rd m}{\rho r}
\,,
\label{e:DOG-Wt}
\end{eqnarray}
additionally to expression~(\ref{e:DOG-Wg}) for the gas-pressure contribution $W_{\mathrm{g}}$, but with
the lower integration boundary replaced by 
the mass $m_{\mathrm{b}}$ at the bottom boundary of the stellar envelope model.
The complex pulsation frequency $\omega=\omega_{\mathrm{r}}+\mathrm{i}\,\omega_{\mathrm{i}}$ is then related to the
work integrals by
\begin{equation}
 \frac{\omega_{\mathrm{i}}}
      {\omega_{\mathrm{r}}}
=\frac{W_{\mathrm{g}}+W_{\mathrm{t}}+F}
      {2\pi\omega^2_{\mathrm{r}}\int_{m_{\mathrm{b}}}^M|\delta r|^2 \rd m}
:=\frac{-\eta_{\mathrm{g}}-\eta_{\mathrm{t}}+\tilde F}{\omega_{\mathrm{r}}}
:=\hat\eta_{\mathrm{g}}+\hat\eta_{\mathrm{t}}+\hat F\,,
\label{e:DOG-Wall}
\end{equation}  
where $\eta_{\mathrm{g}}$ and $\eta_{\mathrm{t}}$ are respectively the linear damping rates 
of the gas and turbulent pressure contributions,
$\hat\eta_{\mathrm{g}}$ and $\hat\eta_{\mathrm{t}}$ are the associated stability coefficients, and
$F:=\left[4\pi^2r^2\Im({\delta p^\ast\delta r})\right]^M_{m_{\mathrm{b}}}$ is typically negligible.

\subsubsection{Expressions for nonradial pulsations}

Expressions of work integrals for nonradial pulsations were reported by,
e.g., \citet{GrigahceneEtal05} in the framework of applying the time-dependent
convection model of Section~{\ref{sec:unno_gabriel_mlm}} to several classes of pulsating stars.
As before, the work integrals can be obtained by 
taking the scalar product of the linearly perturbed average momentum equation
with the displacement vector $\vec{\xi}^\ast$, followed by integration by part 
over the total stellar mass and considering only
the imaginary part of the final result, which is:
\begin{eqnarray}  \label{int1}
\omega_{\mathrm{i}}&=&\frac{\displaystyle{\int_0^M \Im
\left\{ \frac{{\delta\rho}}{\rho}\frac{\delta\hat p^*}{\rho} 
+ \rd W_{\mathrm{Rey}}\right\}\rd m}}
{2\,\omega_{\mathrm{r}}\int_0^M {\left|\vec{\xi}\right|}^2 \rd m}\,,
\end{eqnarray}
where $\Im\{{\delta\rho}\delta\hat p^*/\rho^2\}$ and 
$\rd W_{\mathrm{Rey}}\simeq\left(\xi_{\mathrm{r}}^*\Xi_{\mathrm{r}}+l(l+1)\xi_{\mathrm{h}}^*\Xi_{\mathrm{h}}\right)/\rho$ 
are the work contributions per unit mass 
produced during one pulsation cycle by the total (gas + radiation + turbulence) pressure $\hat p$ 
and by the non-diagonal components of the Reynolds stress tensor, respectively,
and $\Xi_{\mathrm{r}}$ and $\Xi_{\mathrm{h}}$ are defined in Eq.~(\ref{reynoteq}). 
As for the radial case the work produced by the total pressure can be separated into
contributions from the gas and radiation pressure ($p$) and from the turbulent 
pressure ($p_{\mathrm{t}}$), i.e.,
\begin{eqnarray}
\Im\left\{ \frac{{\delta\rho}}{\rho}\frac{\delta\hat p^*}{\rho}\right\}&=&
\Im\left\{ \frac{{\delta\rho}}{\rho}\frac{\delta p^*}{\rho}\right\}+\rd W_{\mathrm{t}}\,,
\end{eqnarray}
where $\rd W_{\mathrm{t}}=\Im\{{\delta\rho}\delta p_{\mathrm{t}}^*/\rho^2\}$
is the work by the turbulent pressure.
From the equations of state and energy conservation we obtain for the contribution 
of the gas (and radiation) pressure
\begin{eqnarray}
\Im\left\{ \frac{{\delta\rho}}{\rho}\frac{\delta p^*}{\rho}\right\}
&=&\rd W_{\mathrm{FR}}+\rd W_{\mathrm{FC}}+\rd W_{\epsilon_{\mathrm{t}}}\,,
\label{e:work-MAD2}
\end{eqnarray}
where $W_{\mathrm{FR}}$, $W_{\mathrm{FC}}$ and 
$W_{\epsilon_{\mathrm{t}}}$ are the contributions from the 
radiative and convective heat flux, and from the dissipation of turbulent 
kinetic energy, respectively.
For simplicity we neglect in the next equations the terms corresponding 
to the horizontal component of the flux divergence. 
These terms are very small because the pressure and temperature scale heights are much smaller 
compared to the term, $r/l$, for modes with low spherical degree $l$, and we 
also neglect the nuclear reactions. We then obtain
\begin{equation}
 \rd W_{\mathrm{FR}}+\rd W_{\mathrm{FC}} =
 -\Re\left\{\frac{{\delta T}^*}{T\omega} \frac{\rd \,\delta
     \left( L_{\mathrm{R}}+L_{\mathrm{c}}\right)
   }{\rd m}\right\}\,, 
\end{equation}
and for the last term in Eq.~(\ref{e:work-MAD2}), which is the contribution from the
viscous dissipation of turbulent kinetic energy into heat, we obtain, after neglecting the 
contributions of the last geometrical term in Eq.~(\ref{ecinperteq}), the work contribution
\begin{equation}
\rd W_{\epsilon_{\mathrm{t}}}
\;\simeq\;-({\gamma_3}-1)\frac{{\Phi}}{2}\Im\left\{ \frac{{\delta\rho}}{\rho}
\frac{\delta p_{\mathrm{t}}^*}{\rho}\right\}\,,
\end{equation}
where $\gamma_3$ is the third adiabatic exponent defined as 
$\gamma_3-1:=(\partial\ln T/\partial\ln\rho)_s$, and $s$ is the specific entropy.
As noted by \citet{LedouxWalraven58} and \citet{GrigahceneEtal05}, the terms $\rd W_{\mathrm{t}}$
and $\rd W_{\epsilon_{\mathrm{t}}}$ cancel exactly for a completely ionized gas 
without radiation (${\gamma_3}-1=2/3$) and for isotropic turbulence (${\Phi}=3$). 

It should be noted that in \citepos{Gough77a, Gough77b} convection model the viscous dissipation by
turbulent kinetic energy into heat is neglected in the thermal energy equation, as 
suggested by \citet{SpiegelVeronis60} for a (static, i.e., $\vec{U}=\vec{0}$) Boussinesq fluid.

In the following sections, we shall review and compare results of stability calculations between 
the two time-dependent convection formulations by \citet{Gough77a, Gough77b} 
and \citet{GrigahceneEtal05} for various classical pulsators and for
stars with stochastically excited oscillations. 
 

\subsection{Intrinsically unstable pulsators}
\label{s:classpuls}

One of the most prominent stability computations in stars has been the modelling of the location of
the classical instability strip in the Hertzsprung--Russell diagram.
Since the seminal work by \citet{Baker-Kippenhahn62,Baker-Kippenhahn65} for modelling
linear stability coefficients in Cepheids, various
attempts have been made to reproduce theoretically the observed location of the instability strip.
The properties of the hotter, blue edge of the instability strip could be explained first 
\citep[e.g.,][and references therein]{Castor70, PetersenJoergensen72, DziembowskiKozlowski74,
Stellingwerf79}, mainly because for these hotter stars the rather thin surface convection zone 
does not affect pulsation dynamics too severely.
The modelling of the return to pulsational stability at the cooler, red edge, however,
has been less successful, despite the first promising attempts by, e.g., \citet{Deupree77},
who solved the nonlinear hydrodynamic equations, using a time-varying eddy viscosity, for
studying the stability properties of RR Lyrae stars. The need for a time-dependent convection 
treatment for modelling the low-temperature, red edge of the instability strip was recognized 
by \citet[][see also \citealp{Cox74}]{Baker-Kippenhahn65}. 

\subsubsection{Cepheids and RR Lyrae stars}
\label{s:Cepheids}

The first theoretical studies describing successfully the location of the cool edge 
of the classical Cepheid instability strip were reported by \citet{BakerGough79} for 
RR Lyrae stars, using linear stability analyses of radial modes and the local time-dependent
convection formulation of Section~\ref{sec:time-dependent_LMLM_1}.
Shortly thereafter, \citet{Xiong80} was successful with Cepheid models, using his own 
local time-dependent convection model 
\citep[][see Section~\ref{sec:reynolds_stress_models}]{Xiong77}.
In the same year \citet{GoncziOsaki80} used 
Unno's~(\citeyear{Unno67}, see Section~\ref{sec:time-dependent_LMLM}) time-dependent convection model for
analysing stability properties of Cepheid models, but only with the inclusion of an additional 
scalar turbulent viscosity, brought about by the small-scale turbulence 
[see Eq.~(\ref{eq:Wnu})], could \citet{Gonczi81} successfully model the return to 
stability near the cool edge of the instability strip.

Later, \citet{Stellingwerf86}, using \citepos{Stellingwerf82}
turbulence formulation with 
a simplified extension for one-zone pulsation models \citep{Baker66, BakerEtal66}, conducted
linear and nonlinear Cepheid stability analyses. He 
reported that the coupling between pulsation and convection can describe a return to 
stability for cooler Cepheid models. In this study, however, the effect of 
turbulent pressure was omitted in the calculations, but later included by 
\citet{MunteanuEtal05}, who concluded that the turbulent pressure appears to be a 
driving mechanism. Nonlinear pulsation modelling of Cepheids, using the nonlocal, time-dependent,
one-equation, convection formulation by \citet{Kuhfuss86}, were reported by
\citet{SmolecMoskalik08, Buchler09}, and \citet{SmolecMoskalik10}. 

Linear stability analyses of radial Cepheid pulsations were also conducted by
\citet{Balmforth-Gough88}, using \citepos{Gough77a} 
local convection model of 
Section~\ref{sec:time-dependent_LMLM}.
\citet{HoudekEtal99b} discussed linear stability analyses and nonadiabatic pulsation-period 
ratios in double-mode Cepheids, using \citepos{Gough77b}
nonlocal convection formulation of 
Section~\ref{sec.TNLMLT}. Both studies reproduced the cool edge of the classical instability strip,
with
the pulsationally perturbed turbulent pressure being the main contributor for stabilizing the
pulsation modes. \citet{YeckoEtal98}, on the other hand, found the damping effect of the 
small-scale turbulent eddy viscosity (see Eq.~\ref{eq:Wnu}) to be the main agent 
for making the pulsation modes stable at the cool side of the instability strip. The authors adopted 
the convection model by \citet{Gehmeyr92}, which is based on 
\citepos{Stellingwerf82}
turbulence model, for their linear stability computations.

\subsubsection{Mira variables}
\label{s:Mira}

Mira variables are long-period variables (LPV) with radial pulsation periods $P\gtrsim80\mathrm{\ days}$ 
located to the red of the classical instability strip with typical surface temperatures between 
2500 and 3500~K and luminosities between $\sim 10^3$ and $\sim 7\times10^3\,\mathrm{L}_{\odot}$.
The detailed driving mechanism of these low-order radial oscillations depends crucially on the
treatment of the coupling of the pulsations to the convection. Several attempts have been made
in the past to model this coupling in both linear and nonlinear calculations with rather
oversimplified descriptions \citep[e.g.,][]{Kamijo62, Keeley70, Langer71, CoxOstlie93}.
The first attempt to describe the coupling in a more realistic way was conducted by 
\citet{Gough66, Gough67}, who included the pulsational perturbations of both the 
convective heat (enthalpy) flux $\delta F_{\mathrm{c}}$ and momentum flux $\delta p_{\mathrm{t}}$ in the
linear stability analyses. Gough concluded that in particular the momentum flux perturbation 
$\delta p_{\mathrm{t}}$ has a stabilizing effect on the pulsations if the pulsation period is 
much shorter than the characteristic time scale of the convection, whereas for long-period
pulsations, such as in Mira variables or supergiants, the turbulent pressure fluctuations 
$\delta p_{\mathrm{t}}$ destabilizes (drives) the stellar pulsations.
It is perhaps interesting to note that a similar effect was noticed 
in linear Delta Scuti stability computations by \citet{Houdek96} and more recently by
\citet{AntociEtal14} (see Section~\ref{s:Delta_Scuti_Stars}), in 
which $\delta p_{\mathrm{t}}$ was found to drive high-order radial pulsations, in agreement with
observations.
Using the local time-dependent formulation by \citet{Gough77a}, \citet{BalmforthEtal90} 
concluded that including the turbulent pressure in the mean model of Mira variables 
modifies the equilibrium structure such as to make the
observed radial pulsations overstable in the pulsation computations which is, however, 
partially offset by the stabilizing influence of $\delta p_{\mathrm{t}}$.
\citet{XiongEtal98a}, on the other hand, found $\delta p_{\mathrm{t}}$, together with the 
turbulent eddy viscosity [see Eq.~(\ref{eq:Wnu})], to be the main stabilizing 
contribution to linear Mira pulsations.
\citet{MunteanuEtal05} 
and \citet{OlivierWood05} conducted nonlinear pulsation models using the one-equation 
turbulence models by \citet{Stellingwerf86} and \citet{Kuhfuss86} respectively, and reported
about the destabilizing effect of $\delta p_{\mathrm{t}}$, i.e., in accordance with the earlier findings
by \citet{Gough66, Gough67}.
It appears that further progress on modelling the interaction between convection 
and pulsations in Mira variables is required.

\epubtkImage{m17_eta_all_A_wint.png}{%
\begin{figure}[htb]
  \centerline{\includegraphics[width=\textwidth]{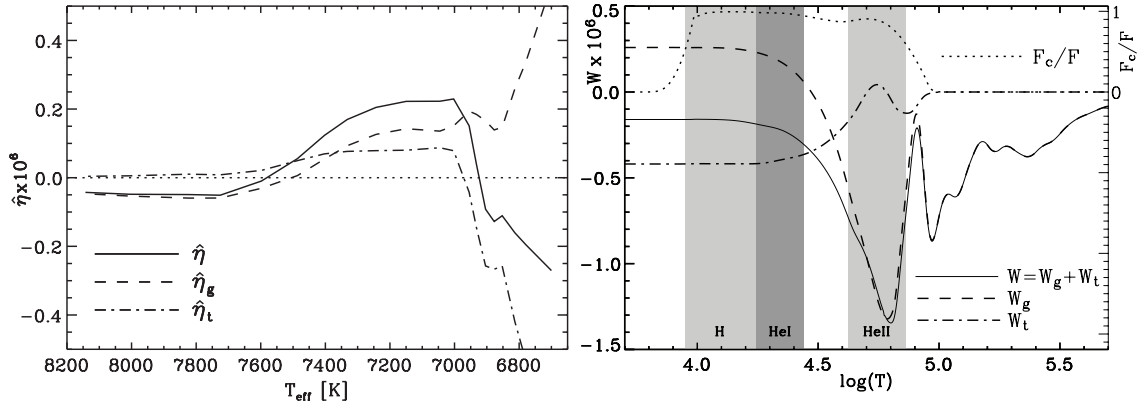}}
  \caption{Mode stability of an $1.7\,M_{\odot}$ Delta Scuti star computed with
               \citepos{Gough77a, Gough77b} convection model. \emph{Left:} Stability coefficient 
               $\hat\eta=\omega_\mathrm{i}/\omega_{\mathrm{r}}$ as a function of surface temperature 
               $T_{\mathrm{eff}}$ across the instability
               strip. Results are shown for the fundamental radial
               mode ($n=1$). Positive $\hat\eta$ values indicate mode instability. 
               The separate contributions to $\hat\eta$ arising from the
               gas pressure perturbations, $\eta_{\mathrm{g}}$, and from the perturbation
               of the momentum flux, $\eta_{\mathrm{t}}$, with $\eta=\eta_{\mathrm{g}}$+
               $\eta_{\mathrm{t}}$, are obtained from the evaluation of the work integral
               (\ref{e:DOG-Wall}).
               \emph{Right:} Integrated work integral $W$ as a function of the depth co-ordinate 
               $\log(T)$ for a model lying just outside the cool edge 
               of the instability strip (surface temperature
               $T_{\mathrm{eff}}=6813\mathrm{\ K}$). Results are
               plotted in units of $\hat\eta$. Contributions to $W$ 
               (solid curve) arising from the gas pressure perturbation, $W_{\mathrm{g}}$ 
               (dashed curve), and the turbulent pressure perturbations, 
               $W_{\mathrm{t}}$ (dot-dashed curve), are illustrated ($W=W_{\mathrm{g}}+W_{\mathrm{t}}$). 
               The dotted curve is the ratio of the convective to the total 
               heat flux ${F_{\mathrm{c}}/F}$. Ionization zones of H and He
               (5\% to 95\% ionization) are indicated \citep[from][]{Houdek00}.}
  \label{fig:hg_stab_delscuti}
\end{figure}}

\subsubsection{Delta Scuti stars}
\label{s:Delta_Scuti_Stars}

Already before the successful space missions CoRoT \citep{BaglinEtal09} and
\textit{Kepler} \citep{JCDEtal09b} several observing campaigns, e.g., the Delta Scuti Network (DNS)
or the Whole Earth Telescope (WET), have been providing excellent
oscillation data of Delta Scuti stars. For example, \citet{BregerEtal99} identified 24 pulsation
frequencies in the Delta Scuti star FG~Vir.
Such high-quality seismic data also provided well-defined observed locations of the lower part
of the classical instability strip \citep[e.g.,][]{RodriguezEtal00}, which modellers could use to test their
time-dependent convection models.

For example, \citet{HoudekEtal99} reported linear stability results about the locations of 
the lower instability strip for the fundamental and first overtone radial modes.
For one of these models,
detailed stability computations were reported by \citet{Houdek00}.
These computations included consistently the Reynolds stresses in both the equilibrium envelope
models [$p_{\mathrm{t}}$ in Eq.~(\ref{eq:mean_momentum_ba})]\epubtkFootnote{Note 
that $p_{\mathrm{t}}$ can be as large as 70\% of the total pressure in these 
Delta Scuti models \citep{Houdek00}, and may even exceed that value in 
hotter Delta Scuti stars \citep{AntociEtal13}.} 
and in the linear pulsation calculations 
($\delta p_{\mathrm{t}}$, Eqs.~\ref{eq:momentum-flux-perturb}, \ref{nlmlt-conv-fluxes-perturb}). 
The left panel of Figure~(\ref{fig:hg_stab_delscuti}) displays the stability coefficient 
$\hat\eta:=\omega_{\mathrm{i}}/\omega_{\mathrm{r}}$ ($\omega=\omega_{\mathrm{r}}+\mathrm{i}\omega_{\mathrm{i}}$)
of the fundamental mode (solid curve) 
for an evolving $1.7\,M_{\odot}$ Delta Scuti star crossing the lower instability strip. 
The individual contributions from the
gas (and radiation) pressure, $\eta_{\mathrm{g}}$ (dashed curve), and (perturbed) turbulent
pressure, $\eta_{\mathrm{t}}$ (dot-dashed curve), are indicated. 
The return to stability ($\hat\eta<0$) near the red edge, at about 6895~K, is entirely 
determined by the turbulent pressure contribution $\eta_{\mathrm{t}}$ to the work integral.
Another way to demonstrate this result is to analyse the work integral $W$.
The right panel of Figure~\ref{fig:hg_stab_delscuti} shows $W$ and its individual contributions
$W_{\mathrm{g}}$ [dashed curve; see Eq.~(\ref{e:DOG-Wg})] and
$W_{\mathrm{t}}$ [dot-dashed curve; see Eq.~(\ref{e:DOG-Wt})]
for a stellar model (with a surface temperature $T_{\mathrm{eff}}=6799\mathrm{K}$) located just outside 
the red edge of the instability strip. 
From this it is obvious that the dominating damping term to the work integral $W$ 
is the contribution from the turbulent pressure perturbations $W_{\mathrm{t}}$
(dot-dashed curve with negative value at the stellar surface), whereas the gas (and radiation) 
pressure contribution $W_{\mathrm{g}}$ (dashed curve) contribute to the driving (positive value at
the stellar surface).

\epubtkImage{mad_stab_delscuti.png}{%
\begin{figure}[htb]
   \centerline{\includegraphics[width=\textwidth]{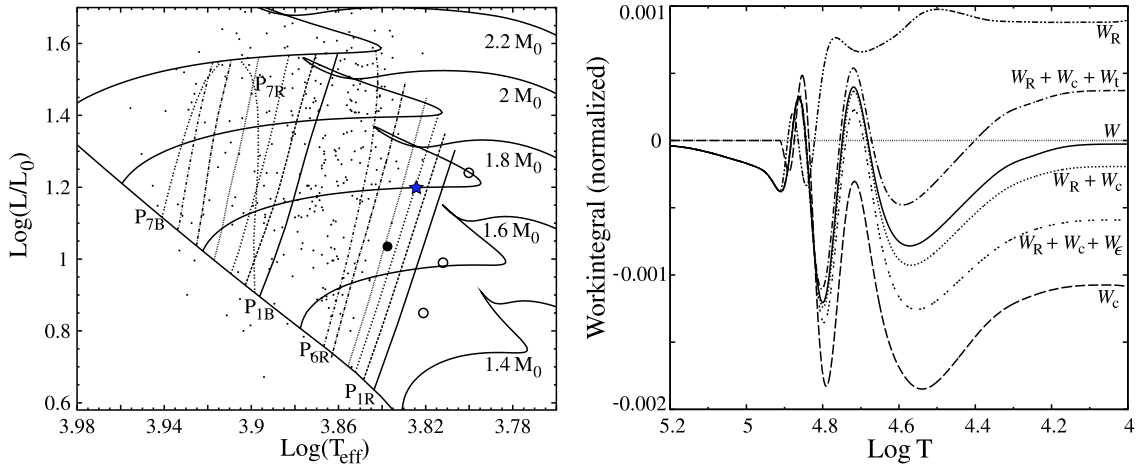}}
  \caption{Stability computations of Delta Scuti stars, which include the viscous
           dissipation rate $\epsilon_{\mathrm{t}}$ of turbulent kinetic energy according 
           to \citet{GrigahceneEtal05}.
           \emph{Left:} Blue and red edges of the instability strip superposed on evolutionary tracks 
           on the theorists Hertzsprung--Russel diagram. The locations of the edges, labelled 
           $\mathrm{p}_{n\mathrm{B}}$ and $\mathrm{p}_{n\mathrm{R}}$, are indicated 
           for radial modes with orders $1\le n\le7$. Results by \citet[]
           [$\alpha=2.0$, see Figure~\ref{fig:hg_stab_delscuti}]{Houdek00} and \citet{XiongDeng01} 
           for the gravest p modes are plotted as filled and open circles respectively.
           The small dots are observed Delta Scuti Stars \citep{RodriguezEtal00}.  
           \emph{Right:} Integrated work integral 
           $W$ as a function of the depth co-ordinate $\log(T)$ for a stable $n=3$ 
           radial mode of a $1.8\,M_{\odot}$ star (see `star' symbol in the 
           left panel).  Contributions to $W$ arising from the radiative 
           flux, $W_\mathrm{R}$, the convective flux, $W_\mathrm{c}$, the turbulent pressure 
           perturbations, $W_\mathrm{t}$ ($W_\mathrm{R}+W_\mathrm{c}+W_\mathrm{t}$), and from the 
           perturbation of the turbulent kinetic energy dissipation, $W_{\epsilon_{\mathrm{t}}}$ 
           ($W_\mathrm{R}+W_\mathrm{c}+W_{\epsilon_{\mathrm{t}}}$), are indicated. 
Images adapted from \citet{DupretEtal05a}.}
  \label{fig:mad_stab_delscuti}
\end{figure}}

\citet{DupretEtal05a} used the local time-dependent convection treatment 
of \citet[][see Section~\ref{sec:unno_gabriel_mlm}]{GrigahceneEtal05},
to study the stability properties of radial and nonradial pulsations in $\delta$~Sct stars.
In these calculations the perturbations of both the convective heat flux $\delta F_{\mathrm{c}}$ and
turbulent pressure $\delta p_{\mathrm{t}}$ were included in the linear pulsation computations,
but the (mean) turbulent pressure $p_{\mathrm{t}}$ was omitted in the construction of the 
equilibrium structure.
\citet{DupretEtal05a} found well defined red edges of the instability strip for both 
radial and nonradial modes 
using Grigahc{\'e}ne \etal's time-dependent convection model.
The authors found that the $\delta$~Sct low-order p modes become stable again with decreasing 
$T_{\mathrm{eff}}$ when the two thin convective zones, associated with the partial ionization of 
hydrogen and helium, merge to form one large surface convection zone.
For a solar-calibrated mixing-length parameter $\alpha=\ell/H$ ($\ell$ is the mixing length
and $H$ is the pressure scale height) 
the return to stability occurs at the observed location in the Hertzsprung--Russell diagram.
For smaller values of $\alpha$ the calculated cool edge of the instability strip is shifted
towards cooler surface temperatures $T_{\mathrm{eff}}$ \citep{DupretEtal05a}, in accordance with
the findings by \citet{Houdek00}, i.e., the observed location of the red edge could be used
to calibrate the mixing-length parameter.

The results of Dupret \etal's stability analysis for Delta Scuti stars are 
depicted in Figure~\ref{fig:mad_stab_delscuti}. The left panel compares the location of the
red edge with results reported by Houdek (\citeyear{Houdek00}, see also Figure~\ref{fig:hg_stab_delscuti}) and 
\citet{XiongDeng01}. The right panel of Figure~\ref{fig:mad_stab_delscuti} displays the 
individual contributions to the accumulated work integral $W$ for a star 
located near the red edge of the $n=3$ mode (indicated by the `star' symbol 
in the left panel). It demonstrates the near cancellation effect between 
the contributions of the turbulent kinetic energy dissipation, $W_{\epsilon_{\mathrm{t}}}$, 
and turbulent pressure, $W_{\mathrm{t}}$, making the contribution from the 
perturbations of the convective heat flux, $W_{\mathrm{c}}$, the dominating damping term.
Moreover, even calculations without the inclusion of $\delta p_{\mathrm{t}}$ in the
stability analysis led to a definition of the theoretical red edge of the instability strip.
This is in contrast to the finding by \citet{HoudekEtal99} and \citet{Houdek00}, who 
reported the turbulent pressure perturbations, $\delta p_{\mathrm{t}}$, as the
main mechanism for defining the red edge of the instability strip.

\epubtkImage{xiong_stab_delscuti.png}{%
\begin{figure}[htb]
  \centerline{\includegraphics[width=0.9\textwidth]{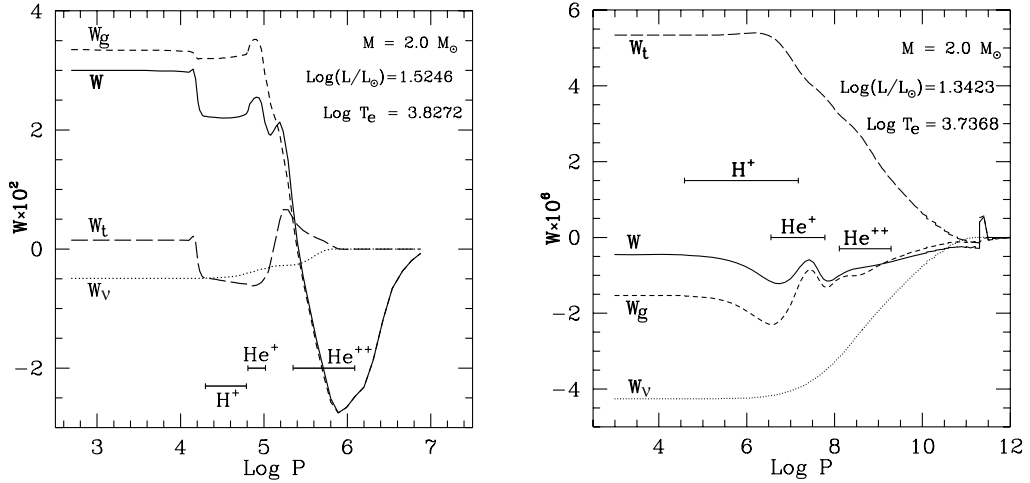}}
  \caption{Accumulated work integral $W$ as a function of the depth co-ordinate $\log(p)$.
               Results are shown for the $n=1$ radial mode of a Delta Scuti star located 
               inside the instability strip (left panel) and outside the red edge of 
               the instability strip (right panel). 
               The stability calculations include viscous dissipation by the 
               small-scale turbulence \citep[][see Eq.~(\ref{eq:Wnu})]{Xiong89}.
               Contributions
               to $W$ (solid curve) arising from the gas pressure perturbations, $W_{\mathrm{g}}$ 
               (dashed curve), the turbulent pressure perturbations, $W_{\mathrm{t}}$ 
               (long-dashed curve), and from the turbulent viscosity, $W_\nu$ (dotted curve),
               are illustrated ($W=W_{\mathrm{g}}+W_{\mathrm{t}}+W_\nu$). The ionization zones of H and 
               He are indicated. 
Images adapted from \citet{XiongDeng07}.}
  \label{fig:xiong_stab_delscuti}
\end{figure}}

The convection model by \citet{Xiong77, Xiong89} uses transport equations for the 
second-order moments of the convective fluctuations (see Section~\ref{sec:reynolds_stress_models}). 
Similar to \citet{Gough77a, Gough77b}, Xiong does not include a work integral 
for the viscous dissipation of turbulent kinetic energy, $W_{\epsilon_{\mathrm{t}}}$ 
(neither does \citet[][$\mathsection$26, $\mathsection$30]{UnnoEtal89}, but includes 
the viscous damping effect of the small-scale turbulence in his model.
All the convection models described in this review consider only the largest, most 
energy-containing, eddies and ignore the dynamics of the small-scale eddies 
lying further down the turbulent cascade. Small-scale turbulence does, however, 
contribute directly to the turbulent fluxes and, under the assumption 
that they evolve isotropically, they generate an effective viscosity 
$\nu_{\mathrm{t}}$, which is felt by a particular pulsation mode as an 
additional damping effect. The turbulent viscosity can be estimated as
(e.g., \citealp{Gough77b}; \citealp{UnnoEtal89}, $\mathsection$20) 
$\nu_{\mathrm{t}}\simeq\lambda(\ob{ww})^{1/2}\ell$, where $\lambda$ is a parameter
of order unity. The associated work integral $W_\nu$ can be written in cartesian
coordinates as \citep[][$\mathsection$63]{LedouxWalraven58} 
\begin{equation}
W_\nu=-2\pi\;\omega_{\mathrm{r}}\int_{m_{\mathrm{b}}}^M\nu_{\mathrm{t}}
\left[e_{ij}e_{ij}-\frac{1}{3}\left(\nabla\cdot\bx\right)^2\right] \rd m\,,
\label{eq:Wnu}
\end{equation}
where $e_{ij}=(\dj\xi_i+\di\xi_j)/2$ and $\bx$ is the displacement eigenfunction.
\citet{XiongDeng01, XiongDeng07} modelled successfully the instability strip of 
Delta Scuti and red giant stars and found the dominating damping effect 
to be the turbulent viscosity~(\ref{eq:Wnu}). This is illustrated in
Figure~\ref{fig:xiong_stab_delscuti} for models of two Delta Scuti stars: one model 
is located inside the instability strip (left panel), the other model is located outside the 
cool edge of the instability strip (right panel).
The contribution from the small-scale turbulence was also the 
dominant damping effect in the stability calculations by 
\citet{XiongEtal00} of radial p modes in the Sun, although the
authors still found unstable modes with radial orders between $11\le n\le23$. 
In contrast, \citet{Balmforth92a}, who adopted the convection model of 
\citet{Gough77a,Gough77b}, found all solar p modes to be stable due mainly to the 
damping of the turbulent pressure perturbations, $W_\mathrm{t}$, and 
reported that viscous damping, $W_\nu$, is about one order of magnitude 
smaller than the contribution of $W_\mathrm{t}$. Small-scale turbulent viscosity 
(\ref{eq:Wnu}) leads always to mode damping, where as the perturbation 
of the turbulent kinetic energy dissipation, $\Ldel\epsilon_{\mathrm{t}}$,
can contribute to both damping and driving of the 
pulsations \citep{Gabriel96}. The driving effect of 
$\Ldel\epsilon_{\mathrm{t}}$ was shown by \citet{DupretEtal05c} for $\gamma$ Doradus stars
(see Section~\ref{sec:gammaDor}).

It is clear from the discussion above that all three time-dependent convection models
\citep{Gough77a, Xiong89, GrigahceneEtal05} are able to reproduce theoretically the red
edge of the instability strip, and about at the same location as observations suggest. 
The very detailed
physical processes, however, that lead to the definition of the red edge are different in all
three convection models: Gough's model predicts that it is the perturbed Reynolds stress,
\citet{Xiong89} the viscous dissipation by the small-scale turbulence, and the model by
\citet{GrigahceneEtal05} predicts that it is the perturbed convective heat flux, which 
is responsible for the return to stability.

Form these results it is obvious that further research is necessary to identify the 
correct processes that define the location of the cool edge of the classical instability strip.

\subsubsection{Gamma Doradus stars}
\label{sec:gammaDor}

$\gamma$~Dor stars are F-type g-mode pulsators located near the 
red edge of the $\delta$~Scuti instability strip. 
A driving mechanism for these modes was proposed by \citet{GuzikEtal00}, who used
a standard time-independent, or frozen, convection model. 
The time scale associated to convective motions is, however,
shorter than the pulsation periods in most of the convective envelope $\gamma$~Doradus stars, 
and the validity of frozen convection models for estimating stability properties of
oscillations is therefore doubtful in these stars.
This motivated \citet{DupretEtal04b} and \citet{DupretEtal05b, DupretEtal05c} to use 
the time-dependent
convection treatment of \citet{GrigahceneEtal05}
for studying the driving mechanisms in $\gamma$~Dor stars. 
The important result was that unstable g modes are also obtained with this 
time-dependent convection
treatment, with a range of frequencies (from $\sim$~0.3 to 3~days) in agreement with 
typical observations. The theoretical instability strip could be computed and good 
agreement with observations was obtained for certain values of the mixing-length parameter $\alpha$. 


In the study of nonadiabatic processes we generally define the 
transition region in a star as the region where the thermal relaxation time 
is of the same order as the pulsation period. This region generally plays the 
major role in the driving or damping \citep[see e.g.,][]{Cox74}.
Efficient driving of $\gamma$~Dor g~modes occurs when much of the region lies just 
above the base of the convective envelope, for there the mode of heat transport 
changes dramatically. Because convection typically carries most of the heat, yet 
the flux 
is presumed to be frozen, it dams up heat when 
the radiative flux from below is relatively high and transmits more when the 
incident flux is low. The radiative component of the flux in the convection 
zone is essentially unchanged, aside from that resulting directly from the 
modification by convection of the mean thermal stratification. The process 
can drive the pulsations, and is often called ``convective blocking'', a 
terminology that could be misleading. A more accurate term would 
be ``convective shunting'', because convection essentially 
redistributes the radiative flux, thereby 
reducing the relative modulation by radiation of the total flux.

For the mixing-length parameter $\alpha=2$, and adopting Grigahc{\`e}ne \etal's 
convection model, the transition region and bottom of the 
convective envelope coincide for stellar models that are located in the 
Hertzsprung--Russell diagram where $\gamma$~Dor stars are observed. 
For smaller values of $\alpha$, stellar models with lower effective temperatures are
required to have a sufficiently deep convective envelopes, i.e., the location of the 
theoretically determined instability strip is shifted to lower temperatures 
in the Hertzsprung--Russell diagram. An important issue that has
not yet been fully studied is the role of the non-diagonal components of the
Reynolds stress in the driving. Preliminary studies using the formulation
of \citet{Gabriel87} indicate that it is important, but numerical instabilities
make this problem very delicate \citep[see also][]{GoughHoudek01}.

\subsubsection{Rapidly oscillating Ap stars}
\label{sec:roAp}

Rapidly oscillating Ap stars (hereafter roAp stars) are main-sequence stars 
with typical masses between 1.5 and $2.0\,M_{\odot}$ and with effective 
temperatures $T_{\mathrm{eff}}$ between 6800 and 8400~K. They are the coolest stars
amongst the chemically peculiar A-type (Ap) stars with high 
overabundances of Sr, Cr and Eu. They show strong, predominantly dipolar,
large-scale magnetic fields with magnitudes varying typically from 
1 to about 25~kG, leading to antipodal spots. The roAp stars have in general
rotation periods larger than about two days. The periods of the light 
variability range from roughly 5 to 21~minutes and are interpreted as 
high-order, low-degree acoustic modes. The first roAp star was discovered 
photometrically by \citet{Kurtz78} and their number has increased today
to about 43 \citep{KurtzEtal11}.
Recent reviews on roAp stars were given by \citet{Gough05, Cunha07}, 
\citet{Shibahashi08} and \citet{Kochukhov09}.

The observed pulsation properties in roAp stars suggest that the pulsation axis is not
aligned with the rotation axis. This had led to the so-called oblique pulsator model
\citep{Kurtz82}, in which the observed cyclically varying oscillation amplitudes are 
explained by dipole oscillations being aligned with the magnetic axis, which itself 
is oblique to the rotation axis of the star. The pulsation eigenfunction 
differs, however, from a simple spherical harmonic 
\citep[e.g.,][]{TakataShibahashi94, TakataShibahashi95, MontgomeryGough03, SaioGautschy04}.
By taking into account the effects of rotation and magnetic field, \citet{BigotDziembowski02}
generalized the oblique pulsator model, suggesting that the pulsation axis can be 
located anywhere between the magnetic and rotation axis.

Several models were suggested for the mechanism that drives the low-degree high-order 
acoustic modes to the relatively low (up to 6~mmag) observed amplitudes 
\citep[for a review see, e.g.,][]{Houdek03}.
In the first theoretical paper on roAp stars by \citet{DolezGough82}, the authors assumed a strong
dipolar magnetic field which inhibits convection totally in the polar spot-like regions, 
whereas in the equatorial region the convection is unaffected. The high-order acoustic 
oscillations are excited by the $\kappa$ mechanism in the hydrogen layers of the radiative 
polar spot-like regions. This model was adopted by \citet{BalmforthEtal01} using
updated opacity tables and the nonlocal, time-dependent convection model by 
\citet{Gough77a, Gough77b}. Depending on the assumed size of the
polar spot-like regions \citet{BalmforthEtal01} did find overstable, high-order, 
axisymmetric dipole modes and other overstable modes with increasing spot size.
   
This encouraging result has led \citet{Cunha02} to model the instability strip for 
roAp stars, but the author concluded that the models cannot explain the presence of 
observed oscillations in the coolest roAp stars. Even if the metallicity is varied
\citep{TheadoEtal09} the agreement between theory and observation could not be improved.

\citet{DolezGough82} also addressed the question why the axisymmetric oscillations should
always be nearly aligned with the spots, even if those spots are located near the 
(rotational) equator. They proposed that the (essentially standing) eigenmodes of 
oscillation suffer retrograde Coriolis precession in a frame of reference rotating 
with the star, and are therefore excited to observable amplitudes by the $\kappa$ mechanism
only if they are nearly aligned with the spots. A more detailed discussion on this
model was recently presented by \citet{Gough12b}.

The theory of roAp stars is further complicated by the still not fully understood mechanism that
limits the pulsation amplitudes to the rather small values of $\lesssim$~6~mmag, compared to the
amplitudes of classical pulsators such as Cepheids or Delta Scuti stars. A possible explanation 
could be energy dissipation in the higher atmospheric layers brought about by shocking
characteristics leading to steepening of the eigenfunctions which can then be thought of as 
a temporally harmonic series \citep{Gough13}, with the high harmonics propagating farther out into
the atmosphere where they dissipate the energy. 
Obviously, there is still much more to investigate in this type of stars.

\subsection{Mode lifetimes in stars supporting solar-like oscillations}
\label{s:solar_damping}

It is now generally accepted that stochastically excited oscillations are intrinsically 
damped.\epubtkFootnote{Note that this is strictly true only if the stochastic excitation 
dominates over any other driving mechanism that may operate at the same time.}
The excellent data from the \textit{Kepler} spacecraft of solar-like oscillations in many 
distant stars
have further strengthened this picture \citep[e.g.,][]{AppourchauxEtal14}. 
Nonadiabatic effects contribute, however, to the destabilization of stochastically 
excited modes, known as the ``general kappa-mechanism'' \citep{Balmforth92a}, which
are believed to be responsible, at least in part, for the local depression in the
linear mode damping rates at an oscillation frequency near the maximum of the
spectral mode heights in the Fourier power spectrum (see also discussion about 
Figure~\ref{fig:solar_dampingrates}). 
Some early studies about solar mode stability
discussed the possibility that stochastically excited modes could be overstable
\citep{Ulrich70, AntiaEtal88}. This idea was reconsidered recently by \citet{XiongDeng13},
but no convincing explanation was given by these authors about a mechanism that could
limit the amplitudes to the observed values. If solar-like acoustic modes were indeed
overstable some nonlinear mechanism must limit their amplitudes. The only convincing
mechanism, reported until today, that could limit the growth of overstable modes is
nonlinear mode coupling proposed by \citet{KumarGoldreich89}. For the rather small
amplitudes of stochastically excited oscillations only three-mode coupling is important.
\citet{KumarGoldreich89} studied the three-mode coupling analytically and concluded that
this nonlinear mechanism cannot limit the growth of unstable modes within 
the observed amplitude values. The remaining discussion on the properties of stochastically 
excited modes will therefore interpret the full width at half maximum (FWHM), or linewidth, of 
the spectral peaks in the Fourier power spectrum as (approximately) twice the 
linear damping rate, $2\eta$, and $\tau:=\eta^{-1}$, where $\eta$ is in units of
angular frequency, 
as the lifetime of the mode amplitude.

\epubtkImage{eta_phys.png}{%
\begin{figure}[htb]
  \centerline{\includegraphics[width=0.8\textwidth]{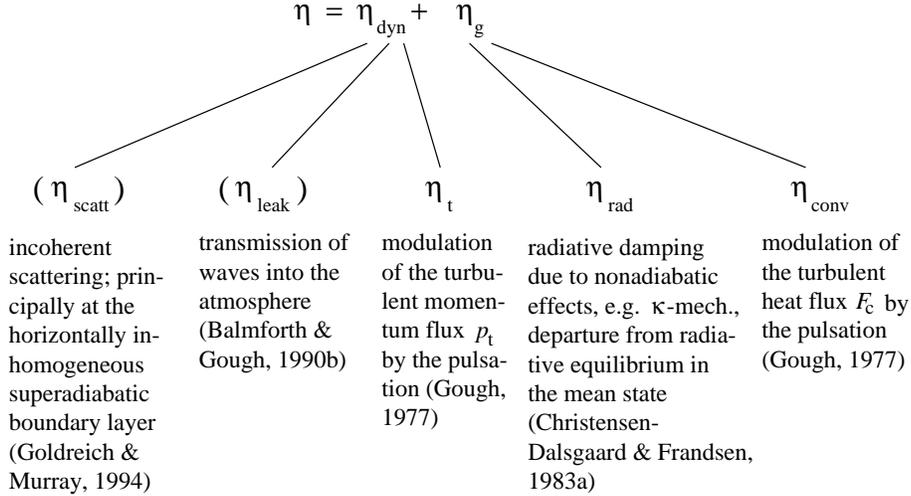}}
  \caption{Physical processes contributing to the linear damping rate $\eta$.
               They can be associated with the effects arising from the
               momentum balance ($\eta_{\mathrm{dyn}}$) and from the thermal
               energy balance ($\eta_{\mathrm{g}}$). The contributions $\eta_{\mathrm{scatt}}$
               and $\eta_{\mathrm{leak}}$ are in parentheses because they have not been
               taken into account in the computations reported in this paper.
               The influence of Reynolds stresses on solar modes, 
               contributing to $\eta_{\mathrm{t}}$, has been treated by \citet{GoldreichKeeley77a}
               in the manner of a time-independent scalar turbulent viscosity.
               The width of the line in the Fourier power spectrum of the oscillations
               is influenced also by nonlinearities, both those coupling a mode to others
               \citep{KumarGoldreich89} and those intrinsic to the mode itself.
Image reproduced with permission from \citet{HoudekEtal99}, copyright by ESO.}
  \label{f:eta_phys}
\end{figure}}

\subsubsection{Solar-type stars}

Damping of stellar oscillations arises basically from two sources: processes 
influencing the momentum balance, and processes influencing the thermal energy 
equation. Each of these contributions can be divided further according to their
physical origin as summarized in Figure~\ref{f:eta_phys}. A more detailed discussion
about the individual contributions to $\eta$ was given by \citet{HoudekEtal99}.

Important processes that influence the thermal energy balance are
nonadiabatic processes attributed to the modulation of the convective heat
flux by the pulsation. This contribution is related to the way that convection
modulates large-scale temperature perturbations induced by the pulsations
which, together with the conventional $\kappa$-mechanism, influences 
pulsational stability.

Current models suggest that an important contribution that influences
the momentum balance is the exchange of energy between the pulsation and 
the turbulent velocity field through dynamical effects of the perturbed 
Reynolds stress. In fact, it is the modulation of the turbulent fluxes by 
the pulsations that seems to be the predominant mechanism responsible for 
the driving and damping of solar-type acoustic modes.
It was first reported by \citet{Gough80}, using his local time-dependent convection
model of Section~\ref{sec:time-dependent_LMLM}, that the dynamical effects, arising 
from the turbulent momentum flux perturbation $\Ldel p_\mathrm{t}$, 
contribute significantly to the damping $\Gamma=2\eta$.
Detailed analyses by \citet{Balmforth92a}, \citet{HoudekEtal99}, and \citet{ChaplinEtal05} revealed 
how damping is controlled largely by the phase difference between the momentum 
and density perturbations. Those authors used the nonlocal generalization
(Section~\ref{sec:nonlocal_mlm}) of Gough's convection model including consistently the
Reynolds stresses (turbulent pressure) in both the equilibrium and pulsation calculations.
Damping arising from incoherent scattering, $\eta_{\mathrm{scatt}}$, 
\citep[][see Figure~\ref{f:eta_phys}]{GoldreichMurray94}
was not modelled in these calculations.

\epubtkImage{eta_solar-model.png}{%
\begin{figure}[htb]
  \centerline{\includegraphics[width=0.6\textwidth]{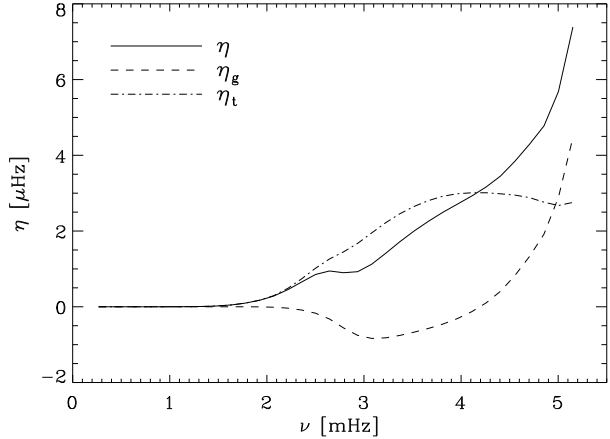}}
  \caption{Linear damping rates $\eta$ for a solar model as a function of
               cyclic frequency. Contributions 
               from the modulation of the radiative and convective heat flux, $\eta_{\mathrm{g}}$, 
               and turbulent moment flux (turbulent pressure), $\eta_{\mathrm{t}}$, 
               to the total damping rate (solid curve), $\eta$, are
               indicated by the dashed and dot-dashed curves respectively.
Image reproduced with permission from \citet{HoudekEtal99}, copyright by ESO.}
  \label{f:eta_solar-model}
\end{figure}}

Results with Gough's convection model are shown in Figure~\ref{f:eta_solar-model}. It plots the 
total linear damping rate $\eta$ as a function of cyclic frequency for a solar model. 
At a cyclic frequency of $\nu\simeq2.8\mathrm{\ mHz}$ the
net damping rate is characteristically flat. This local reduction in $\eta$ is predominantly
determined by radiative processes in the upper superadiabatic boundary layer of the convection
zone, which are locally destabilizing. It is interesting to note that the thermal relaxation time
of the solar superadiabatic boundary layer is of the same order than the period of a 
$\sim 2.8\mathrm{\ mHz}$ mode \citep{Balmforth92a}. 
Moreover, Figure~\ref{f:eta_solar-model} illustrates how
$\eta$ is determined by the delicate balance between the driving (destabilizing) and damping 
processes responsible for the energy exchange between the pulsations
and (turbulent) background state.

The analysis by \citet{DupretEtal04a} also included the pulsational perturbations of 
both the turbulent pressure and the convective heat in the pulsation computations 
using the local time-dependent convection formulation by \citet{GrigahceneEtal05}.
The mean turbulent pressure in the hydrostatic equilibrium model was, however, omitted.
Interestingly, \citet{DupretEtal04a} found  the perturbed convective 
heat flux $\Ldel\vec{F}_{\mathrm{c}}$ as the main mechanism for making solar oscillations stable,
similarly to the results found in Delta Scuti stars by \citet{DupretEtal05a}
(Section~\ref{s:Delta_Scuti_Stars}). The turbulent momentum flux perturbation $\Ldel p_{\mathrm{t}}$,
however, acts as a driving agent in these calculations.
Obviously, turbulent pressure perturbations must not be neglected 
in stability analyses of solar-type p modes. 

\epubtkImage{solar_dampingrates_v2.png}{%
\begin{figure}[htb]
  \centerline{\includegraphics[width=0.96\textwidth]{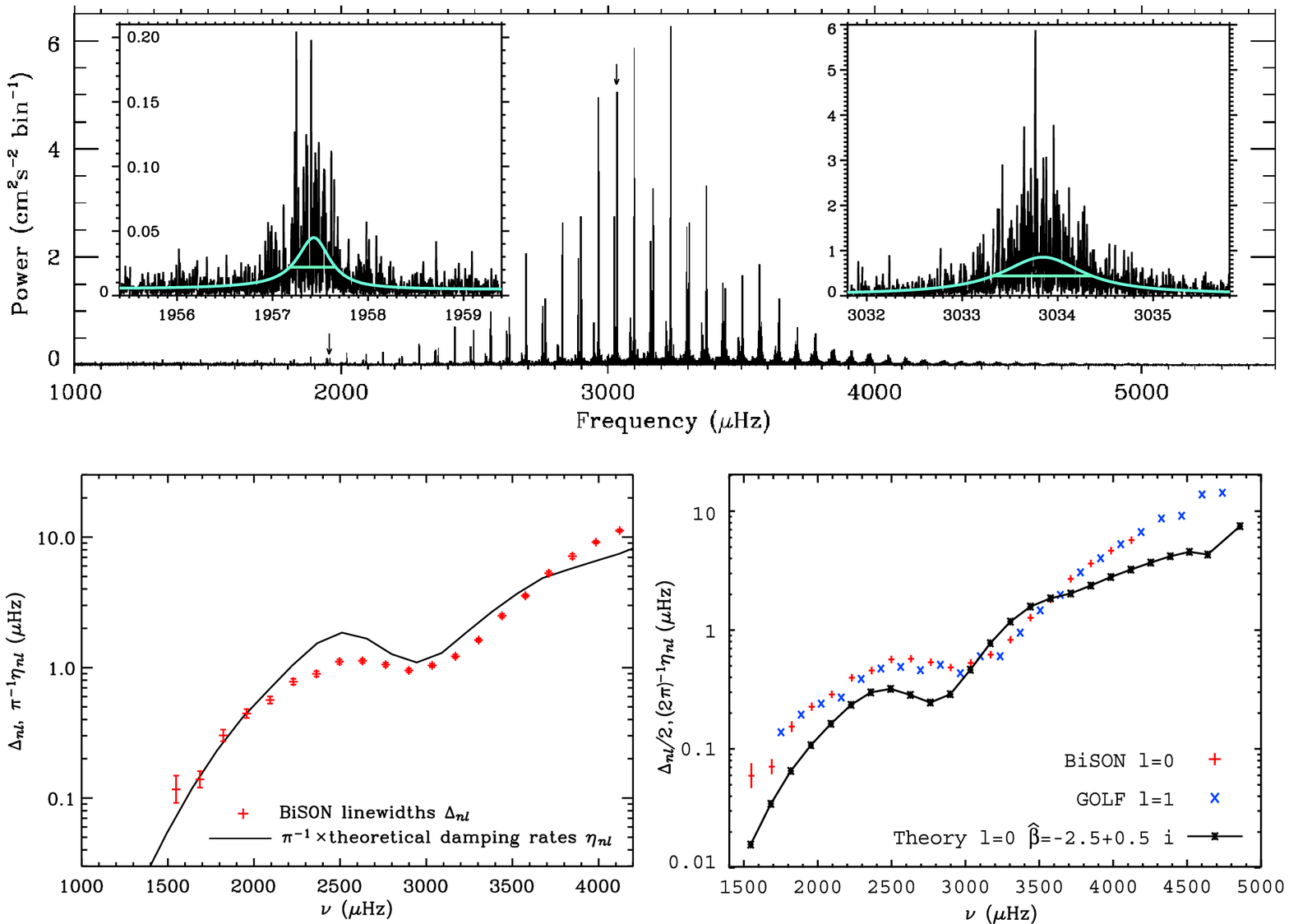}}
  \caption{\emph{Top:} Power spectrum of solar low-degree p modes obtained from a 3456-d
           data set collected by BiSON \citep{ChaplinEtal05}. 
           The two insets show Lorentzian profile fits (and their full-width at 
           half-maximum (FWHM); solid, blue curves) to the spectral peaks of 
           radial modes with order $n=13$ (left) and $n=21$ (right);
           both spectral peaks are indicated by vertical arrows in the power spectrum. 
           \emph{Bottom:} The symbols in the left-hand panel are the measured BiSON linewidths
           (we denote the FWHM in units of cyclic 
           frequency by $\Delta_{nl}$) which are 
           compared with the theoretical damping rates $\pi^{-1}\eta_{nl}$ (connected 
           by the solid curve) obtained with the nonlocal convection model of 
           Section~\ref{sec:nonlocal_mlm} (from \citealp{ChaplinEtal05}). 
           In the right-hand panel theoretical results of
           $(2\pi)^{-1}\eta_{nl}$ (solid curve) by \citet{DupretEtal04a} are compared 
           with observations of $\Delta_{nl}/2$ from the BiSON (red pluses) and the
           GOLF instrument (blue crosses). 
           }
  \label{fig:solar_dampingrates}
\end{figure}}

A comparison between solar linewidth measurements from the BiSON
(Birmingham Solar Oscillation Network) and from the GOLF (Global Oscillations at Low Frequency) instrument on board of the SOHO (SOlar and Heliospheric Observatory) spacecraft
and theoretical damping rates, computed with both time-dependent convection models 
by \citet[][left panel]{Gough77a, Gough77b} and \citet[][right panel]{GrigahceneEtal05},
is given in Figure~\ref{fig:solar_dampingrates}. 
The top panel shows the Fourier power spectrum of the observational time series
from BiSON \citep{ChaplinEtal05} obtained from a data set of length $T_{\mathrm{obs}} = 3456\mathrm{\ d}$
between 1991 and 2000.
Because the linewidths $\Delta_{nl}$ (FWHM) of this temporal power spectrum extend over many
frequency bins $(2T_{\mathrm{obs}})^{-1}$, $\Delta_{nl}$ is related, in units of cyclic frequency, to 
the damping rate $\eta$, which is in units of angular frequency, according to
\begin{equation} 
\Delta_{nl}=\pi^{-1}\eta_{nl}\,.
\label{eq:relation_dampingrates-linewidths}
\end{equation} 
\citepos{ChaplinEtal05} linewidth estimates, 
using \citepos{Gough77a, Gough77b} 
convection model, of solar radial modes (solid curve) are compared with BiSON data 
(red pluses) in the lower left panel 
of Figure~\ref{fig:solar_dampingrates}.
The outcome of \citepos{DupretEtal04a} stability 
computations for the Sun, 
using the local time-dependent mixing-length formulation by 
\citet[][see Section~\ref{sec:unno_gabriel_mlm}]{GrigahceneEtal05}, is 
illustrated in the lower right panel of Figure~\ref{fig:solar_dampingrates}
for a calibrated value of the complex parameter $\hat\beta$ [see Eq.~(\ref{pclose1})].
The value of $\hat\beta$ needs additional calibration 
(see also Section~\ref{sec:unno_gabriel_differences}) for it affects the 
estimated linewidths, which can vary up to a factor of four
(M.~Grosjean, personal communication).
In a nonlocal treatment, such as that given in the left panel 
of Figure~\ref{fig:solar_dampingrates}, 
no additional parameter is needed to suppress the rapid spatial oscillations in the pulsation
eigenfunctions (see discussions in Sections~\ref{sec:time-dependent_LMLM} and \ref{convpert}).

Although both calculations provide similar results, the very physical mechanism for
defining the frequency-dependent function of the estimated linear damping rates is
very different between these two calculations:
whereas \citet[][right panel of Figure~\ref{fig:solar_dampingrates}]{DupretEtal04a} 
reports that it is predominantly the
perturbed convective heat flux that stabilizes the solar p modes,  the results
from \citet[][left panel of Figure~\ref{fig:solar_dampingrates}; see also 
Figure~\ref{f:eta_solar-model}]{ChaplinEtal05} suggest 
that it is the perturbed turbulent pressure (Reynolds stress) that makes all modes stable.

\citet{BelkacemEtal12} repeated the pulsation calculations with a (simplified) 
nonlocal generalization (see Section~\ref{sec.TNLMLT})
of \citepos{GrigahceneEtal05} convection model, still omitting the turbulent pressure in the
equilibrium structure. Interestingly, these authors report, in agreement with the
previous finding by e.g., \citet{Balmforth92a}, that it is the turbulent pressure perturbations
that stabilize the modes in the Sun (see also discussion further down about
Figure~\ref{f:eta_Kepler}).
 
Estimates of linear damping rates in other solar-type stars were reported by \citet{Houdek96},
\citet{HoudekEtal99, ChaplinEtal09} and more recently by \citet{BelkacemEtal12}.
\citet{Houdek97} and \citet{HoudekEtal99} discussed the frequency-dependence of linear 
damping rates in main-sequence models with masses $(0.9\mbox{\,--\,}2.0)\,M_{\odot}$. 
\citet{ChaplinEtal09} discussed mean linear damping rates and linewidths around the 
maximum pulsation mode height in several solar-type stars.
\citet{BelkacemEtal12} compared linear damping rates at the maximum pulsation mode height
with linewidth measurements from the CoRoT (Convection and RoTation) and \textit{Kepler} 
space crafts.

\epubtkImage{Belkacem_etal_fig2-guenter1_log2c.png}{%
\begin{figure}[htb]
  \centerline{\parbox[t]{6.20cm}{\vspace{0pt}\includegraphics[width=6.20cm]{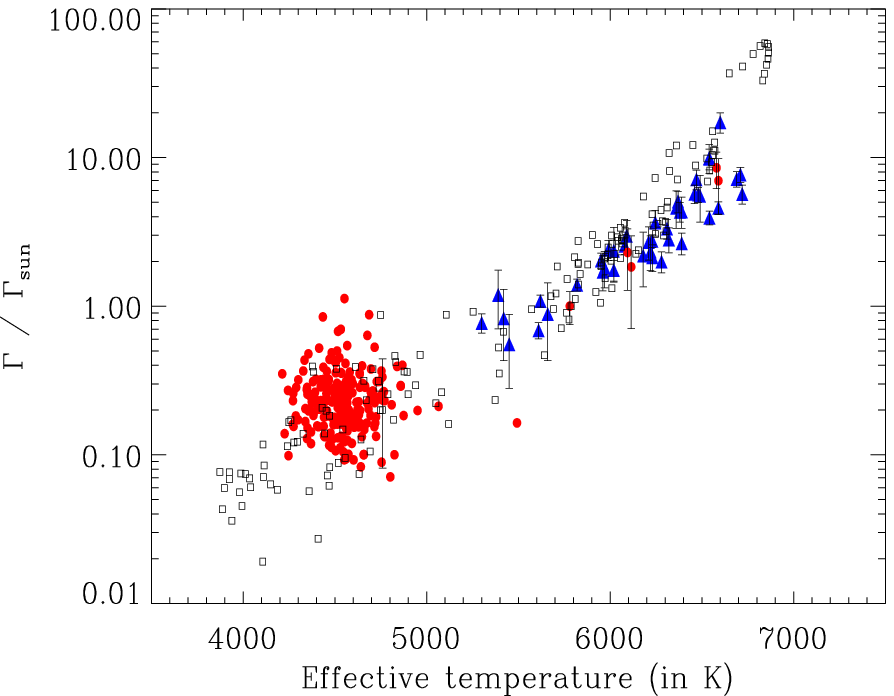}}\quad
              \parbox[t]{7.47cm}{\vspace{-10.5pt}\includegraphics[width=7.47cm]{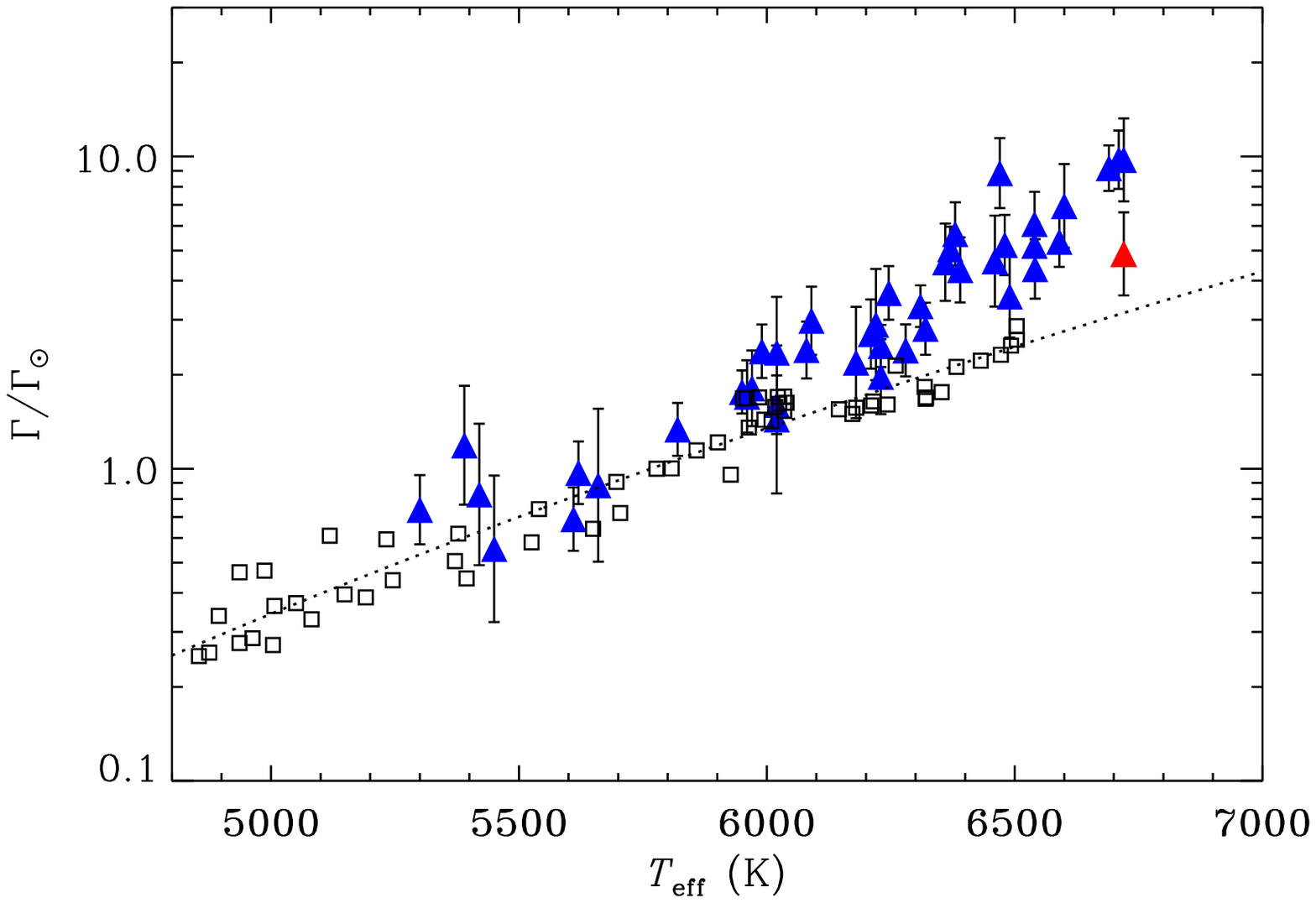}}}
  \caption{Comparison of CoRoT (red-filled circles) and \textit{Kepler} (blue-filled triangles) 
               linewidths $\Gamma$ \citep{AppourchauxEtal12} at maximum mode heights, 
               in units of the solar linewidth $\Gamma\simeq 0.95~\mu\mathrm{Hz}$,
               with theoretical stability calculations (open rectangles)
               as a function of surface temperature $T_{\mathrm{eff}}$.
               The 3-$\sigma$ error bars are indicated.
               \emph{Left panel:} Results 
               using Grigahc{\'e}ne \etal's time-dependent convection model, who
               obtained a power-law fit, $\Gamma\propto T_{\mathrm{eff}}^{10.8}$, to the
               model results (open rectangles).  
               Image reproduced with permission from \citet{BelkacemEtal12}, copyright by ESO.
               \emph{Right panel:} results are shown from Houdek et al.\, (in preparation)
               using \citepos{Gough77a, Gough77b} time-dependent convection model. 
               In this panel only \textit{Kepler} data (blue-filled triangles) are shown and only 
               from a single group (IAS) of mode fitters using a maximum likelihood estimator.  
               The dotted curve is a power-law fit, 
               $\Gamma\propto T_{\mathrm{eff}}^{7.5}$, to the model results (open rectangles). 
               The red-filled triangle indicates half the measured linewidth value for 
               the \textit{Kepler} star KIC 3733735.}
  \label{f:eta_Kepler}
\end{figure}}

Beside from testing and improving time-dependent convection models, the 
comparison of damping rate estimates with measured linewidths may also provide general 
scaling relations for mode linewidths (or lifetimes) of solar-like oscillations.
A first attempt was made by \citet{ChaplinEtal09}, using a limited number of estimated damping
rates and measured linewidths from predominantly ground-based instruments in solar-type stars. 
The authors reported that the largest dependence
of the linewidths is given by the star's surface temperature and proposed the scaling relation
$\eta\propto\Delta_{nl}\propto T_{\mathrm{eff}}^4$. This scaling relation was challenged 
later by measurements from the high-quality \textit{Kepler} data.
\citet[][see also \citealp{BaudinEtal11}]{AppourchauxEtal12} measured linewidths at both the 
maximum mode amplitude and mode height in 42 \textit{Kepler} stars, supporting solar-like oscillations, 
and reported a steeper surface-temperature dependence of $\Delta_{nl}\propto T_{\mathrm{eff}}^{13}$. 
\citet{BelkacemEtal12} compared these \textit{Kepler} measurements with theoretical 
estimates, $\Gamma=2\eta$, and reported reasonable agreement between observations 
and model computations (see left panel of Figure~\ref{f:eta_Kepler}). Using a nonlocal
generalization \citep[][see also Section~\ref{sec.TNLMLT}]{DupretEtal06c} of 
Grigahc{\'e}ne \etal's time-dependent convection model in the pulsation 
computations only,
\citet{BelkacemEtal12} found a surface-temperature dependence of 
$\Gamma=2\eta\propto T_{\mathrm{eff}}^{10.8}$, which is in reasonable agreement with 
the measurements by \citet{AppourchauxEtal12}.
The hydrostatic equilibrium models
were constructed with the local mixing-length formulation of Section~(\ref{sec:conveq}) 
neglecting the mean turbulent pressure $p_{\mathrm{t}}$ in the equation of hydrostatic support.
Moreover, \citepos{BelkacemEtal12} computations suggest that in the stability computations
the main contribution to mode damping is now the turbulent pressure perturbation $\delta p_{\mathrm{t}}$.
The use of a nonlocal treatment of the turbulent fluxes, though still only in the pulsation
computations and in a simplified manner (see Section~\ref{sec.TNLMLT}), has changed the effect of 
$\delta p_{\mathrm{t}}$ from a driving agent in Grigahc{\`e}ne \etal's local convection 
model to a damping agent in Belkacem \etal's nonlocal stability analyses. 
The damping effect of $\delta p_{\mathrm{t}}$ is in accordance with the previously reported findings by 
\citet{Gough80}, \citet{Balmforth92a}, \citet{HoudekEtal99}, and \citet{ChaplinEtal05}.
In Belkacem \etal's calculations a new strategy was adopted for selecting a value
for the parameter $\hat\beta$ [see Eq.~(\ref{pclose1})]: it was calibrated such as to make
the frequency of the local reduction (depression) in the linear damping rate $\eta$
(see, e.g., Figure~\ref{f:eta_solar-model} for solar model) coincide with the 
frequency $\nu_{\max}$ at which the power in the oscillation Fourier spectrum is largest, 
using the linear scaling relation by \citet[][see also \citealp{BrownEtal91}]{KjeldsenBedding95} 
between $\nu_{\max}$ and (isothermal) cutoff frequency.

Preliminary results by Houdek \etal\ (in preparation), using \citepos{Gough77a, Gough77b}
nonlocal convection model and with the Reynolds stresses consistently included in both the 
equilibrium and pulsation calculations, suggest a less steep surface-temperature dependence
of $\Gamma=2\eta\propto T_{\mathrm{eff}}^{7.5}$ (see right panel of Figure~\ref{f:eta_Kepler}).
The \textit{Kepler} data suggest a steeper surface-temperature dependence of 
about 13 \citep{AppourchauxEtal12}
in the considered temperature range $5300 < T_{\mathrm{eff}} < 6400\mathrm{\ K}$. It is, however, interesting 
to note that the observed mode linewidths may be affected by a short-periodic (magnetic) 
activity cycle, which modulates (periodically shifts) the mode frequencies and thereby 
effectively augments the mode linewidths when measured over a period longer than the activity cycle.
Possible evidence of such an effect was recently reported by R.~Garc{\'{\i}}a and T.~Metcalfe \citep[personal
communication, see also][]{GarciaEtal10} for the \textit{Kepler} star KIC 3733735 with a 
preliminary estimated effective linewidth
broadening of up to a factor of two. If this is indeed the case, a substantial amount of active
stars would then have smaller intrinsic linewidths than those plotted in Figure~\ref{f:eta_Kepler}
by the blue-filled triangles, thereby bringing the observations closer to the model estimates 
(open rectangles) in the right panel of this figure as illustrated, for example, by the
red-filled triangle for the \textit{Kepler} star KIC~3733735. 
The remaining discrepancy between theory and observation indicate that most likely a physical 
mechanism is still missing in our current theory. One such crucial mechanism is incoherent 
scattering at the inhomogeneous upper boundary layer 
\citep[][see also Figure~\ref{f:eta_phys}]{GoldreichMurray94}, which 
becomes increasingly more important for stars with higher effective 
temperatures \citep{Houdek12}.



\subsubsection{Red-giant stars}


From the scaling relations for stochastically excited modes
\citep[e.g.,][see also \citealp{JCDFrandsen83}]{KjeldsenBedding95, 
HoudekEtal99, Houdek06, SamadiEtal07}
it is expected to observe such modes with even larger pulsation amplitudes  
in red-giant stars.
First evidence of stochastically excited oscillations in red-giant stars were
reported by \citet{SmithEtal87, InnisEtal88} and \citet{EdmondsGilliland96}.
A comprehensive review about asteroseismology of red giants was recently
provided by \citet{JCD11}.

The first convincing detection of solar-type oscillations in a red-giant star was 
announced by an international team of astronomers \citep{FrandsenEtal02} for the star
$\xi$ Hydrae (HR~4450). 
\citet{HoudekGough02} calculated mode properties for the 
red-giant star $\xi$ Hydrae, using the nonlocal time-dependent formulation
by \citet[][see Section~\ref{sec.TNLMLT}]{Gough77a, Gough77b},
and reported velocity amplitudes that were in good 
agreement with the observations. Moreover, these authors also predicted theoretical 
mode lifetimes and reported for the most prominent p modes a lifetime $\tau$ of 15\,--\,17~days.
Their theoretical predictions of the radial damping rates and mode lifetimes for
$\xi$ Hydrae are shown Figure~\ref{f:xihya_pt_dampingrates}.

\epubtkImage{xiHya_dampingrates_ai3-xiHya_tau_smoothed_bw_7_v4.png}{%
\begin{figure}[htb]
  \centerline{\parbox[t]{6.18cm}{\vspace{0pt}\includegraphics[width=6.18cm]{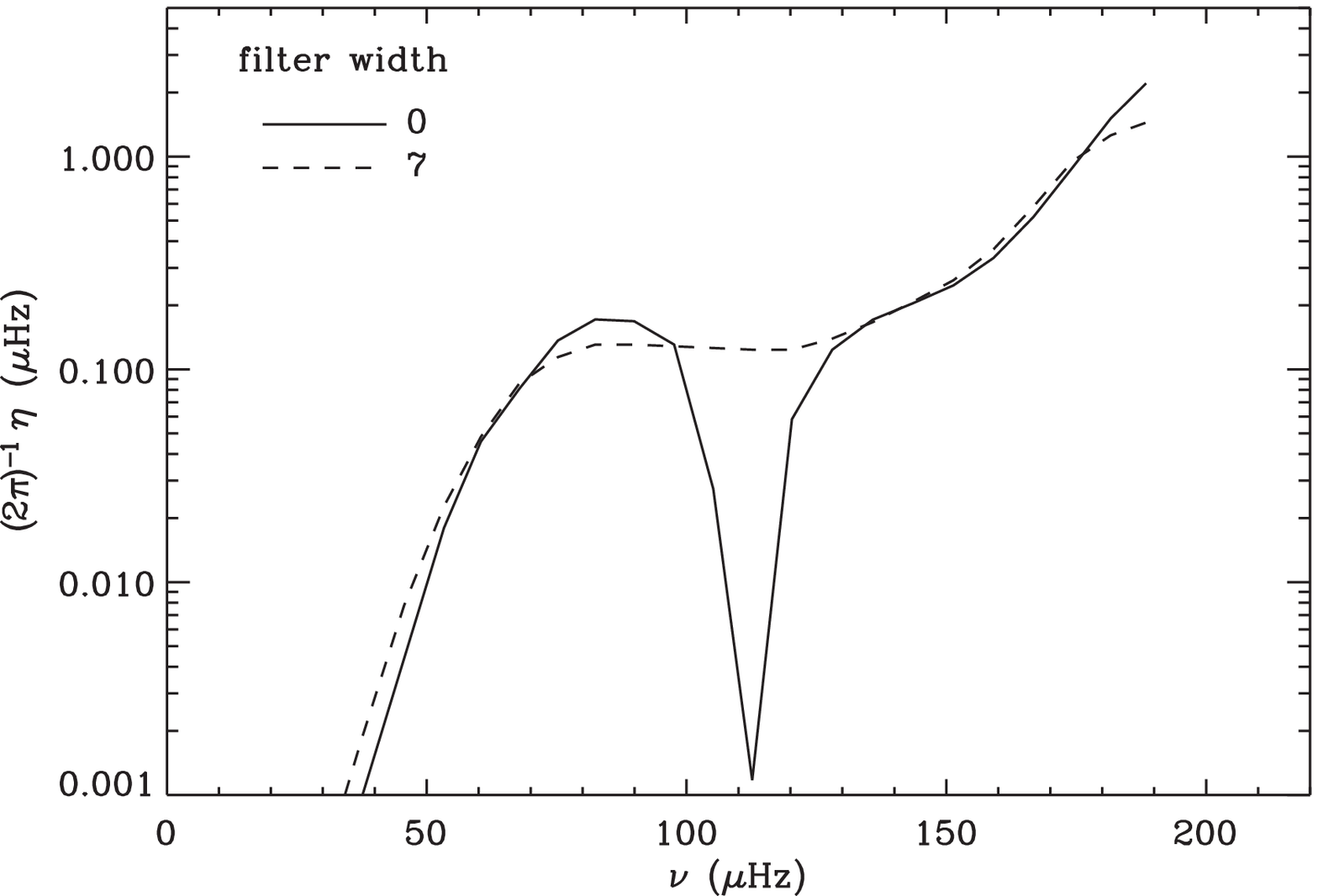}}\qquad
              \parbox[t]{6.6cm}{\vspace{0pt}\includegraphics[width=6.6cm]{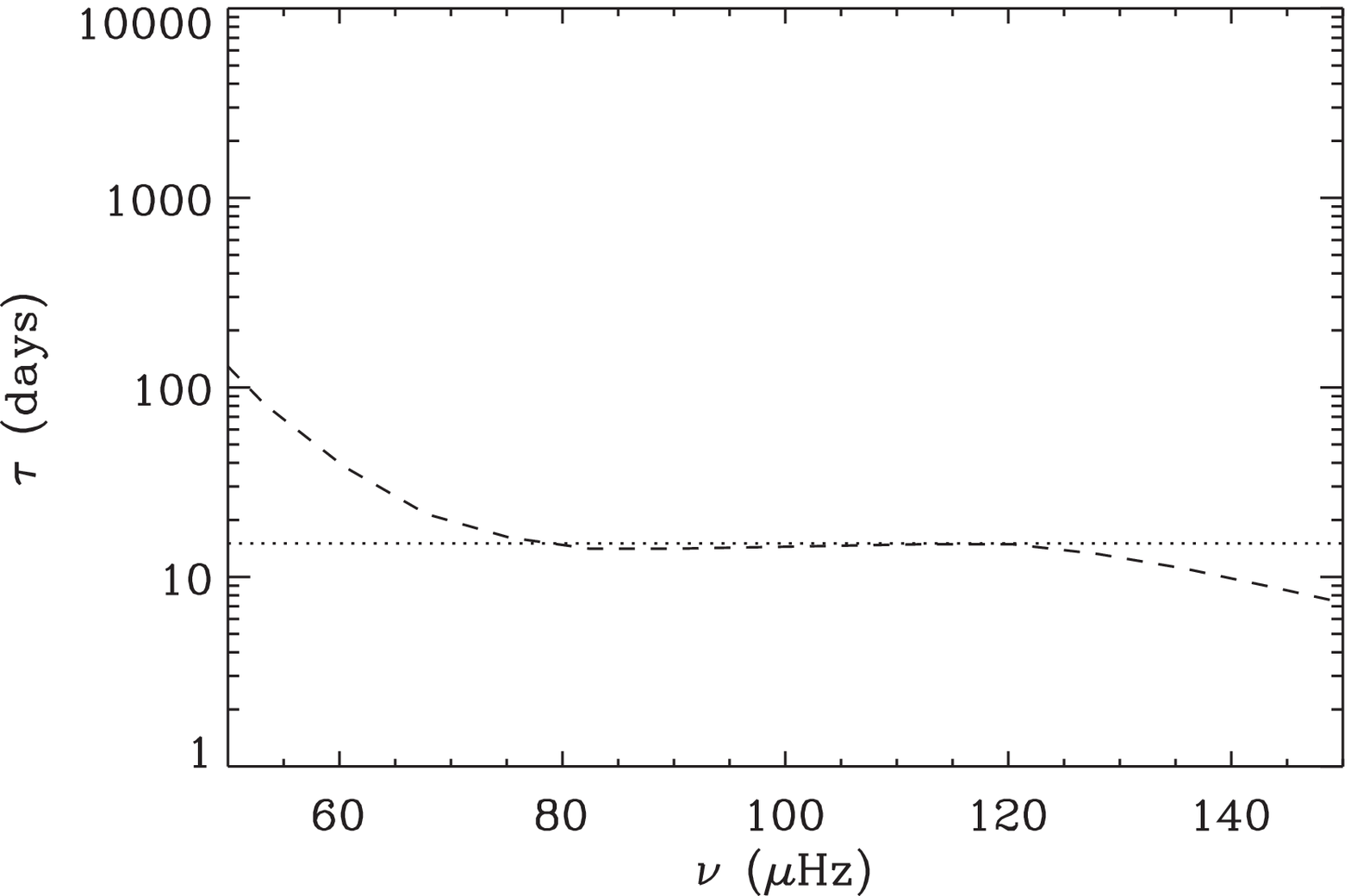}}}
  \caption{Model computations for the red giant star $\xi$ Hydrae (from \citealp{HoudekGough02}).
           \emph{Left panel:} The solid curve represents the theoretically predicted linear 
           damping rates as a function of cyclic pulsation frequency.
           The dashed curve is the result after applying
           a median smoothing filter with a width in frequency corresponding to 
           seven radial modes.
           \emph{Right panel:} Predicted mode lifetime $\tau=\eta^{-1}$ (where $\eta$ is in 
           units of angular frequency)
           for the median-smoothed damping rate (dashed curve) in the left panel. The horizontal
           dotted line indicates a mode lifetime of 15~days.}
  \label{f:xihya_pt_dampingrates}
\end{figure}}

Detailed structure modelling of $\xi$ Hydrae was carried out by, e.g., \citet{TeixeiraEtal03},
who concluded that $\xi$ Hydrae could either be in the ascending phase on the red giant branch
or in the later phase of stable helium-core burning, i.e., located in the so-called 
`red clump' in the Hertzsprung--Russell diagram. Because the stable helium-core burning phase lasts
by far much longer than the ascending phase, it is more likely that $\xi$ Hydrae is a
`red clump' star. Regardless of its detailed evolutionary phase, the model's mean large 
frequency separation $\Delta\nu:=\langle\nu_{n+1l}-\nu_{nl}\rangle$ was identified to be similar 
to the frequency separation between two consecutive peaks in the observed Fourier power spectrum,
i.e., identifying all the observed modes to be of only one single spherical degree, presumably of
radial order. Supported by previous arguments by \citet{Dziembowski77a} and \citet{DziembowskiEtal01},
\citet{JCD04a} discussed qualitatively the possibility that all nonradial modes in red-giant stars
are strongly damped and therefore have small amplitudes and peaks in the
Fourier power spectrum. Adopting this idea, \citet{StelloEtal04,StelloEtal06} 
developed a new method for measuring mode lifetimes 
from various properties of the observed oscillation power spectrum and reported mode 
lifetimes of only about 2\,--\,3 days for the star $\xi$ Hydrae. This is in stark contrast 
to the predicted values of 15\,--\,17~days by 
\citet[][see Figure~\ref{f:xihya_pt_dampingrates}]{HoudekGough02}. 

This discrepancy was resolved later by more detailed observations of red-giant stars.
Spectroscopic observation of oscillation modes in red-giant stars by \citet{HekkerEtal06}
reported first  evidence of the presence of nonradial pulsation modes
and \citet{KallingerEtal08}  reported possible nonradial oscillations in
a red-giant star using data from the Canadian spacecraft MOST (Microvariability and
Oscillations of STars).
It was, however, the high-quality data from the CoRoT satellite
that showed clear evidence of nonradial oscillations in several hundreds of red-giant stars
\citep[][see also \citealp{MosserEtal11}]{DeRidderEtal09} and later also from the NASA 
spacecraft \textit{Kepler} \citep{HuberEtal10}.
Lifetime measurements from these high-quality space data provided values of about 15 days
\citep{CarrierEtal10, HuberEtal10, BaudinEtal11} which are in good agreement with the 
earlier predictions for radial modes by \citet{HoudekGough02}.

\epubtkImage{DupretEtal09_Fig7_B-DupretEtal09_Fig11_E.png}{%
\begin{figure}[htb]
  \centerline{\parbox[t]{6.50cm}{\vspace{0pt}\includegraphics[width=6.50cm]{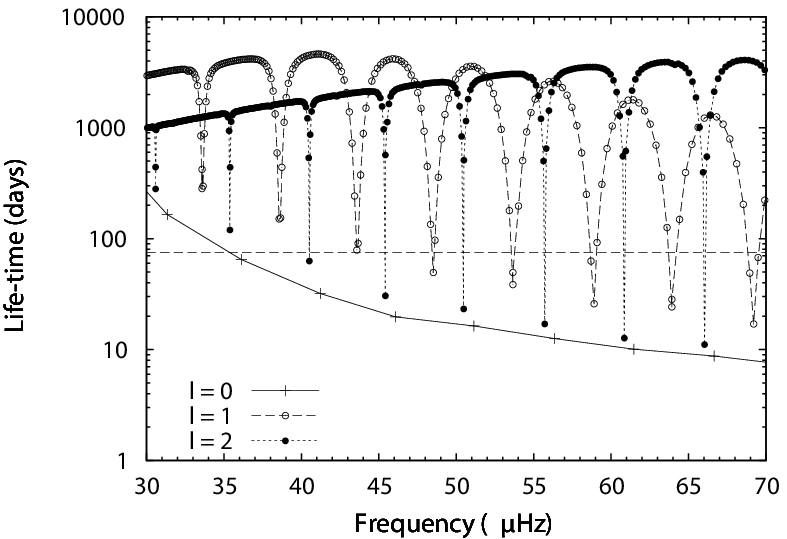}}\qquad
              \parbox[t]{6.50cm}{\vspace{0pt}\includegraphics[width=6.50cm]{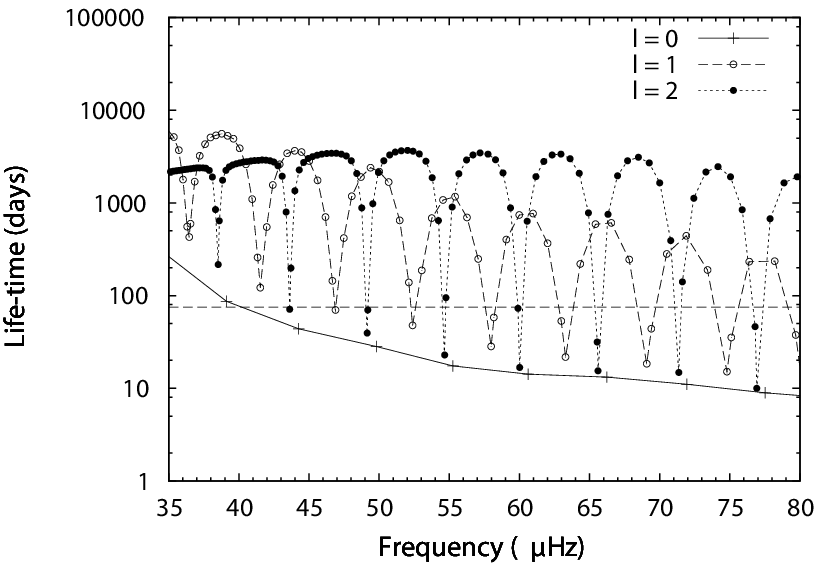}}}
  \caption{Estimated mode lifetimes for red-giant models.
               Results are shown for radial modes (pluses connected by solid lines) and
               for nonradial modes of spherical degree $l=1$ (open circles connected by dashed lines)
               and $l=2$ (filled circles connected by dotted lines). The horizontal dashed line
               at 75~days indicates the border between unresolved and resolved modes for the
               CoRoT long run of ${\cal T}$\,=\,150~days. 
               \emph{Left panel:} Results for a $2\,M_{\odot}$ red-giant model located approximately
               half-way up the ascending red-giant branch in the Hertzsprung--Russell diagram, 
               i.e., in the phase of hydrogen-shell burning.
               \emph{Right panel:} Results for a $3\,M_{\odot}$ red-giant model in the phase
               of central-helium burning.
Image reproduced with permission from \citet{DupretEtal09}, copyright by ESO.}
  \label{f:dupret09}
\end{figure}}

A theoretical explanation of the observed properties of nonradial oscillations in red-giant stars
was provided by \citet{DupretEtal09}. In the pulsation calculations the authors used 
\citepos{DupretEtal02} simplified, nonlocal 
generalization (see Section~\ref{sec.TNLMLT}) 
of \citepos{GrigahceneEtal05} convection model. The equilibrium model was constructed
with the local mixing-length formulation of Section~(\ref{sec:conveq}) omitting
the mean turbulent pressure $p_{\mathrm{t}}$ in the equation of hydrostatic support.
Figure~\ref{f:dupret09}
shows theoretical mode lifetimes of radial ($l=0$) and nonradial ($l=1,2$) modes for two red-giant
models in different evolutionary phases. The pluses connected by the solid lines are the solutions
for radial modes, with values of about 15\,--\,20~days at mid-frequency values, and may be compared 
with the theoretical results in the right panel of 
Figure~\ref{f:xihya_pt_dampingrates} by \citet{HoudekGough02}.
Except at lowest frequencies the lifetimes of radial modes are shorter than half of the
assumed observing time, ${\cal T}/2$, of 75~days (horizontal dashed line in Figure~\ref{f:dupret09}),
corresponding to a long-run observation of ${\cal T}=150\mathrm{\ days}$ by CoRoT. 
The `observational' damping time of ${\cal T}/2=75\mathrm{\ days}$ represents the border between unresolved 
and resolved oscillation modes. 
The radial modes in Figure~\ref{f:dupret09}) are therefore well
resolved with mode heights $H=2\tau V^2_{\mathrm{rms}}$ \citep[][see also \citealt{JCD11}]{ChaplinEtal05}
of the spectral peaks in the observed Fourier power spectrum, where $V_{\mathrm{rms}}$ is the 
root-mean-square of the velocity amplitude of the oscillation modes.
Because the mode height $H$ is independent of mode inertia \citep[e.g.,][]{DupretEtal09} one would
expect all modes of any degree $l$ in a given frequency interval to have similar heights $H$.
This behaviour was, however, not observed in $\xi$ Hydrae, for example. The explanation for this
behaviour is provided by the properties of the nonradial-mode lifetimes 
($l=1$: open circles connected by dashed lines; $l=2$: filled circles connected by dotted lines)  
as demonstrated for two red-giant models
in Figure~\ref{f:dupret09}, calculated by \citet{DupretEtal09}.
The nonradial modes of these red-giant models are mixed modes with g-mode character in 
the core and p-mode character near the surface \citep[see also][]{DziembowskiEtal01, JCD04a}. 
Nonradial modes that are dominated by p-mode characteristics have 
the shortest lifetimes with values only slightly larger than those of the radial modes.
The majority of nonradial modes in the models depicted in Figure~\ref{f:dupret09}, however,
are dominated by g-mode characteristics with lifetimes substantially larger 
than the `observational' damping time of ${\cal T}/2$ of 75~days. 
These g-dominated nonradial modes are,
therefore, unresolved with spectral peaks in the Fourier power spectrum resembling that of a 
simple undamped sine wave. The width of the spectral peaks is than proportional to ${\cal T}^{-1}$, 
and the corresponding mode height is rather small because $H\propto{\cal T}V^2_{\mathrm{rms}}$ for 
${\cal T}\ll\tau$ \citep[e.g.,][]{ChaplinEtal05, FletcherEtal06, DupretEtal09}, making the 
modes very likely invisible. 
Another reason for the non-detection of g-dominated modes is radiative 
damping near the bottom of the hydrogen-burning shell, such as in the model depicted
in the left panel of Figure~\ref{f:dupret09}, which becomes stronger with
increasing density contrast between core and envelope.
The Fourier power spectrum of red-giant stars can, therefore, be
of large complexity \citep{DupretEtal09, GrosjeanEtal14}.

\newpage
\section{Multi-colour Photometry and Mode Identification}
\label{sec:multi-colour}

Mode identification is an important but difficult problem for classical pulsators.
If the effect of rotation is neglected and some other approximations
are done, a simple semi-analytical formula can be obtained \citep{Dziembowski77, BalonaStobie79, 
StamfordWatson81, Watson88, CugierEtal94}
for the monochromatic magnitude variations $\delta m_\lambda$ of nonradial oscillations 
in different colour passbands of wavelength $\lambda$. Because of its importance, we 
give it here in the form proposed by \citet{DupretEtal03}

\begin{eqnarray}
\label{deltaml}
\delta m_\lambda &=&
-\:\frac{2.5}{\ln 10}\;\epsilon\,\:P_l^m(\cos i)\:b_{l\lambda} 
\nonumber \\ 
& & \Big[\,-\:(l-1)(l+2)\:\cos(\omega\,t) \nonumber \\ 
& & \hspace*{0.3cm}+\;f_{T}\:\cos(\omega\,t + \psi_T)
\,\left(\alpha_{T\lambda}\,+\,\beta_{T\lambda}\,\right)
\nonumber \\
& & \hspace*{0.3cm}-\;f_g\:\cos(\omega\,t)\,\left(\alpha_{g\lambda}\,+\,
\beta_{g\lambda} \,\right)\,\Big]\,,
\end{eqnarray}
where $\epsilon$ is the relative amplitude of the radial displacement $\xi_{\mathrm{r}}$,
$P_l^m$ is the associated Legendre function of degree $l$ and azimuthal order $m$, 
$i$ is the inclination angle between the stellar axis and the observer's line of sight,
and $\omega$ is the angular oscillation frequency.
This equation depends directly on the spherical 
degree $l$ of the modes. Therefore, a comparison between  theoretical and observed 
amplitude ratios and phase differences provides a possibility to identify $l$. 
Some coefficients of Eq.~(\ref{deltaml}) depend on the equilibrium atmosphere model, i.e.,
\begin{equation}
b_{l\lambda}= \int_0^1\,h_\lambda\: \mu \: P_l \rd \mu\,,
\end{equation}
where $h_\lambda(\mu)$ is the normalized monochromatic limb-darkening law, and
\begin{eqnarray}
\alpha_{T\lambda}&=&\partial \ln  F_{\lambda}/\partial \ln T_{\mathrm{eff}}|_g\,,\cr
\alpha_{g\lambda}&=&\partial \ln  F_{\lambda}/\partial \ln g|_{T_{\mathrm{eff}}}\,,\cr
\beta_{T\lambda} &=&\partial \ln\,b_{l\lambda}/\partial\ln T_{\mathrm{eff}}|_g\,,\cr
\beta_{g\lambda} &=&\partial \ln\,b_{l\lambda}/\partial\ln g|_{T_{\mathrm{eff}}}\,.
\end{eqnarray}
The two important theoretical ingredients in Eq.~(\ref{deltaml}) are the amplitude, 
$f_{T}$, and 
phase, $\psi_{T}$, of local effective temperature variation of a 
pulsation mode.
Several authors 
\citep[e.g.,][but see results from \citealp{DaszynskaEtal05} in Section~\ref{s:delscuti_multi-colour}]{DaszynskaEtal03}
represent these nonadiabatic parameter by a simple complex quantity $f$.
The relation between the two is
\begin{equation}
f=4\,f_{T}\,\exp(i\psi_{T})\,.
\end{equation}
The quantity $f_{g}$ is the relative amplitude of 
effective gravity variation of a 
pulsation mode.
Linear pulsation models do not provide absolute amplitudes. Theoretical amplitude 
ratios and phase differences between different photometric passbands can be 
determined by integrating Eq.~(\ref{deltaml}) over the passbands and 
taking the complex ratios.

The quantities $f_{T}$ and $\psi_{T}$ can only be rigorously obtained from
nonadiabatic computations.
Mode identification methods based on multi-colour photometric observations are
thus model dependent. This is particularly important for stars with convective envelopes
(e.g., $\delta$~Sct and $\gamma$~Dor stars). For these stars the nonadiabatic predictions are
very sensitive to the treatment of convection and its interaction with oscillations.

\subsection{Delta Scuti stars}
\label{s:delscuti_multi-colour}

Mode identification based on multi-colour photometry has been widely applied to
$\delta$~Sct stars. First studies considered $f_{T}$ and $\psi_{T}$ as 
free parameters \citep[e.g.,][]{GarridoEtal90}.
Later, nonadiabatic computations were performed
but with a time-independent convection treatment 
(frozen convection approximation), predominantly  with  local mixing-length models 
\citep{BalonaEvers99, DaszynskaEtal03, MoyaEtal04},
and with Full Spectrum of Turbulence models \citep{MontalbanDupret07}.
The frozen convection approximation, however,  is often not justified 
in $\delta$~Sct stars.
\citet{DupretEtal05b} used the local time-dependent treatment of
\citet{GrigahceneEtal05} to determine the nonadiabatic photometric observables 
in $\delta$~Sct stars and compared their theoretical results with the observations 
for several stars
(see also \citealp{Houdek96}, who used \citepos{Gough77a} nonlocal time-dependent convection model 
to predict the complex $f$ quantity and phases in the $\delta$ Scuti star FG Vir).
\citet{DupretEtal05b} found that from the middle to the red border of the 
instability strip 
models with the time-dependent convection treatment provide 
significantly different predictions for the photometric amplitudes and
phases compared to models in which the perturbation of the turbulent fluxes were
neglected (frozen convection). 
The largest differences are found for models with values for the mixing-length 
parameter $\alpha$ of the order of the solar-calibrated value. With the frozen convection and a
large value for $\alpha$ a significant phase lag is obtained in the hydrogen ionization zone. 
This phase lag is related
to the huge time variations of the temperature gradient in this region.
Using models with time-dependent convection, large variations of the entropy gradient 
(and thus the temperature gradient) are not allowed because of the control by the 
convective flux, and smaller phase-lags in the hydrogen zone are predicted.
Therefore, a time-dependent treatment of the turbulent fluxes is required in the 
stellar model calculations for photometric mode identification in cooler 
$\delta$~Sct stars.

\citet{DaszynskaEtal05} compared the real and imaginary parts of the complex $f$
parameter
between observations of the relatively cool $\delta$~Scuti star FG~Vir
\citep{BregerEtal99} and model computations using the nonlocal, time-dependent 
convection model by \citet[][see Section \ref{s:nonlocal-mlm}]{Gough77b}. 
The outcome is presented in Figure~\ref{fig:FGVir_f}.
Best agreement with observations are
found for rather small values of the mixing-length parameters $\alpha$, i.e., 
for values that are small compared to the solar-calibrated value of $\alpha=2.03$ 
for the adopted 
nonlocal, time-dependent convection model by \citet{Gough77b}. Using a standard, local, 
time-independent convection formulation leads to even smaller values for $\alpha$
relatively to a solar-calibrated value for the standard mixing-length formulation.

\epubtkImage{FGVir_f_nlmlm.png}{%
\begin{figure}[htb]
  \centerline{\includegraphics[width=0.8\textwidth]{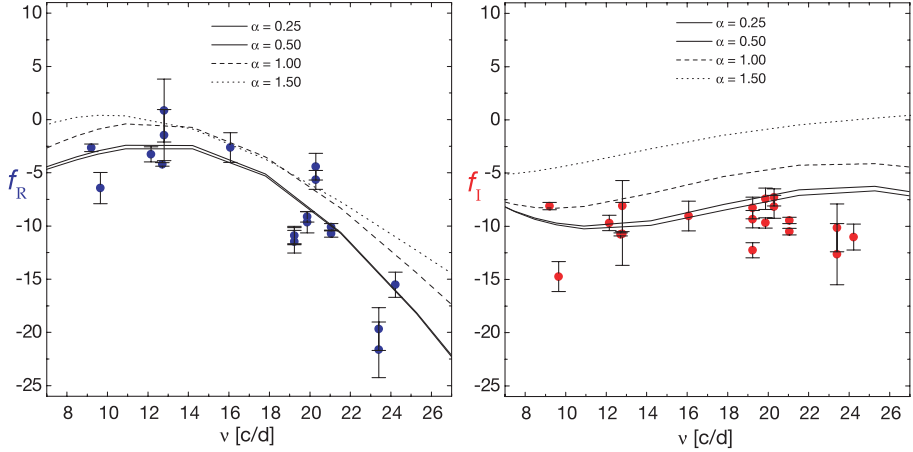}}
  \caption{Comparison of the observable $f$ between observations in the
  Delta Scuti star FG~Vir (filled circles 
  with errorbars) and theory (curves). Results are plotted as a function of frequency 
  in units of cycles/day. The model results are shown for four different values of the 
  mixing-length parameter $\alpha$. The turbulent fluxes were obtained from the nonlocal, 
  time-dependent mixing-length formulation of Section~\ref{sec:nonlocal_mlm}. 
  The real and 
  imaginary parts of $f$ are shown in the left and the right panels,
  respectively. 
Image reproduced with permission from \citet{DaszynskaEtal05}, copyright by ESO.}
  \label{fig:FGVir_f}
\end{figure}}

\subsection{Gamma Doradus stars}

Mode identification based on multi-colour photometry  can also be considered 
for $\gamma$~Dor
stars. \citet{DupretEtal05c} showed that frozen-convection models give phase-lags 
in complete disagreement
with observations. Time-dependent convection models give a better agreement with observations
and are thus
required for photometric mode identification. In frozen-convection models
the $\kappa$-mechanism plays some
role in the He and H ionization zones, implying the wrong phase-lags. 
In time-dependent convection models, the
control by the convective flux does not allow significant phase-lags inside the
convective zone, which leads to a better agreement with observations.
However, it must be mentioned that rotation through the action of the Coriolis force could 
affect significantly the geometry of the modes in $\gamma$~Dor stars, because their long 
pulsation periods are not much smaller than the rotation periods. 
Moreover, the Reynolds stress tensor perturbations, which were not included in 
\citet{DupretEtal05c}, could also significantly affect the nonadiabatic predictions.
Hence, we must not be surprised
that some disagreements between theoretical and observed amplitude and phases can be found
when the effects of rotation and Reynolds stress perturbations are neglected.

\newpage
\section{Brief Discussion and Prospects}
\label{sec:prospects}

This review provides only a small cross section of the complex physics of how stellar
pulsations are coupled to the convection and how simplified convection models can describe most 
of the relevant processes of this interaction.
The discussions concentrated preliminarily on one-dimensional (1D) modelling, yet we know 
that convection is an inherently three-dimensional process, such as
vortex-stretching which is believed to be the 
major nonlinear mechanism for transferring turbulent kinetic energy from larger to smaller
scales, at least within the so-called inertial range of the turbulent kinetic energy spectrum.
Three-dimensional (3D) hydrodynamical simulations do become more accessible now, 
thanks to the ever increasing calculation speed of modern computers, with many
astrophysical applications such as modelling star formation, accretion disks, or 
supernovae explosions.
In the context of stellar structure and dynamics, several 3D numerical codes 
are now available to simulate either the outer atmospheric stellar layers in a rectangular box 
\citep[e.g.,][]{SteinNordlund00, WedemeyerEtal04, TrampedachEtal14a, MagicEtal15}, 
cone-like geometries \citep[e.g.,][]{MuthsamEtal10, MundprechtEtal15}, or even the whole star
\citep[e.g.,][]{ElliottEtal00, BrunEtal11, BrunEtal14}.

A promising approach today, and also for the near future, is the use 
of 3D simulation results in 1D stellar calculations. For example, 
an interesting approach is to replace the outer layers of 1D (equilibrium) model calculations
by 3D simulations after applying appropriate averages (in space and time) to the 3D results. 
Such a procedure was
adopted, for example, by \citet{RosenthalEtal95, RosenthalEtal99} for estimating the 
so-called surface effects (see Section~\ref{sec:frequency_effects}) on the solar 
acoustic oscillation frequencies, and is now being applied also to other stars by 
various research groups with interesting results to be expected soon.
Another promising approach is to calculate a grid of 1D stellar atmospheric layers as a 
function of surface gravity and (effective) temperature 
obtained from properly averaged 3D simulations \citep{TrampedachEtal14a, TrampedachEtal14b}.
Here the atmospheric structure is provided as a $T-\tau$ relation 
($T$ being temperature and $\tau$ the frequency-averaged optical depth) together with a calibrated 
value of the mixing length for a particular version of the mixing-length 
formulation. This allows a relatively simple integration of 3D simulation
results into 1D stellar evolutionary calculations together with the selection of the correct
adiabat in the deeper convectively unstable surface layers through the adoption of the 
3D-calibrated mixing length. A first application of this approach was recently reported
by \citet{SalarisCassisi15}.
\citet{MundprechtEtal15} used 2D hydrodynamical simulation results to constrain the 
convection parameters of a 1D nonlinear stability analyses of (short-period) Cepheid 
pulsations by adopting the 1D time-dependent convection model by \citet{Kuhfuss86}. 
The interesting outcome of this simulation study 
is that constant assumed convection parameters in 1D models can vary 
up to a factor of 7.5 over the pulsation cycle. The 2D 
simulation results can then be included in the 1D nonlinear stability computations 
by varying the (otherwise constant assumed) 1D convection parameters over the 
pulsation cycle according to the simulations. Hydrodynamical simulations of
pulsations in classical variable stars were also conducted by \citet{GerouxEtal13}.
These are a few of the examples that point towards the direction in which this 
complex field of convection-pulsation interaction is heading.

Although the move to 3D hydrodynamical simulations for describing stellar convection is 
the most promising path to go, we must remain aware of its current shortcomings. 
Firstly, the spatial dimensions of stellar simulations in a box are typically of  
the order of 10~Mm \citep[e.g.,][]{TrampedachEtal14a}, which therefore makes it difficult
to describe the coupling of turbulent convection to oscillation modes of the lowest radial order. 
Secondly, the typical Reynolds numbers $R_{\mathrm{e}}\simeq10^{12}$ in stars 
suggest that 
the ratio $l_{\mathrm{L}}/l_\eta\sim R_{\mathrm{e}}^{3/4}$ of the largest ($l_{\mathrm{L}}$) to the 
smallest ($l_\eta$) scales would require a total meshpoint number $N\simeq10^{27}$ in 
the 3D numerical simulations. With today's super computers, however, the maximum
achievable number of meshpoints $N_{\max}\simeq10^{12}$, and is therefore some 15 magnitudes
too small for what is required to resolve all the turbulent scales of stellar convection.
Consequently, only a very limited range of scales are resolved by today's 3D hydrodynamical
simulations, which are therefore called large-eddy simulations or just LES. LES require  
so-called sub-grid models  for describing the dynamics of the 
numerically unresolved smaller scales of the turbulent cascade. Various models are available. 
The most commonly used models are hyperviscosity and Smagorinsky models.
All sub-grid modes assume that the turbulent transport is a diffusive process.
Hyperviscosity models, for example, use higher derivatives, a purely mathematical device, 
for the diffusion operator 
in the momentum equation, thereby extending the inertial range, which also leads to a better
representation of the dynamics of the larger scales. More detailed discussions on 
large-eddy simulations and sub-grid models in the context of stellar convection 
can be found in, e.g., \citet{Elliott03} and \citet{Miesch05}. 
It is also important to note that the Prandtl number in 3D simulations
is currently about 0.01\,--\,0.25 \citep[e.g.,][]{ElliottEtal00, MieschEtal08}. 
It is therefore substantially larger than the Prandtl number in solar-type 
stars. 

3D hydrodynamical simulations are the best tools we currently have at hand for extending our
knowledge of stellar convection and for
calibrating semi-analytical 1D convection models. 1D models will still be necessary, for 
many years to come, for stellar evolutionary calculations and for both linear and 
nonlinear stability analyses of stellar pulsations.

\newpage
\begin{appendix}
\section{Gough's Model: The Turbulent Fluxes}
\label{app:A}

In order to construct the expressions for the turbulent fluxes,
we should have the following specific model in mind.
The growth of the fluid parcels can be considered to be linear,
at least initially, whereas nonlinear processes may become important
at the end of the parcel's lifetime and eventually responsible for
its breakup. If, however, this final stage of the parcel's existence
is treated as occurring instantaneously, then we may approximate the 
entire evolution of the parcel by its linear growth rate, and use
some mathematical device to account for the nonlinear destruction of
the fluid parcel.
Such a mathematical device can be established in terms of an
eddy survival probability ${\cal P}(r,t,t_0)$, where $t_0$ is the
time at which the eddy was created. Depending on the cause what may
break up an eddy different probabilities can be derived. For example,
\citet{Gough78} sets the disruption probability proportional to the
internal rms vorticity of an convective element, whereas
\citet{Gough12a} and \citet{SmolecEtal13} set this probability 
proportional to the eddy's internal rate of shear.
Here we may follow \citepos{Spiegel63} idea, 
where the probability of
disruption of an eddy that has turned over by a distance $\rd x$ along
its trajectory of length $\ell$ is $\rd x/\ell$. Thus the probability
that the eddy will survive  until a time $t$,  can be set to
\begin{equation}
{\cal P}(r,t,t_0)=
\exp\left[-\int_{t_0}^t\frac{W(t^\prime;t_0)\rd t^\prime}{\ell}\right]\,.
\label{eq:survival-probability}
\end{equation}
Assuming that the initial conditions, or convective fluctuations
at the eddy's creation time, do not significantly contribute to the 
final heat flux, the time dependence of $W$ and $\Theta$ is described
only by the linear growth rate, i.e.,
\begin{equation}
W=\widehat W_0\exp\left[\sigma_{\mathrm c}(t-t_0)\right]
\,,\qquad\ 
\Theta=\widehat\Theta_0\exp\left[\sigma_{\mathrm c}(t-t_0)\right]\,,
\label{eq:flucquant-timedependent}
\end{equation}
and consequently the eddy's survival probability can be approximated as
\begin{equation}
{\cal P}(r,t,t_0)=
\exp\left[-\frac{\widehat W_0\,\mathrm{e}^{\sigma_{\mathrm c}(t-t_0)}}{\sigma_{\mathrm c}\ell}\right]\,,
\label{eq:survival-probability-approx}
\end{equation}
where $\widehat W_0$ and $\widehat\Theta_0$ are determined from the 
Eqs.~(\ref{eq:dog-lmlt-moment-expand}) and (\ref{eq:dog-lmlt-engy-expand}), respectively.

The turbulent fluxes are constructed in terms of the survival probability
by the following integral expressions
\begin{equation}
F_{\mathrm c}=n\,m\,c_p\int_{-\infty}^tW\Theta{\cal P} \rd t_0\,,
\label{eq:heat-flux-integral}
\end{equation}
\begin{equation}
p_{\mathrm t}=n\,m\int_{-\infty}^tW^2{\cal P} \rd t_0\,,
\label{eq:momentum-flux-integral}
\end{equation}
where $n$ is the creation rate per unit volume of the convective
eddies, each having mass $m$. In a statistically steady state,
where as many eddies are created as destroyed, the following
relation holds
\begin{equation}
n\,m\int_{-\infty}^t{\cal P}\rd t_0=\rho\,.
\label{eq:eddy-creation-rate}
\end{equation}
Because the initial conditions of the eddies are unimportant, i.e., 
the amplitudes of $W_0$ and $\Theta_0$ are small compared to the average
values of $W$ and $\Theta$, and because $\exp(\sigma_{\mathrm c}\tau)\gg 1$, with $\tau$ being the 
mean lifetime of the eddy, the eddy creation rate $n\,m$ can be expressed with
the help of \eq{eq:survival-probability-approx} as
\begin{equation}
n\,m=\rho\tau^{-1}=\rho\sigma_{\mathrm c}\chi\,,
\label{eq:eddy-creation-rate1}
\end{equation}
where $\chi$ is a numerical constant, which can be calibrated from
the turbulent fluxes, obtained from the 
Eqs.~(\ref{eq:dog-lmlt-moment-expand}) and (\ref{eq:dog-lmlt-engy-expand}), 
yielding $\chi=1/2$.

\newpage
\section{Gough's Model: Coefficients for the Perturbed Convection Equations}
\label{app:Gough-coefficients}

Below, we summarize the coefficients for the linearized perturbations of the turbulent
fluxes (\ref{eq:heat-flux-perturb}, \ref{eq:momentum-flux-perturb}) and eddy shape parameter
(\ref{eq:shape-param-perturb}), following \citet{Gough77a} and \citet{BakerGough79}
(overbars of mean values are omitted):
\begin{eqnarray}
\hseq
W_{10}=&&
\frac{\delta g}{g_0}+\frac{\delta(\hat\delta)}{\hat\delta_0}+\frac{\delta(\Delta)}{\Delta_0}
-\epsilon^2\kappa_{10}-(1-\epsilon^2)\mu_{10}
+\frac{2\mathrm{i}}{\tilde\sigma}
\frac{(1+\epsilon)\mu_{11}-\epsilon(1-\mathrm{i}\tilde\sigma)\kappa_{11}}
      {2-\mathrm{i}\tilde\sigma(1-\epsilon)}\cr
&&+2(\Phi_0-1)\frac{\delta\chi}{\chi_0} \,,
\hspace{298.6pt}
\label{e:W10}
\end{eqnarray}
\smallskip
\begin{equation}
\hseq
W_{11}=-\frac{2\mathrm{i}}{\tilde\sigma}\,
\frac{(1+\epsilon)\mu_{11}-\epsilon\kappa_{11}}
     {2+\mathrm{i}\tilde\sigma(1-\epsilon)} \,,
\hsdw
\label{e:W11}
\end{equation}
\begin{equation}
\hseq
W_{12}=(1+\epsilon)\mu_{10}-\epsilon\kappa_{10} \,,
\hsdw
\label{e:W12}
\end{equation}
\begin{equation}
\hseq
W_{21}=
-2\left[\left(W_{11}-\frac{\delta\ell}{\ell_0}\right)\left(1+\mathrm{i}\frac{\omega}{\tilde\sigma}\right)^{-1}-\dror\right]\,,
\hsdw
\label{e:W21}
\end{equation}
\begin{equation}
\hseq
\Theta_{10}=W_{10}+W_{12}+\Phi_{10}\,,
\hsdw
\label{e:T10}
\end{equation}
\begin{equation}
\hseq
\Theta_{11}=(1+\mathrm{i}\tilde\sigma)W_{11}+\Phi_{11}+{\dtot}
              -\frac{\delta g}{g_0}-\frac{\delta(\hat\delta)}{\hat\delta_0}
              -\mathrm{i}\tilde\sigma\Phi_0^{-1}\left(\ddod+2\dror\right)\,,
\hsdw
\label{e:T11}
\end{equation}
\begin{equation}
\hseq
\Theta_{12} = W_{12}\,,
\hsdw
\label{e:T12}
\end{equation}
\begin{equation}
\hseq
\Phi_{10}=-2\left(1-\Phi_0^{-1}\right)\left(3\dror+\ddod+\Phi_0\frac{\delta\chi}{\chi_0}\right)\,,
\hsdw
\label{e:P10}
\end{equation}
\begin{equation}
\hseq
\Phi_{11}= 2\left(1-\Phi_0^{-1}\right)\left(3\dror+\ddod\right)\,,
\hsdw
\label{e:P11}
\end{equation}
where
\begin{equation}
\hseq
\frac{\delta g}{g_0}=-\left(2+\frac{\omega^2\,r_0^3}{G\,m}\right)\dror\,,
\hsdw
\label{e:g1}
\end{equation}
\begin{equation}
\hseq
2\mu_{10} = -\Phi_{10}\,,
\hsdw
\label{e:mu10}
\end{equation}
\begin{eqnarray}
\hseq
2\mu_{11}=&&
 \hat\delta_p\dpop+(\hat\delta_T-1)\dtot-\Phi_0^{-1}\tilde\sigma(1+\epsilon)^{-1}
 [\tilde\sigma(1-\epsilon)-2\mathrm{i}\epsilon]\left(\ddod+2\dror\right)\cr
&&-2\left(1-\Phi_0^{-1}\right)\left(\ddod+3\dror\right)
 +\frac{\delta\beta}{\beta_0}+\frac{\delta g}{g_0}\,,
\hspace{192.0pt}
\label{e:mu11}
\end{eqnarray}
\begin{equation}
\hseq
2\mu_{12}=\frac{\delta g}{g_0}+\frac{\delta(\hat\delta)}{\hat\delta_0}
+\frac{\delta\beta}{\beta_0}-\dtot+2(\Phi_0-1)\frac{\delta\chi}{\chi_0}\,,
\hsdw
\label{e:mu12}
\end{equation}
\begin{equation}
\hseq
\kappa_{10}=-2\phi_0\left[\left(2-3\Phi_0^{-1}\right)\dror+\left(1-\Phi_0^{-1}\right)\ddod
               +\frac{\delta H_{\mathrm{p}}}{H_{\mathrm{p0}}}-\frac{\delta\chi}{\chi_0}\right]\,,
\hsdw
\label{e:k10}
\end{equation}
\begin{eqnarray}
\hseq
\kappa_{11}&=&
2\phi_0\left[\left(2-3\Phi_0^{-1}\right)\dror-\Phi_0^{-1}\ddod\right]+3\dtot+(1-2\phi)\dkok\cr
&&-\frac{\delta c_{\mathrm{p}}}{c_{\mathrm{p0}}}+\frac{\mathrm{i}}{2}\tilde\sigma\left(\epsilon^{-1}-1\right)
\Biggl\lbrack(1-\hat\delta_0+c_{p_{T}})\dtot-\hat\delta_{T}\frac{\nabla_{ad}}{1-\nu_1}\dpop-\frac{\delta g}{g_0}\cr
&&-\frac{\delta(\hat\delta)}{\hat\delta_0}+2(3-4\Phi_0^{-1})\dror+(2-3\Phi_0^{-1})\ddod
\Biggr\rbrack
\,,
\hspace{182.14pt}
\label{e:k11}
\end{eqnarray}
\begin{equation}
\hseq
\kappa_{12}=3\dtot-2\phi\left(\frac{\delta\ell}{\ell_0}+\ddod
            -\frac{\delta\chi}{\chi_0}\right)+(1-2\phi)\dkok
            -\frac{\delta c_{\mathrm{p}}}{c_{\mathrm{p0}}}\,,
\hsdw
\label{e:k12}
\end{equation}
\begin{equation}
\hseq
\frac{\delta(\Delta)}{\Delta_0}
=\frac{\delta\ell}{\ell_0}
+\frac{\delta\beta}{\beta_0}-\dtot
+\epsilon(1-\epsilon)(\mu_{12}-\kappa_{12})\,,
\hsdw
\label{e:DOD}
\end{equation}
\begin{eqnarray}
\hseq
\frac{\delta\chi}{\chi_0}&=&
\frac{1}{8}
\Biggl\lbrack
  \frac{\delta(\hat\delta)}{\hat\delta_0}-\left(2+\frac{\omega^2}{\Omega^2}\right)\dror
  +2\left(
    \frac{\delta c_{\mathrm{p}}}{c_{\mathrm{p0}}}-\frac{\delta\varkappa}{\varkappa_0}
    \right)
  +4\left(\alpha_0\dpgopg-\hat\delta_0\dtot\right)\cr
 &&+4\frac{\delta\ell}{\ell_0}+\frac{\delta\beta}{\beta_0}-7\dtot
\Biggr\rbrack\,,
\hspace{278.7pt}
\label{e:flcshp}
\end{eqnarray}
\begin{eqnarray}
\hseq
\frac{\delta\beta}{\beta_0}&=&
\left(1+\frac{\hat\delta_0 g_0}{\mut\beta_0 c_{\mathrm{p0}}}\right)
\left(2\dror+\alpha_0\dpgopg-\hat\delta_0\dtot\right)+
\frac{\hat\delta_0 g_0}{\mut\beta_0 c_{\mathrm{p0}}}
\Biggl\lbrace
   \left(
      \frac{\rho_0c_{\mathrm{p0}}T_0}{\hat\delta_0p_0}\nabla+\nu_2\delta_0
   \right)\dtot\cr
  &&+\frac{\rho_0c_{\mathrm{p0}}T_0}{\hat\delta_0p_0}
   \frac{\partial}{\partial\ln p_0}\left(\dtot\right)+(1-\nu_2)
   \left(
     \frac{\delta c_{\mathrm{p}}}{c_{\mathrm{p0}}}-\frac{\delta(\hat\delta)}{\hat\delta_0}
   \right)
   +\mut\left(2+\frac{\omega^2r_0^3}{Gm}\right)\dror\cr
  &&+\nu_1\frac{\partial}{\partial\ln p_0}(\dptopt)
   +\nu_2(\dptopt-\alpha_0\dpgopg)
\Biggr\rbrace\cr
&&-\frac{\nu_1\hat\delta_0p_0}{\beta_0\rho_0c_{\mathrm{p0}}r_0}
\left[(3-\Phi_0)\left(\dptopt-\alpha_0\dpgopg+\hat\delta_0\dtot-\dror\right)-\delta\Phi\right]
\,,
\hspace{107.0pt}
\label{e:flcbet}
\end{eqnarray}
\begin{eqnarray}
\hseq
\frac{\delta H_{\mathrm{p}}}{H_{\mathrm{p0}}}&=&
\dpop-\mut
\Biggl\lbrace
  \alpha_0\dpgopg-\hat\delta_0\dtot-\left(2+\frac{\omega^2r_0^2}{Gm}\right)\dror\cr
 &&+\nu_1\frac{p_0r_0}{Gm\rho_0}
  \left[(3-\Phi_0)\left(\dptopt-\dror\right)-\delta\Phi\right]
\Biggr\rbrace\,,
\hspace{175pt}
\label{e:flcmlt}
\end{eqnarray}
and where $\epsilon:=(1+\eta_0^2S)^{-1/2}$ 
[see Eq.~(\ref{eq:dog-lmlt-conv-sigma})], $\alpha_0:=(\partial\ln\rho/\partial\ln p)_T$, 
$\tilde\mu:=[1+(3-\Phi)p_{\mathrm{t}}/g\rho r]^{-1}$, $\tilde\sigma:=\omega/\sigma_{\mathrm{c}}$,
$\nu_1:=p_{\mathrm{t}}/\tilde p$, $\nu_2: =\rd p_{\mathrm{t}}/\rd \tilde p$ with
$\tilde p:=p+p_{\mathrm{t}}$, and $\delta\ell/\ell_0$ is obtained from
Eq.~(\ref{eq:verti-wavenumber-puls}) using definition~(\ref{eq:vertical-wavenumber}).
\bigskip

\noindent
The remaining functional expressions ${\cal F},\;{\cal G}$, and ${\cal H}$ 
in Eqs.~(\ref{eq:heat-flux-perturb}), (\ref{eq:momentum-flux-perturb}) and 
(\ref{eq:shape-param-perturb}) are defined as
\begin{equation}
\hseq
{\cal F}={\cal I}\Gamma(2-\mathrm{i}\tilde\sigma)\,,
\hsdw
\label{e:scrF}
\end{equation}
\begin{equation}
\hseq
{\cal G} = {\cal J/I}+\digamma(2-\mathrm{i}\tilde\sigma)\,,
\hsdw
\label{e:scrG}
\end{equation}
\begin{equation}
{\cal H}=
\left[2\dror+(1+\epsilon)\mu_{12}-\epsilon\kappa_{12}\right]{\cal F}
-(2-\mathrm{i}\tilde\sigma)(W_{10}-W_{12}){\cal F}
-W_{12}{\cal F}[1+(2-\mathrm{i}\tilde\sigma){\cal G}]\,,
\label{e:scrH}
\end{equation}
where
\begin{equation}
\hseq
{\cal I}= 107\{E_1[2.88(1+\mathrm{i}\tilde\sigma)]-320E_1[2.88(3+\mathrm{i}\tilde\sigma)]\}\,,
\hsdw
\label{e:scrI}
\end{equation}
\begin{equation}
\hseq
{\cal J}=\mathrm{i}\frac{\rd {\cal J}}{\rd \tilde\sigma}=
\frac{12}{(1+\mathrm{i}\tilde\sigma)(3+\mathrm{i}\tilde\sigma)}\,\left(\frac{5^{1/2}s}{2}\right)^{\mathrm{i}\tilde\sigma}\,,
\hsdw
\label{e:scrJ}
\end{equation}
and where $\Gamma$, $\digamma$ are the gamma and digamma functions, $E_1$ the
exponential integral of first order, and $s=0.05$.

\newpage
\section{Grigahc{\`e}ne \etal's Model: Perturbation of the Mean Structure}
\label{app:meanpert}

In this section, we perturb the equations of the mean structure, which gives
the linear nonradial nonadiabatic pulsation equations. As in
Eqs.~(\ref{eq13}) and (\ref{eq18}), coupling terms between convection
and pulsation will appear here (e.g., perturbation of the convective
flux). They will be evaluated in Section~\ref{convpert}. 
The Lagrangian variation of any quantity $y$ is denoted, for a given
spheroidal mode, by
\[
\delta y\left( \vec{r},t\right) =\delta y\left( r\right) \exp \left( -\mathrm{i}\omega
t\right) Y_{l }^{m}\left( \theta ,{\varphi} \right), 
\]
where $\omega$ is the angular frequency and $Y_{l }^{m}$ the spherical
harmonic of spherical $l$ and azimuthal order $m$.
In order to be able to distinguish global perturbations from convective
motion, we must consider $l$ values small enough so that $r/l \gg \ell$.
From now on we shall omit overbars of mean quantities where possible.

The perturbed equation of mass conservation is (overbars of mean values are omitted)
\begin{equation}
\frac{\delta \rho}{\rho}+\frac{1}{r^{2}}\frac{\partial }{\partial r}\left(
r^{2}\,\xi_{\mathrm{r}}\right) =l \left( l +1\right) \frac{\xi_{\mathrm{h}}}{r}\,,
\label{mass}
\end{equation}
where
$\vec{\xi}=(\xi_{\mathrm{r}},{\xi}_{\mathrm{h}})$ is the displacement vector with
$\xi_{\mathrm{r}}$, $\xi_{\mathrm{h}}$ being the amplitude functions as defined by
\citet{UnnoEtal89}.

The equation of motion is obtained from perturbing Eq.~(\ref{eq13}), i.e.,
\begin{eqnarray}
-\omega ^{2}\rho\,\vec{\xi} &=&-\delta \rho\nabla {\Psi} -\nabla \left( \delta
p_{\mathrm{g}}+\delta p_{\mathrm{R}} +\delta p_{\mathrm{t}%
}\right)  \nonumber \\
&&-\nabla \vec{\xi} \cdot\nabla \cdot \tens{\sigma}_{\mathrm{t}}-\rho
\,\nabla \delta {\Psi} -\delta \left( \nabla \cdot \tens{\sigma}_{\mathrm{t}%
}\right)\,
\end{eqnarray}
where $\Psi$ is the gravitational potential obtained from the Poisson equation.
From the definition of $\Phi$ [Eq.~(\ref{anisotropyeq})], we have
$\nabla \cdot \tens{\sigma}_{\mathrm{t}}=(3-{\Phi}) p_{\mathrm{t}}/r\;\vec{e}_{r}$.
For the perturbation of the divergence of the Reynolds stress tensor we use the
notation 
\begin{eqnarray}
\delta \left( \nabla \cdot \tens{\sigma}_{\mathrm{t}}\right) &=& \Xi
_{\mathrm{r}}\left( r\right) Y_{l }^{m}\left( \theta ,{\varphi} \right)\vec{e}_{\mathrm{r}} \nonumber\\
&+&\Xi_{\mathrm{h}}\left( r\right) \left(r\, \vec{\nabla}_{\mathrm{h}}Y_{l}^{m}\left( \theta
,{\varphi} \right) \right),\label{reynoteq}
\end{eqnarray}
where $\vec{\nabla}_{\mathrm{h}}$ is the transverse component of the gradient.
The radial component of the equation of motion then becomes
\begin{eqnarray}
\omega ^{2}\xi_{\mathrm{r}}&=&\frac{\rd \delta {\Psi} }{\rd r}
+\frac{1}{%
\rho }\frac{\rd }{\rd r}\left( \delta p_{\mathrm{g}}+\delta p_{%
\mathrm{R}}+\delta p_{\mathrm{t}}\right)  \nonumber \\
&&+\, g\frac{\delta \rho }{\rho }+(3-{\Phi})\frac{p_{\mathrm{t}}}{r\rho }%
\frac{\rd \xi_{\mathrm{r}}}{\rd r}+\frac{\Xi_{\mathrm{r}}}{\rho } \,,  \label{pmm}
\end{eqnarray}
where $g={\rd\Psi/\rd r}$ is the gravitational acceleration. 
The transverse component of the equation of motion is 
\begin{eqnarray}
\omega ^{2}r\,\xi_{\mathrm{h}}&=& \delta {\Psi} +\frac{r\,\Xi_{\mathrm{h}}}{\rho}+\frac{\delta
p_{\mathrm{g}}+\delta p_{\mathrm{R}}+\delta p_{\mathrm{t}}%
}{\rho}  \nonumber \\
&&+\, (3-{\Phi})\frac{p_{\mathrm{t}}}{\rho }\left( \frac{\xi_{\mathrm{r}}}{r}-%
\frac{\xi_{\mathrm{h}}}{r}\right)\,.  \label{pmmt}
\end{eqnarray}

From the perturbation of the energy equation~(\ref{eq18}),
we obtain
\begin{eqnarray}
-{\mathrm{i}}\omega T\delta s &=&-\,\frac{\rd \,\delta \left( L_{%
\mathrm{R}}+L_{\mathrm{c}}\right) }{\rd m} 
+\left[ \frac{\delta \varepsilon }{\varepsilon }+l \left( l +1\right) 
\frac{\xi_{\mathrm{h}}}{r}\right] \varepsilon  \nonumber \\
&+&\frac{l \left( l +1\right) }{4\pi r^{3}\rho }\left[ L_{%
\mathrm{R}}\left( \frac{\delta T}{r\left( \rd T/%
\rd r\right) }-\frac{\xi_{\mathrm{r}}}{r}\right) -L_{\mathrm{c}}\frac{\xi_{\mathrm{h}}%
}{r}\right]  \nonumber \\
&+&\frac{l \left( l +1\right) }{\rho r}\delta F_{\mathrm{c, h}}+\delta
\left( \epsilon _{\mathrm{t}}+\frac{\overline{\vec{u}\cdot\nabla \left( p_{\mathrm{g}%
}+p_{\mathrm{R}}\right) }}{\overline{\rho }}\right)\,,
\label{eq32}
\end{eqnarray}
where $L(r)=4\pi r^2F(r)$. In the diffusion approximation we have
\begin{equation}
\frac{\delta L_{\mathrm{R}}}{L_{\mathrm{R%
}}}=2\frac{\xi_{\mathrm{r}}}{r}+3\frac{\delta T}{T}-\frac{\delta \kappa }{\kappa }-%
\frac{\delta \rho }{\rho }+\frac{\rd \delta T/\rd r}{\rd T/%
\rd r}-\frac{\rd \xi_{\mathrm{r}}}{\rd r}.  \label{diff}
\end{equation}

The diffusion approximation is not valid in stellar atmospheres. 
Other approximations must therefore be adopted. For example, the Eddington approximation
\citep{UnnoSpiegel66} is often 
used in pulsating atmospheres \citep[e.g.,][]{Gough77a}. 
Another possibility was proposed by \citet{DupretEtal02}.

For the convective heat flux, we use the following notation: 
\begin{eqnarray}
\delta \vec{F}_{\mathrm{c}} &=&\delta F_{\mathrm{c, r}}\left( r\right)
\,Y_{l }^{m}\left( \theta ,{\varphi} \right) \,\vec{e}_{r}  \nonumber \\ 
&&+\:\delta F_{\mathrm{c, h}}\left( r\right) \,\,\left( r\nabla _{h}Y_{l
}^{m}\left( \theta ,{\varphi} \right) \right) \,.  \label{dfcdef}
\end{eqnarray}
%
\newpage
\section{Grigahc{\`e}ne \etal's Model: Convection Equations}
\label{app:G05_conv}

Subtracting the mean equations~(\ref{eq8}), (\ref{eq13}) and (\ref{eq18}) from the
instantaneous 
equations~(\ref{eq4}), (\ref{eq5}) and (\ref{eq6}), provide the equations 
for the (Eulerian) convective fluctuations.

The difference between Eqs.~(\ref{eq4}) and (\ref{eq8}) provides the fluctuating
continuity equation 
\begin{equation}
\overline{\rho }\frac{\rd }{\rd t}\left( \frac{\rho' }{%
\overline{\rho }}\right) +\nabla \cdot \left( \rho \vec{u}\right) =0 .
\label{eq:conti_equ_ba}
\end{equation}
In the Boussinesq approximation the density fluctuations are neglected
in the continuity equation, leading to \eq{eq*}, i.e., $\nabla\cdot\vec{u}=0$. 


Taking the difference between Eqs.~(\ref{eq5}) and (\ref{eq13}), and using
Eq.~(\ref{eq8}), leads to the fluctuating momentum equation 
\begin{eqnarray}
\overline{\rho }\frac{\rd }{\rd t}\left( \frac{\rho \vec{u}}{%
\overline{\rho }}\right) 
&=&-\rho (\vec{u}\cdot \nabla) \vec{U}%
    -\nabla \cdot \left( \rho \vec{uu} - \bm{\uptau} \right)
    +\frac{\rho}{\ob{\rho}}\nabla\cdot\ob{\left( \rho \vec{uu} - \bm{\uptau} \right)}
    + \rho^\prime\left(\vec{{g}}-\frac{\rd\vec{U}}{\rd t}\right)\nonumber\\
&\simeq&
   -\rho (\vec{u}\cdot \nabla) \vec{U}-\nabla\cdot(\rho\vec{u}\vec{u}-\ob{\rho\vec{u}\vec{u}})
   -\nabla p^\prime + \rho^\prime\left(\vec{{g}}-\frac{\rd\vec{U}}{\rd t}\right) \,,
\label{mvtconv}
\end{eqnarray}
where we neglected, as we did earlier in the mean momentum Eq.~(\ref{eq13}),
the divergence of $\bm{\upsigma}$ and its average.


Taking the difference between Eqs.~(\ref{eq6}) and (\ref{eq18}), we
obtain the fluctuating thermal energy equation 
\begin{equation} 
\ob{\rho }\frac{\rd }{\rd t}\left( \frac{\rho e}{\ob{\rho }}-\ob{e}\right) 
+\nabla \cdot \left( \rho { h}\vec{u}-\ob{\rho { h}\vec{u}}\right)
-\vec{u}\cdot \nabla p+\ob{\vec{u}\cdot \nabla p}
+p^\prime \nabla \cdot \vec{{ U}}
{-{\bm{\upsigma}\!:\!\nabla\vec{u}}+\ob{\bm{\upsigma}\!:\!\nabla\vec{u}}}
=\rho \varepsilon -\overline{\rho \varepsilon }
-\nabla \cdot \vec{F}^\prime_{\mathrm{R}}\,, \nonumber
\end{equation}
where $\ob{\bm{\upsigma}:\nabla\vec{u}}=:\ob{\rho} \epsilon_{\mathrm{t}}$.
In the Boussinesq approximation
\begin{equation}
\ob{\rho}\,\ob{T}\frac{\rd s^\prime}{\rd t}
+\left( \rho T\right)^\prime\frac{\rd\ob{s}}{\rd t}
{
+\ob{\rho}\left[\vec{u}\cdot(T\nabla s)^\prime-\ob{\vec{u}\cdot(T\nabla s)^\prime}\right]
-{\bm{\upsigma}\!:\!\nabla\vec{u}}+\ob{\bm{\upsigma}\!:\!\nabla\vec{u}}
+\ob{\rho}\,\ob{T}\nabla\ob{s} \cdot \vec{u}
}
=\rho \varepsilon-\ob{\rho \varepsilon}-\nabla \cdot \vec{F}^\prime_{\mathrm{R}}\,,
\label{enconv}
\end{equation}
where the radial component of 
$\ob{\rho}\,\ob{T}\nabla\ob{s} \cdot \vec{u}=-\ob{\rho}\,\ob{c_p}\beta w$ 
($w$ is the vertical component of the turbulent velocity field $\vec{u}$ and $\beta$ is defined
by Eq.~(\ref{eq:beta})).
The nonlinear terms in Eqs.~(\ref{mvtconv}) and (\ref{enconv}) need to be approximated
by appropriate closure assumptions in order to obtain a closed system of equations for the
convective fluctuations $\vec{u}$ and $s^\prime$.
 
Following \citet{Unno67}, we adopt approximation~(\ref{eq20}) for the nonlinear 
terms in \eq{mvtconv} leading to expression~(\ref{eq23}), where we also used
the Boussinesq equation of state
\begin{equation}
\frac{\rho^\prime}{\ob{\rho}}=-\ob{\hat\delta}\:\frac{s'}{\ob{c_p}}
=-\ob{\hat\delta}\:\frac{T' }{\ob{T}},
\label{dtds}
\end{equation}
with $\hat\delta=-({\partial \ln \rho }/{\partial \ln T})_{p}$ being the isobaric 
expansion coefficient.
{
Similarly, the nonlinear terms in the thermal energy equation can be approximated 
by expression (\ref{closen}) \citep{Gabriel96} leading to
\eq{eq:fluc_thermal_nonradial}.
}

\newpage
\section{Grigahc{\`e}ne \etal's Model: Perturbation of the Convection Equations}
\label{sec:nonradial-confluc}

Linear perturbation of Eq.~(\ref{omr}) leads to (overbars of mean values are omitted)
\begin{equation}
\label{pomr}
\frac{\delta \omega _{\mathrm{R}}}{\omega _{%
\mathrm{R}}}=3\frac{\delta T}{T}-\frac{\delta c_{\mathrm{p}%
}}{c_{\mathrm{p}}}-\frac{\delta \kappa }{\kappa}-2\frac{\delta \rho }{\rho }%
-2\frac{\delta\ell}{\ell}.
\end{equation}
From now on, we shall use the notations 
\begin{eqnarray*}
B &=& \frac{-{\mathrm{i}}\omega \tau _{\mathrm{c}}+\Lambda }{\Lambda }, \\
C &=& \frac{\omega _{\mathrm{R}}\tau _{\mathrm{c}}+1}{%
-{\mathrm{i}}\omega \tau _{\mathrm{c}}+\omega _{\mathrm{R}}\tau _{%
\mathrm{c}}+1}, \\
D &=& \frac{1}{-{\mathrm{i}}\omega \tau _{\mathrm{c}}+\omega _{\mathrm{R%
}}\tau _{\mathrm{c}}+1}.
\end{eqnarray*}
We isolate $(\delta s')/s'$ in Eq.~(\ref{enconv0}), 
 multiply the equation by $u_j$, and integrate over all $k$ such that
$k_{\theta }^{2}+k_{{\varphi} }^{2}=k_{\mathrm{r}}^{2}/({\Phi}-1)$. After some algebra we obtain
\begin{eqnarray}
\frac{\overline{u_{j}\delta s'}}{\overline{u_{\mathrm{r}}s'}}&=&D\frac{\overline{%
u_{\mathrm{r}}u_{j}}}{\overline{u_{\mathrm{r}}^{2}}} \left[ \mathrm{i}\omega \tau _{\mathrm{c}}(1-\tdel)%
\frac{\delta {s}}{c_{\mathrm{p}}}+\frac{\delta \tau _{\mathrm{c}}}{%
\tau _{\mathrm{c}}}-\omega _{\mathrm{R}}\tau _{\mathrm{c}}%
\frac{\delta \omega _{\mathrm{R}}}{\omega _{%
\mathrm{R}}}\right]   \nonumber \\
&&+\:C\left[ \left( \frac{\nabla _{k}\left( \delta {s}\right) }{%
\left( \rd s/\rd r\right) }-\nabla _{k}\xi _{\mathrm{r}}\right) \frac{%
\overline{u_{j}u_{k}}}{\overline{u_{\mathrm{r}}^{2}}}+\frac{\overline{u_{j}\delta u_{\mathrm{r}}}}{\overline{u_{\mathrm{r}}^{2}}}\right] ,  \label{dsh}
\end{eqnarray}
where we use the Einstein convention for repeated indices ($j, k$) summation.

The perturbation of the radial component of the turbulent velocity is
\begin{eqnarray}
\frac{\overline{u_{\mathrm{r}}\delta u_{\mathrm{r}}}}{\overline{u_{\mathrm{r}}^2}} 
&=& \frac{1}{B+(-{\mathrm{i}}\omega \tau _{\mathrm{%
c}}+1)D}  \nonumber \\
&&\times \left\{ -\frac{\delta c_{\mathrm{p}}}{c_{\mathrm{p}}}+\frac{\delta (\tdel)}{%
\delta}-\frac{\delta \rho }{\rho }+\frac{\rd \delta (p+p_{\mathrm{t}})}{\rd (p+p_{\mathrm{t}})}-\frac{%
\rd \xi _{\mathrm{r}}}{\rd r} \right.  \nonumber \\
&&{+}{\mathrm{i}}\omega \tau _{\mathrm{c}} D (1-\delta)\frac{\delta s}{%
c_{\mathrm{p}}} +C\left[ \frac{\rd \delta s}{\rd s}-\frac{%
\rd \xi _{\mathrm{r}}}{\rd r}\right]  \nonumber \\
&&+\frac{{\mathrm{i}}\omega \tau _{\mathrm{c}}}{\Lambda\,{\Phi} } \left( \frac{%
\rd \xi _{\mathrm{r}}}{\rd r}+({\Phi}-1)\frac{\xi _{\mathrm{r}}}{r}-l
\left( l +1\right)({\Phi}-1)\frac{\xi _{\mathrm{h}}}{2 r}\right)  \nonumber \\
&&-\omega _{\mathrm{R}}\tau _{\mathrm{c}} D \left( 3\frac{%
\delta T}{T}-\frac{\delta c_{\mathrm{p}}}{c_{\mathrm{p}}}-\frac{\delta
\kappa }{\kappa }-2\frac{\delta \rho }{\rho }\right)  \nonumber \\
&& + \left. (-{\mathrm{i}}\omega \tau _{\mathrm{c}}+3\omega _{\mathrm{R}}
\tau _{\mathrm{c}}+2)D\:\frac{\delta\ell}{\ell} \right\}\,.  \label{eq44}
\end{eqnarray}
The equations separate into spherical harmonics 
\begin{eqnarray}
\frac{\overline{{u}_{\theta}\delta u_{\mathrm{r}}}}{\overline{u_{\mathrm{r}}^2}}
\:+\:\frac{\overline{{u}_{{\varphi}}\delta u_{\mathrm{r}}}}{\overline{u_{\mathrm{r}}^2}}
&=&\frac{\overline{u_{\mathrm{h}}\delta u_{\mathrm{r}}}}{\overline{u_{\mathrm{r}}^2}} \,\left( r\nabla _{\mathrm{h}}Y_{l
}^{m}\left( \theta ,{\varphi} \right) \right) \,,\nonumber\\
\frac{\overline{u_{\mathrm{r}}\delta {u}_{\theta}}}{\overline{u_{\mathrm{r}}^2}}
\:+\:\frac{\overline{u_{\mathrm{r}}\delta{u}_{{\varphi}}}}{\overline{u_{\mathrm{r}}^2}}
&=&\frac{\overline{u_{\mathrm{r}}\delta u_{\mathrm{h}}}}{\overline{u_{\mathrm{r}}^2}} \,\left( r\nabla _{\mathrm{h}}Y_{l
}^{m}\left( \theta ,{\varphi} \right) \right) \,.\nonumber\\
\end{eqnarray}
Calculating moments of Eq.~(\ref{eq34}) provide
\begin{eqnarray}
\frac{\overline{u_{\mathrm{h}}\delta u_{\mathrm{r}}}}{\overline{u_{\mathrm{r}}^2}}
&=& (B-C)^{-1}  \nonumber \\
&&\times\left\{\frac{{\Phi}-1}{2}\left[ \left(\frac{\delta (p+p_{\mathrm{t}})}{\rd (p+p_{\mathrm{t}})/\mathrm{%
d} \ln r} -\frac{\xi_{\mathrm{r}}}{r}+\frac{\xi_{\mathrm{h}}}{r}\right)\right. \right.  \nonumber
\\
&&+\left.C \left(\frac{\delta s}{\rd s/\rd \ln r} -\frac{\xi_{\mathrm{r}}%
}{r}+\frac{\xi_{\mathrm{h}}}{r}\right)\right]  \nonumber \\
&&\left.\frac{{\Phi}-1}{2{\Phi}}\frac{{\mathrm{i}}\omega \tau _{\mathrm{c}}}{\Lambda } \left(%
\frac{\rd \xi_{\mathrm{h}}}{\rd r}+\frac{\xi_{\mathrm{r}}}{r}-\frac{\xi_{\mathrm{h}}}{r}%
\right)\right\}\,,  \label{vhdvr}
\end{eqnarray}
\begin{eqnarray}
\frac{\overline{u_{\mathrm{r}}\delta u_{\mathrm{h}}}}{\overline{u_{\mathrm{r}}^2}}
&=& (2\,B)^{-1} \left[C({\Phi}-1)\left(\frac{\delta s}{\rd s/\rd \ln r} -\frac{\xi_{\mathrm{r}}}{r}+\frac{\xi_{\mathrm{h}}}{r}\right)\right. 
\nonumber \\
&&+\:(2{\Phi}-1)\left(\frac{\delta (p+p_{\mathrm{t}})}{\rd (p+p_{\mathrm{t}})/\rd \ln r} -%
\frac{\xi_{\mathrm{r}}}{r}+\frac{\xi_{\mathrm{h}}}{r}\right)  \nonumber \\
&&\left.+\:\frac{{\mathrm{i}}\omega \tau _{\mathrm{c}}({\Phi}-1)}{\Lambda\,{\Phi}}\left(\frac{\xi_{\mathrm{r}}}{r%
}-\frac{\xi_{\mathrm{h}}}{r} +\frac{2{\Phi}-1}{{\Phi}-1}\frac{\rd \xi_{\mathrm{h}}}{\rd r}\right)\right] 
\nonumber \\
&&+\:\frac{C}{B}\:\frac{\overline{u_{\mathrm{h}}\delta u_{\mathrm{r}}}}{\ob{u_{\mathrm{r}}^2}}\,.
\label{dvh}
\end{eqnarray}

Expressions of the form $\overline{u_k u_l\,\delta u_{j}/u_{\mathrm{r}}^3}$
can be obtained by considering appropriate moments of Eq.~(\ref{eq34}).
We also note the 
useful result: 
\begin{equation}
\label{dvhh}
\frac{\overline{u_{\theta}\,\delta u_{\theta}}}{\ob{u_{\mathrm{r}}^2}} +\frac{\overline{%
u_{{\varphi}}\,\delta u_{{\varphi}}}}{\ob{u_{\mathrm{r}}^2}} = ({\Phi}-1)\frac{\overline{u_{\mathrm{r}}\delta
u_{\mathrm{r}}}}{\overline{u_{\mathrm{r}}^2}}\,.
\end{equation}

\end{appendix}


\newpage
\section*{Acknowledgements}
\label{section:acknowledgements}

We thank J{\o}rgen Chistensen-Dalsgaard and Douglas Gough for many helpful discussions.
GH is grateful to Neil Balmforth for providing a running version of the computer programme
for Gough's nonlocal mixing-length model. We also thank the two anonymous referees for their
very useful comments which improved the manuscript substantially.
Funding for the Stellar Astrophysics Centre is provided by The Danish National Research Foundation
(Grant agreement no.: DNRF106). The research is supported by the ASTERISK project (ASTERoseismic
Investigations with SONG and Kepler) funded by the European Research Council (Grant agreement no.:
267864).

\newpage



\bibliography{LivRevSolar}

\end{document}